%% file: main.tex
\documentclass[preprint,12pt]{elsarticle}
\pdfoutput=1
\usepackage{graphicx,epstopdf,amsmath,amsfonts,amssymb,appendix,comment}
\usepackage{color,slashed,subfigure,braket,multirow}
\usepackage{epsfig,mathrsfs,latexsym,color,url,etoolbox}
\usepackage{bbm}
\usepackage{ulem}
\usepackage{mathtools}
\usepackage{footnote}
\usepackage[usenames,dvipsnames,table]{xcolor}
\usepackage{siunitx}
\usepackage[T1]{fontenc}
\usepackage{bookmark}
\usepackage{microtype}
\usepackage{xspace}
\usepackage{cancel}

\newcommand{\meV}{\ensuremath{\mathrm{meV}}\xspace}
\newcommand{\MeV}{\ensuremath{\mathrm{MeV}}\xspace}
\newcommand{\eV}{\ensuremath{\mathrm{eV}}\xspace}
\newcommand{\Neff}{\ensuremath{N_\mathrm{eff}}\xspace}
\newcommand{\yp}{\ensuremath{Y_\mathrm{P}}\xspace}

\journal{Physics of the Dark Universe}

\begin{document}

\begin{frontmatter}

\date{\today}

\title{Synergy between cosmological and laboratory searches in neutrino physics\tnoteref{t1}}
\tnotetext[t1]{This article is registered under preprint number: arXiv:2203.07377 [hep-ph]}

\author[1]{Martina~Gerbino}
\ead{gerbinom@fe.infn.it}
\author[2]{Evan~Grohs\corref{cor1}}
\ead{ebgrohs@ncsu.edu}
\cortext[cor1]{Corresponding author}
\author[1]{Massimiliano~Lattanzi}
\ead{lattanzi@fe.infn.it}
\author[3]{Kevork~N.~Abazajian}
\author[4]{Nikita~Blinov}
\author[1,5]{Thejs~Brinckmann}
\author[3]{Mu-Chun~Chen}
\author[6]{Zelimir~Djurcic}
\author[7]{Peizhi~Du}
\author[8]{Miguel~Escudero}
\author[9]{Steffen~Hagstotz}
\author[10,11]{Kevin~J.~Kelly}
\author[12]{Christiane~S.~Lorenz}
\author[13]{Marilena~Loverde}
\author[14,15]{Pablo~Mart\'inez-Mirav\'e}
\author[14]{Olga~Mena}
\author[16]{Joel~Meyers}
\author[17]{Walter~C.~Pettus}
\author[18,19]{Ninetta~Saviano}
\author[20,21]{Anna~M.~Suliga}
\author[25,26,22]{Volodymyr~Takhistov}
\author[14,15]{Mariam~T\'ortola}
\author[14]{Jos\'e~W.~F.~Valle}
\author[23,24]{Benjamin~Wallisch}

\affiliation[1]{organization={Istituto Nazionale di Fisica Nucleare (INFN), Sezione di Ferrara},
            postcode={44122}, 
            city={Ferrara},
            country={Italy}}

\affiliation[2]{organization={Department of Physics, North Carolina State University},
            city={Raleigh},
            state={North Carolina},
            postcode={27695}, 
            country={USA}}

\affiliation[3]{organization={Department of Physics and Astronomy, University of California, Irvine},
            city={Irvine},
            state={California},
            postcode={92697}, 
            country={USA}}

\affiliation[4]{organization={Department of Physics and Astronomy, University of Victoria},
            city={Victoria},
            province={British Columbia},
            postcode={V8P 5C2}, 
            country={Canada}}

\affiliation[5]{organization={Dipartimento di Fisica e Scienze della Terra, Universit\`a di Ferrara, Polo Scientifico e Tecnologico},
            postcode={44122}, 
            city={Ferrara},
            country={Italy}}

\affiliation[6]{organization={Argonne National Laboratory},
            city={Argonne},
            state={Illinois},
            postcode={60439}, 
            country={USA}}

\affiliation[7]{organization={New High Energy Theory Center, Rutgers University},
            city={Piscataway},
            state={New Jersey},
            postcode={08854}, 
            country={USA}}

\affiliation[8]{organization={Physik-Department, Technische Universit{\"{a}}t M{\"{u}}nchen},
            postcode={85748}, 
            city={Garching},
            country={Germany}}

\affiliation[9]{organization={Universit\"ats-Sternwarte, Fakult\"at  f\"ur Physik, Ludwig-Maximilians Universit\"at M\"unchen},
            postcode={81679}, 
            city={M\"unchen},
            country={Germany}}

\affiliation[10]{organization={Fermi National Accelerator Laboratory},
            city={Batavia},
            state={Illinois},
            postcode={60510}, 
            country={USA}}

\affiliation[11]{organization={CERN},
            city={Geneva},
            postcode={23}, 
            country={Switzerland}}

\affiliation[12]{organization={Institute for Particle Physics and Astrophysics, ETH Z\"urich},
            postcode={8093}, 
            city={Z\"urich},
            country={Switzerland}}

\affiliation[13]{organization={Department of Physics, University of Washington},
            city={Seattle},
            state={Washington},
            postcode={98195}, 
            country={USA}}

\affiliation[14]{organization={Instituto de F\'isica Corpuscular, CSIC-Universitat de Val\`encia},
            postcode={46980}, 
            city={Paterna},
            country={Spain}}

\affiliation[15]{organization={Departament de F\'isica Te\`orica, Universitat de Val\`encia},
            postcode={46100}, 
            city={Burjassot},
            country={Spain}}

\affiliation[16]{organization={Department of Physics, Southern Methodist University},
            city={Dallas},
            state={Texas},
            postcode={75275}, 
            country={USA}}

\affiliation[17]{organization={Department of Physics, Indiana University},
            city={Bloomington},
            state={Indiana},
            postcode={47405}, 
            country={USA}}

\affiliation[18]{organization={Istituto Nazionale di Fisica Nucleare (INFN), Sezione di Napoli, Complesso Univ.\ Monte S.~Angelo},
            postcode={80126}, 
            city={Napoli},
            country={Italy}}

\affiliation[19]{organization={Scuola Superiore Meridionale, Universit\`a degli studi di Napoli ``Federico~II''},
            postcode={80138}, 
            city={Napoli},
            country={Italy}}

\affiliation[20]{organization={Department of Physics, University of California Berkeley},
            city={Berkeley},
            state={California},
            postcode={94720}, 
            country={USA}}

\affiliation[21]{organization={Department of Physics, University of Wisconsin--Madison},
            city={Madison},
            state={Wisconsin},
            postcode={53706}, 
            country={USA}}

\affiliation[25]{organization={International Center for Quantum-field Measurement Systems for Studies of the Universe and Particles (QUP, WPI), High Energy Accelerator Research Organization (KEK)}, 
            city={Tsukuba},
            province={Ibaraki},
            postcode={305-0801}, 
            country={Japan}}

\affiliation[26]{organization={Theory Center, Institute of Particle and Nuclear Studies (IPNS), High Energy Accelerator Research Organization (KEK)},
            city={Tsukuba},
            province={Ibaraki},
            postcode={305-0801}, 
            country={Japan}}

\affiliation[22]{organization={Kavli Institute for the Physics and Mathematics of the Universe (WPI), UTIAS, The University of Tokyo},
            city={Kashiwa},
            province={Chiba},
            postcode={277-8583}, 
            country={Japan}}

\affiliation[23]{organization={School of Natural Sciences, Institute for Advanced Study},
            city={Princeton},
            state={New Jersey},
            postcode={08540}, 
            country={USA}}

\affiliation[24]{organization={Department of Physics, University of California San Diego},
            city={La Jolla},
            state={California},
            postcode={92093}, 
            country={USA}}

\begin{abstract}

The intersection of the cosmic and neutrino frontiers is a rich field where much discovery space still remains. Neutrinos play a pivotal role in the hot big bang cosmology, influencing the dynamics of the universe over numerous decades in cosmological history.  Recent studies have made tremendous progress in understanding some properties of cosmological neutrinos, primarily their energy density.  Upcoming cosmological probes will measure the energy density of relativistic particles with higher precision, but could also start probing other properties of the neutrino spectra.  When convolved with results from terrestrial experiments, cosmology can become even more acute at probing new physics related to neutrinos or even Beyond the Standard Model (BSM).  Any discordance between laboratory and cosmological data sets may reveal new BSM physics and/or suggest alternative models of cosmology.  We give examples of the intersection between terrestrial and cosmological probes in the neutrino sector, and briefly discuss the possibilities of what different laboratory experiments may see in conjunction with cosmological observatories.

\end{abstract}

\end{frontmatter}

\tableofcontents

\newpage

\input{introduction}

\input{nu_cosmology}

\input{lab_probes}

\input{synergy}

\input{conclusions}

\section*{Acknowledgments}
We would like to thank Martin Hirsch for useful discussion about BSM contributions to the $0\nu2\beta$ decay rate, and Alex Drlica-Wagner for discussion on sterile neutrino constraints from small-scale structure.
M.G.~and M.L.~acknowledge financial support from the INFN InDark project and from the COSMOS network~(www.cosmosnet.it) through the ASI~(Italian Space Agency) Grants 2016-24-H.0 and 2016-24-H.1-2018.
E.G.~is supported by the Department of Energy Office of Nuclear Physics award DE-FG02-02ER41216 and by the National Science Foundation grant No.\ PHY-1430152.
K.N.A.~is supported in part by NSF Theoretical Physics Grant PHY-1915005. 
P.D.~is supported by the US Department of Energy under grant DE-SC0010008. 
M.E.~was supported by a Fellowship of the Alexander von Humboldt Foundation. 
P.M.M., M.T.~and J.W.F.V.~are supported by the Spanish grants PID2020-113775GB-I00 (AEI/10.13039/501100011033) and CIPROM/2021/054~(Generalitat Valenciana).
P.M.M.~is also supported by grant FPU18/04571~(MICIU).
O.M.~is supported by the Spanish grants PID2020-113644GB-I00, PROMETEO/2019/083 and by the European ITN project HIDDeN (H2020-MSCA-ITN-2019//860881-HIDDeN).
J.M.~is supported by the US~Department of Energy under Grant~\mbox{DE-SC0010129}.
W.C.P.\ is supported in part by the US National Science Foundation Grant PHY-2209530.
N.S.~is supported by the research grant number 2017W4HA7S ``NAT-NET: Neutrino and Astroparticle Theory Network'' under the program PRIN~2017 funded by the Italian Ministero dell'Universit\`a e della Ricerca~(MUR) and by the research project TAsP~(Theoretical Astroparticle Physics) funded by the Istituto Nazionale di Fisica Nucleare~(INFN).
A.M.S.~acknowledges the support from the US~National Science Foundation~(Grant No.~PHY-2020275).
V.T.~is supported by the World Premier International Research Center Initiative~(WPI), MEXT, Japan, and the Japanese Society Promotion of Science Kakenhi grant No. 23K13109.
B.W.~was supported by the US~Department of Energy under Grants~\mbox{DE-SC0009919} and~\mbox{DE-SC0019035}, and the Simons Foundation under Grant~SFARI~560536.
This manuscript was originally prepared for submission to the Snowmass 2021 community planning exercise.

\bibliographystyle{elsarticle-num-names} 
\bibliography{biblio}

\end{document}

%% file: introduction.tex
\section{Introduction}

Neutrino physics is both the most elusive corner of the Standard Model of particle physics and a rich field of potential ground-breaking discoveries for the coming decade. To elucidate the origin of neutrino masses, neutrino nature, mass ordering, and interactions represents a major endeavor of a wide spectrum of research fields in physics, and a multi-avenue approach is key to study the neutrino sector. A huge effort is put forward by laboratory searches such as neutrino flavor oscillation experiments, beta-decay and neutrinoless double-beta decay (0$\nu$2$\beta$) experiments. A program is also under way to directly detect the cosmic neutrino background (CNB), i.e., remnant neutrinos from the Big Bang. Neutrino telescopes looking for signatures of astrophysical neutrino sources are spanning the energy range from low scales ($\mathcal{O}$(keV)) up to ultra-high energy events (beyond $\mathcal{O}$(PeV)). In this vibrant landscape, cosmology holds promising chances to be the first to make landmark measurements of neutrino properties in this decade. Any cosmological detection will have to be confirmed in the laboratory in order to claim a robust discovery.

In this article, we provide a snapshot of this multi-probe investigation of neutrino properties and advocate for the need for strengthening the links and foster collaboration opportunities across diverse fields. We present the observational windows of the various probes of neutrino properties and summarize the individual constraining power. We clearly discuss the two main scenarios we might be facing in the coming decade: 1) a concordance scenario where all probes provide results that are in overall agreement; 2) a scenario where two or more probes provide results that are in tension with each other. We stress how, in both scenarios, the comparison and -- where statistically allowed -- combination of diverse probes is key to unveil still unknown neutrino properties. What is more, the synergic approach we advocate for is the primary tool we have to convince ourselves of the robustness of experimental findings, especially those that are more prone to model-dependency and/or instrumental systematic issues. 

The aim of this article is twofold. First, provide an easy-access document to inform different communities on strengths and opportunities in disparate fields. Second, advocate for support to networking and cross-cutting research activities. 

The structure of the article is as follows. We begin with an overview of the outstanding issues in the theory of neutrinos in Sec.~\ref{sec:theory}. In Sec.~\ref{sec:intro_cosmo}, we focus on the cosmological imprints of neutrinos and summarize cosmological constraints on the sum of neutrino masses, the number of neutrino families (including possible sterile states), as well as a variety of beyond-standard-model properties, such as neutrino interactions beyond weak and gravitational couplings. In Sec.~\ref{sec:lab_probes}, we turn to laboratory probes of neutrino physics. We review the status of neutrino flavor oscillation, beta-decay and neutrinoless double-beta decay experiments and summarize experimental results from these searches on the neutrino mass scale, mixing parameters and interactions. In Sec.~\ref{sec:synergy}, we bring together the information presented in the previous sections and discuss at length the possible scenarios mentioned above. We draw our conclusions in Sec.~\ref{sec:concl}.

\section{Outstanding issues in the theory of neutrino physics}\label{sec:theory}

Neutrinos were introduced as massless fermions in the Standard Model (SM)
of particle physics. In particular, left-handed neutrinos appeared in SU(2) doublets together with the corresponding leptons.\footnote{Helicities are
swapped in these considerations for antineutrinos.} At the time, since there was no direct indication for
their mass available, right-handed neutrinos were not introduced (not even as gauge singlets). As such, no gauge invariant renormalizable mass term can be
constructed. Therefore, within the SM picture, there is no leptonic 
mixing nor charge parity (CP) violation.\footnote{CP violation is of intense interest in its connection to the Baryon Asymmetry of the Universe, where CP-violating interactions are one of the three necessary Sakharov conditions to produce such an asymmetry in the early universe~\cite{Sakharov:1967dj}.} However, as we will discuss at length in Sec.~\ref{sec:oscpreliminary}, the observation of neutrino flavor oscillations firmly establishes that neutrinos are massive particles, leaving a variety of open issues in the theory of neutrino physics. 

In the SM extension to incorporate massive neutrinos, neutrinos are produced and observed in a given interaction (flavor) state, which is a quantum superposition of massive eigenstates. The mixing of different mass states is described via the Pontecorvo-Maki-Nagakawa-Sakata (PMNS) matrix $U_{\alpha i}$ (equivalent of the CKM mixing matrix in the quark sector):

\begin{equation}
\label{eq:PMNS}
    \nu_\alpha = \sum_{i=1}^3 U_{\alpha i}^* \nu_i, \quad \alpha=(e,\mu,\tau);
\end{equation}

Over the last two decades, the measurements of neutrino oscillation parameters (mixing matrix elements and squared mass differences) have entered from the discovery phase into precision phase. Albeit unable to tell the absolute mass scale, data from neutrino oscillation experiments constrain oscillation parameters at a few percent level or better and put a lower bound to the allowed mass sum of $\sim 0.06\,\mathrm{eV}$ (see Sec.~\ref{sec:oscpreliminary} for details)~\cite{deSalas:2020pgw,Esteban:2020cvm,Capozzi:2021fjo}. On the other hand, measurements of the end-point spectrum of beta-decay allow for an upper bound on the mass sum\footnote{This number is obtained as three times the current upper bound from KATRIN on the effective electron (anti)neutrino mass, which only negligibly differs from the mass of an individual eigenstate in the quasi-degenerate mass region probed by KATRIN.} of $\sim 2.4\,\mathrm{eV}$~\cite{KATRIN:2022hbd}, which can be further reduced by cosmological observations down to $\sim 0.12\,\mathrm{eV}$~\cite{Planck:2018vyg}. 

Current data posts the following theoretical puzzles: (i) Why neutrino masses are so much smaller compared to the charged fermion masses; (ii) why neutrino mixing angles are large while quark mixing are small? A variety of approaches based on different new physics frameworks have been proposed to address these challenges. In addition to addressing the neutrino mass generation and flavor puzzle, these models can also afford solutions to other issues in particle physics and have implications for cosmology. 

The scale of new physics at which neutrino mass generation occurs is still unknown. It can range from the electroweak scale all the way to a grand-unified-theory scale. Depending on the new physics, it is possible to obtain naturally small neutrino masses both of the Majorana type and of the Dirac type. If one assumes that the Standard Model is a low energy effective theory, the dimension-5 Weinberg operator turns out to be the lowest higher dimensional operator. Given that the Weinberg operator breaks the lepton number by two units, neutrinos are Majorana fermions. There are three possible ways to UV-complete the Weinberg operator depending on whether the portal particle is a SM gauge singlet fermion, a complex weak triplet scalar, or a weak triplet fermion. These are dubbed the Type-I~\cite{Minkowski:1977sc}, II~\cite{Schechter:1980gr,Mohapatra:1980yp}, and III~\cite{Foot:1988aq} seesaw mechanism, respectively. Beyond the three types of seesaw mechanisms, small neutrino masses can also be generated radiatively~\cite{Zee:1985rj, Babu:1988ki}, or through the $R$-parity breaking $B$-term in the minimal supersymmetric model~\cite{deCampos:2007bn, Ji:2008cq}, in addition to the so-called inverse-seesaw mechanism~\cite{Mohapatra:1986bd}. 

For Dirac neutrinos, it is also possible to generate their small masses naturally. In fact, in many new physics models beyond the SM aiming to address the gauge hierarchy problem, suppression mechanisms for neutrino masses can be naturally incorporated. These include warped extra dimension models~\cite{Grossman:1999ra}, supersymmetric models~\cite{Arkani-Hamed:2000oup}, and more recently the clockwork models~\cite{Hong:2019bki}. Even though in some of these models~\cite{Chen:2012jg}, neutrinos are Dirac fermions and all lepton number violating operators with $\Delta L = 2$ are absent to all orders, having been protected by symmetries, there can exist lepton number violation by higher units, leading to new experimental signatures~\cite{Heeck:2013rpa}. 

To address the flavor puzzle, generally there are two approaches. One is the so-called ``Anarchy'' scenario~\cite{Hall:1999sn,deGouvea:2003xe} which assumes that there is no parametrically small parameter, and the observed large mixing angles and mild hierarchy among the masses are consequences of statistics. Even though at low energy the anarchy scenario appears to be rather random, predictions from UV physics, such as warped extra dimension~\cite{Grossman:1999ra,Huber:2002gp} as well as heterotic string models where the existence of some $\mathcal{O}(100)$ right-handed neutrinos are predicted~\cite{Buchmuller:2007zd}, very often can mimic the results of anarchy scenario~\cite{Feldstein:2011ck}. An alternate approach is to assume that there is an underlying symmetry, whose dynamics governs the observed mixing pattern and mass hierarchy. The observed large values for the mixing angles have motivated models based on non-Abelian discrete flavor symmetries\footnote{The three generations of neutrinos transform under the
discrete symmetries, which give rise to relations among the entries of the neutrino mass
matrix, and thus enable non-trivial predictions for the mixing parameters.}. Symmetries that have been utilized include $A_{4}$~\cite{Babu:2002dz,Ma:2005qf}, $A_{5}$~\cite{Kajiyama:2007gx}, $T^{\prime}$~\cite{Chen:2007afa}, $S_{3}$~\cite{Harrison:2003aw}, $S_{4}$~\cite{Altarelli:2009gn}, $\Delta(27)$~\cite{Ma:2006ip}, $Z_{7} \ltimes Z_{3}$~\cite{Luhn:2007sy} and $Q(6)$~\cite{Babu:2004tn}. Certain non-Abelian discrete symmetries also afford a novel origin for CP violation. Specifically, CP violation can be entirely group theoretical in origin~\cite{Chen:2009gf}, due to the existence of complex Clebsch-Gordan coefficients in certain non-Abelian discrete symmetries~\cite{Chen:2014tpa}. In addition, these discrete (flavor) symmetries may originate from extra dimension compactification~\cite{Altarelli:2005yx, Criado:2018thu}. More recently, models based on modular flavor symmetries~\cite{Feruglio:2017spp} have been proposed to understand the pattern of neutrino mixing. This approach has been shown to be promising due to its improved predictivity as compared to the traditional flavor symmetries.

In addition to predictions that can be tested at particle physics experiments, models of neutrino masses and mixing also have interesting implications for cosmology. In particular, baryogenesis via leptogenesis is closely connected to the neutrino mass generation mechanisms. The realization of leptogenesis depends on whether neutrinos are Majorana or Dirac fermions: for Majorana neutrinos, leptogenesis proceeds through the decays of right-handed neutrinos~\cite{Fukugita:1986hr,Luty:1992un}, where for Dirac neutrinos, the so-called Dirac leptogenesis~\cite{Dick:1999je} is achieved due to the late-time left-right equilibration of neutrinos, as dictated by the small neutrino Yukawa couplings. In addition, the amount of the generated asymmetry is sensitive to the scale of neutrino mass generation, and may be correlated with other neutrino oscillation parameters, including the CP phases and mixing angles, in predictive models of neutrino masses, such as those based on symmetries.

In addition to leptogenesis, new physics associated with neutrino masses may also have its imprints in cosmic neutrino background. Specifically, if neutrinos are Dirac fermions, there can exist non-thermal contribution to the cosmic neutrino background, in addition to the standard thermal background, with compatible number density~\cite{Chen:2015dka}. Such non-thermal relic neutrino background might be detected by future experiment, such as PTOLEMY~\cite{PTOLEMY:2018jst}. 

To answer all the questions still open in neutrino theory, it is of paramount importance to corner unknown regions of the neutrino parameter space from different phenomenological and experimental angles.

%% file: nu_cosmology.tex
\section{Cosmological probes}
\label{sec:intro_cosmo}

\subsection{Cosmological imprints of neutrinos}
\label{ssec:imprints}

Cosmology provides a unique window into the physics of neutrinos due to the existence of the cosmic neutrino background.
Cosmic neutrinos decoupled from the thermal plasma just seconds after the onset of hot Big Bang expansion when the temperature of the plasma was around 1~MeV.
Cosmic neutrinos accounted for a significant fraction of the energy budget of the universe until non-relativistic matter came to dominate about \num{50000} years later.
The cosmic neutrino background has left observable imprints in the primordial abundances of light elements, the fluctuation spectra of the cosmic microwave background (CMB), and the formation of cosmic structure.

Shortly after neutrino decoupling, neutrinos were diluted relative to photons primarily due to the transfer of entropy from electron-positron pairs to photons (which the bulk of the neutrinos did not share). 
The energy density $\rho_\nu$ of the cosmic neutrino background is commonly expressed in terms of the effective number of neutrino species,
\begin{equation}
    N_\mathrm{eff} = \frac{8}{7}\left(\frac{11}{4}\right)^{4/3} \frac{\rho_\nu}{\rho_\gamma} \, ,
    \label{eq:Neff_definition}
\end{equation}
where $\rho_\gamma$ is the energy density of photons. The normalization is chosen such that if the three families of Standard Model neutrinos had decoupled instantaneously prior to electron-positron annihilation, then $N_\mathrm{eff}=3$.  
In fact, Eq.~\ref{eq:Neff_definition} can be extended in such a way to quantify the contribution of any very light and inert dark radiation species: $\rho_\nu \rightarrow \rho_\mathrm{rad}-\rho_\gamma$, with $\rho_{\rm rad} = \rho_\gamma + \rho_\nu + \rho_{\rm DR}$. Clearly, since $\rho_\gamma$ is well known from COBE-FIRAS measurements of the CMB average temperature~\cite{Mather:1993ij, Fixsen:2009ug}, $\Neff$ provides information about the energy density of decoupled relics, including neutrinos. In the Standard Model, $\Neff$ can be accurately computed by explicitly solving for the process of neutrino decoupling in the early Universe when $T\sim 2\,{\rm MeV}$~\cite{Dolgov:2002wy}. The most recent analysis~\cite{Cielo:2023bqp} gives $N_\mathrm{eff}^\mathrm{SM} = 3.0432(2)$, with the very tiny uncertainty (much smaller than current and forecasted experimental sensitivity, see Sec.~\ref{subsec:cosmo_bounds}) coming from the contribution of three factors in quadrature: numerical accuracy, measurement errors in the determination of the solar mixing matrix elements, radiative corrections in neutrino-electron scattering rates~\cite{EscuderoAbenza:2020cmq,Bennett:2019ewm}. Previous analyses did not include radiative corrections in the $e^{+}e^{-}\rightarrow \nu\bar{\nu}$ reactions and reported $N_\mathrm{eff}^\mathrm{SM} = 3.0440(2)$~\cite{EscuderoAbenza:2020cmq,Akita:2020szl,Froustey:2020mcq,Bennett:2020zkv}. A measurement of $N_\mathrm{eff}$ that differs from the Standard Model prediction could indicate the presence of additional neutrino species or a deviation from the standard thermal history~\cite{Snowmass2021:LightRelics}.

The primary pathway by which cosmic neutrinos impact cosmological observables is through their gravitational influence.
Their mean energy density contributes to the expansion rate, and this contribution plays a particularly important role during the radiation-dominated evolution of the early universe.
The energy density of cosmic neutrinos affects the relationship between time and temperature in the early universe, and thereby impacts the yield of primordial light element abundances~\cite{Peebles:1966zz, Dicus:1982bz, Cyburt:2015mya} and the scale of diffusion damping of acoustic fluctuations seen in the CMB and the matter power spectrum~\cite{Bashinsky:2003tk, Hou:2011ec, Planck:2018vyg}.
Additionally, since cosmic neutrinos freely stream after decoupling, perturbations to the cosmic neutrino density propagate at the speed of light while they are relativistic.
Neutrino fluctuations therefore travel at a speed exceeding the sound speed in the photon-baryon plasma, leading to a characteristic phase shift in the spectra of acoustic oscillations~\cite{Bashinsky:2003tk, Follin:2015hya, Baumann:2015rya, Baumann:2017lmt, Baumann:2019keh, Green:2020fjb}.
A measurement of this phase shift can be used to place limits on neutrino self-interactions~\cite{Cyr-Racine:2013jua, Lancaster:2017ksf} or couplings of neutrinos to other fields.

Cosmology also offers a means to measure the absolute mass scale of neutrinos.
While cosmic neutrinos were relativistic at early times, the low present temperature of the cosmic neutrino background ($T_{\nu,0}=1.95$~K) implies that at least two neutrino mass eigenstates are non-relativistic today~\cite{deSalas:2017kay}.
Assuming a standard thermal history, the total energy density of non-relativistic neutrinos is
\begin{equation}
    \Omega_\nu h^2 =  6.2 \times 10^{-4} \, \left( \frac{\sum m_\nu}{58 \, {\rm meV}} \right) \ ,
    \label{eq:Omega_nu_h2}
\end{equation}
where $\Sigma m_\nu$ is the sum of the individual masses of the neutrino mass eigenstates:
\begin{equation}\label{eq:summnu}
\Sigma m_\nu \equiv m_1 + m_2 + m_3 \ .
\end{equation}
Along with a measurement of the total matter density~\cite{Planck:2018vyg}, this implies that non-relativistic neutrinos make up about 0.4 to 1 percent of the total matter density today.
Cosmology thereby acts as a natural source of a significant density of non-relativistic neutrinos, offering an opportunity to study neutrinos in a regime where the effects of their rest mass have observational consequences.

A universe with massive neutrinos experiences a suppression of cosmic structure growth on small scales compared to a universe with massless neutrinos.
Massive neutrinos act like hot dark matter; while non-relativistic neutrinos redshift like matter in the late universe, their velocity remains sufficiently large to prevent them from falling into the gravitational wells created by cold dark matter on scales smaller than their effective Jeans scales.  
Non-relativistic neutrinos contribute to the matter density and the expansion rate but not to the clustering of matter, so probes of density fluctuations on scales smaller than the Jeans scale will exhibit a smaller amplitude of clustering when compared to a universe with massless neutrinos~\cite{Lesgourgues:2006nd,Wong:2011ip,Lesgourgues:2012uu,Lattanzi:2017ubx}.  
Observational probes of this clustering include gravitational lensing of the CMB~\cite{Kaplinghat:2003bh}, clustering of large-scale-structure tracers and weak lensing of galaxies~\cite{Hu:1997mj,Cooray:1999rv}, and the number density of galaxy clusters~\cite{Abazajian:2011dt,Carbone:2011by,Ichiki:2011ue}.

Before moving to discuss the cosmological phenomenology of neutrinos in detail, a few words concerning sterile neutrinos in cosmology are in order. BSM physics required to explain the origin of neutrino masses may include the existence of right-handed neutrino states that are sterile, i.e., not participating in standard interactions. At present, there is neither theoretical limitation to the number of possible sterile states, nor to their mass~\cite{Dasgupta:2021ies}. The cosmological phenomenology of sterile neutrinos is very similar to that of active states. Once the production mechanism and mass of sterile neutrinos are specified, their possible contribution to $N_\mathrm{eff}$ or to the total matter density can be assessed. Depending on their mass, sterile neutrinos can modify structure formation as well by suppressing small-scale features in a similar way as active neutrinos do.
Decaying and annihilating sterile neutrinos during specific cosmological epochs can inject radiation and deplete the amount of matter density, thus providing additional observational handles of their phenomenology.  
In this article, we will briefly review the main aspects of active neutrino cosmology, keeping in mind that similar considerations also apply to sterile neutrinos, unless otherwise stated.

\subsection{Big Bang Nucleosynthesis era and light elements abundances} 

The synthesis of the primordial light elements is highly sensitive to the expansion rate of the Universe at the time of Big Bang Nucleosynthesis (BBN), $10 \,{\rm keV}\lesssim T \lesssim 1 \,{\rm MeV}$, see~\cite{Sarkar:1995dd,Iocco:2008va,Pospelov:2010hj}. At $T > T_D \simeq 0.075\,{\rm MeV}$ (threshold for deuterium photodissociation~\cite{Mukhanov:2003xs}, also known as ``BBN bottleneck''), the baryonic sector of the plasma essentially consists of free neutrons and protons but as the Universe expands and $T < T_D$ deuterium nuclei are not dissociated anymore. With the accumulation of significant amount of deuterium, the BBN chain can proceed rapidly and, eventually, all the free neutrons bind to form helium-4 and very small quantities of deuterium, helium-3, and lithium. The number of free neutrons $n_n$ is strongly sensitive to the age of the Universe at the time, $n_n \propto e^{-t/\tau_n}$, where $\tau_n$ is the neutron lifetime. Therefore, BBN represents a powerful probe of the number of ultrarelativistic species at the time of nucleosynthesis, since $t\propto \rho_\mathrm{rad}^{-1/2}$. 

Cosmological neutrinos (including possible sterile states) influence the production of the primordial light elements in two ways. 
First, $\nu_e$ and $\overline{\nu}_e$  directly participate in the charged current weak interactions which rule the neutron/proton chemical equilibrium:  
\begin{align}
(a)\;\; \nu_e + n \rightarrow e^- + p\;\;, & \;\;\;\;\;\;\;\;\;\; & (d)\;\; \overline{\nu}_e + p \rightarrow e^+ + n  \nonumber \\
(b)\;\; e^- + p \rightarrow \nu_e + n\;\;,  & \;\;\;\;\;\;\;\;\;\; &(e) \;\; n \rightarrow e^-+ \overline{\nu}_e + p   \\
(c)\;\; e^+ + n \rightarrow \overline{\nu}_e + p  \;\;, &\;\;\;\;\;\;\;\;\;\; &(f)\;\; e^- + \overline{\nu}_e+ p\rightarrow n \nonumber
\label{e:reaction}
\end{align}
To get an accurate theoretical prediction for light-element
abundances, the processes $(a)-(f)$ require a careful and accurate treatment.
Since $\nu_e$ and $\overline{\nu}_e$ enter the BBN equations at a fundamental level, any change in the neutrino momentum distributions (e.g., non-zero chemical potentials, spectral distortions) can shift the neutron-to-proton ratio freeze-out temperature and then modify the primordial $^4$He  abundance.
Second, cosmological neutrinos of each flavor gravitate and contribute as relativistic species to  the total  radiation energy density that governs the expansion rate of the Universe before and during BBN epoch. As seen in Sec.~\ref{ssec:imprints}, this effect is encoded into the parameter $\Neff$. 
Changing the expansion rate alters the $n/p$ ratio at the onset of BBN and hence the light element abundances.

The primordial element abundance that is most sensitive to $\Neff$ is the helium abundance, typically parameterized by the fraction of the mass in baryons in the form of ${}^4{\rm He}$: $\yp$. The theoretical prediction for the primordial helium abundance is free from nuclear reaction uncertainties\footnote{The ${}^4{\rm He}$ abundance basically depends on the neutron number at BBN. As such, the dominant source of uncertainty in the predicted amount of primordial helium comes from measurements of the neutron lifetime.} and is very sensitive to $\Neff$, while being only logarithmically dependent upon the baryon energy density.
Another primordial element abundance that can be of relevance to determine $\Neff$ is the deuterium abundance, typically parameterized by the number ratio of deuterium with respect to hydrogen, ${\rm D/H}$. In contrast to $\yp$, the predicted primordial deuterium abundance does depend strongly upon the baryon energy density, ${\rm D/H} \propto \omega_b^{-1.6}$~\cite{Fields:2019pfx}, and its prediction is currently limited by the lack of detailed knowledge of two nuclear reactions -- see~\cite{Pisanti:2020efz,Pitrou:2020etk,Yeh:2020mgl} for recent global analyses and~\cite{Mossa:2020gjc} for a recent experimental result from the LUNA collaboration that helped significantly reduce the theoretical uncertainty. 

We note that sterile neutrino states produced
before the decoupling of active neutrinos could acquire quasi-thermal
distributions (depending on their temperature) and behave as extra degrees
of freedom at the time of primordial nucleosynthesis. This would
anticipate weak interaction decoupling leading to a larger neutron-to-proton
ratio, eventually resulting into a larger $^4$He fraction. Furthermore,
sterile neutrinos can distort the $\nu_e$ phase space distribution via
flavor oscillations with the active ones, leading to a possible effect on
the helium and deuterium abundances.
The decay and/or annihilation of sterile states around the BBN epoch would also affect the production of light elements via e.g., photodissociation by high-energy photons injected in the primordial plasma~\cite{Dienes:2018yoq,Poulin:2016anj,Salvati:2016jng}.

\subsection{Recombination era and the CMB}
\label{sec:recombination}

Neutrinos make up 41\% of the radiation density in the standard model of cosmology and, therefore, also that amount of the energy budget of the universe during the radiation era before recombination. Consequently, relativistic neutrinos have a significant gravitational influence on the expansion of the universe and the perturbations, in particular in photons and baryons~(see e.g.\ the recent reviews~\cite{Lesgourgues:2013sjj, Wallisch:2018rzj, Green:2019glg}). Since the evolution of both the background and the fluctuations are imprinted in the CMB and the distribution of matter after recombination, we can extract this influence of neutrinos from cosmological datasets. In this way, neutrinos contribute through (i)~their mean energy density~\cite{Peebles:1966zz, Dicus:1982bz, Hou:2011ec, Bashinsky:2003tk} and (ii)~their fluctuations, which propagate at the speed of light in the early universe due to the free-streaming nature of neutrinos~\cite{Bashinsky:2003tk, Baumann:2015rya}.\medskip

Neutrinos lighter than about \SI{0.6}{eV} are still relativistic at recombination, so their effect on the primary anisotropies (those generated directly at the last scattering surface) is mainly independent of their mass. A finite neutrino mass can still affect the primary anisotropies through projection effects, i.e.\ by changing the distance to the last-scattering surface. These effects are however small and easily degenerate with other parameters that affect the late-time evolution and, therefore, provide very limited constraining power in this mass range. Secondary anisotropies, on the other hand, are more directly affected by neutrino masses. In flat $\Lambda$CDM models, the additional energy density provided by massive neutrinos at late times can be compensated by a change in the density associated with the cosmological constant. This will induce a late-time integrated Sachs-Wolfe~(ISW) effect, which is however visible at the largest scales where the error budget is dominated by cosmic variance. This therefore again provides very little constraining power for current and future experiments. The effect that neutrinos have on structure formation at late times, due to their free-streaming nature, is much more important. Affecting the clustering of matter, neutrinos leave an imprint on the weak lensing of the CMB. We discuss this and other late-time effect related to the large-scale structure of the Universe in Sec.~\ref{sec:LSS}. In the remainder of this subsection, we neglect the effects of neutrino masses and focus instead on the effects related to the neutrino energy density at early times, when these particles were relativistic.

\begin{figure}
	\centering
	\includegraphics{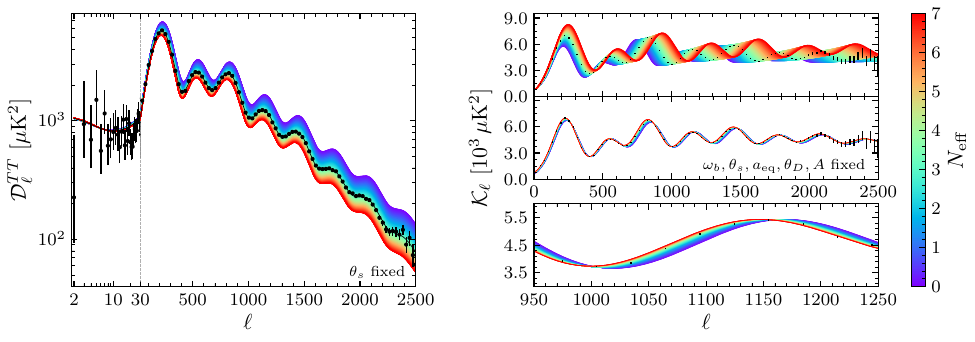}
	\caption{Effect of free-streaming radiation on the CMB temperature power spectrum (adapted from~\cite{Wallisch:2018rzj}). To illustrate the sensitivity of the Planck~2018 temperature data, we also display their $1\sigma$~error bars~\cite{Planck:2019nip}.  \textit{Left}: Variation of the CMB~temperature power spectrum~$\mathcal{D}_\ell^{TT} \equiv \ell (\ell+1)/(2\pi)\,C_\ell^{TT}$ as a function of~$\Neff$ for fixed angular size of the sound horizon~$\theta_s$. The dominant exponential damping of~$\mathcal{D}_\ell^{TT}$ is clearly visible and would be the same if the radiation was non-free-streaming, in contrast to the phase shift illustrated in the right panel. 
	\textit{Right}: Variation of the undamped CMB~temperature power spectrum $\mathcal{K}_\ell =  d^{-1}_\ell \mathcal{D}_\ell^{TT}$, with exponential damping function~$d_\ell$, as a function of~$\Neff$. Following~\cite{Follin:2015hya}, the physical baryon density~$\omega_b$, the scale factor at matter-radiation equality $a_\mathrm{eq} \equiv \omega_m/\omega_r$, the angular size of the sound horizon~$\theta_s$ and the angular size of the damping scale~$\theta_D$ are held fixed in the second panel. In addition, the spectra are normalized at the fourth peak. The remaining variation is the phase shift~$\phi$ with a zoom-in shown in the bottom panel. We refer to~\cite{Wallisch:2018rzj} for additional details.}
	\label{fig:cmbSpectrum_neffVariation}
\end{figure}

At the level of the background cosmology, relativistic neutrinos contribute to the radiation density, i.e.\ their presence increases the expansion rate at a given photon temperature. This affects the CMB~anisotropies through changes to the damping and sound horizon length scales. When fixing the scale of matter-radiation equality and the location of the first acoustic peak~(quantities that are well measured), the effect on the damping tail is the dominant imprint of relativistic particles, such as neutrinos~(see the left panel of Fig.~\ref{fig:cmbSpectrum_neffVariation})~\cite{Hou:2011ec}. They modify the damping tail through their contribution to the Hubble rate, which, in turn, changes the amount of photon diffusion in the pre-recombination universe resulting in an exponential suppression of short-wavelength modes~\cite{Zaldarriaga:1995gi}. This effect is degenerate with other cosmological parameters~(in particular the helium fraction $\yp$) in the CMB~temperature power spectrum~\cite{Bashinsky:2003tk}, but the degeneracy with $\yp$ can be broken by including BBN~information and/or CMB~polarization data. While this is the leading effect to constrain additional radiation in the~CMB, it does not discriminate between free-streaming (i.e.~non-interacting) and non-free-streaming (i.e.~interacting) neutrinos since the Hubble rate only depends on the background energy density~\cite{Baumann:2015rya}.

Perturbations in neutrinos and other free-streaming radiation also affect the photon-baryon fluid in the early universe through their gravitational influence leading to imprints that allow to distinguish between free-streaming and non-free-streaming radiation: a shift in the amplitude, frequency and the phase of the acoustic oscillations, $A(k) \sin(\omega k + \phi)$, in both temperature and polarization~(see the right panel of Fig.~\ref{fig:cmbSpectrum_neffVariation})~\cite{Bashinsky:2003tk}. First, the presence of free-streaming radiation leads to a suppression of the superhorizon gravitational potential which implies that more energy in free-streaming radiation reduces the initial amplitude of adiabatic fluctuations~\cite{Bashinsky:2003tk,Ma:1995ey}. This effect is however somewhat degenerate with the primordial power spectrum amplitude. The frequency can be affected by many physical processes. The third imprint is however not degenerate with other cosmological parameters~\cite{Bashinsky:2003tk, Baumann:2015rya} and is unique to free-streaming radiation, providing a direct connection to the underlying particle properties~\cite{Baumann:2015rya}. The key property of standard neutrinos that distinguishes them from non-free-streaming radiation is their supersonic propagation: while sound waves in the photon-baryon fluid travel at $c_s \approx 1/\sqrt{3}$, SM~neutrinos free-stream at nearly the speed of light. The neutrinos therefore propagate ahead of the sound horizon of the photon-baryon fluid and exert a gravitational pull that shifts the photon and baryon perturbations to larger distances. In the CMB~temperature and polarization power spectra, this effect manifests itself as a phase shift of the CMB~peaks to larger physical scales (i.e.~to smaller multipoles~$\ell$)~\cite{Bashinsky:2003tk, Baumann:2015rya}. While this shift is $\delta \ell \sim 20$ at high multipoles in the standard model of cosmology~\cite{Baumann:2015rya, Pan:2016zla}, it can be smaller or larger in models with non-free-streaming neutrinos or additional free-streaming radiation, respectively. This phase shift from neutrinos has been directly measured in the Planck temperature data~\cite{Follin:2015hya}~(see~\cite{Baumann:2015rya, Brust:2017nmv, Blinov:2020hmc} for complementary analyses) providing the most direct evidence to date for free-streaming radiation consistent with the cosmic neutrino background.

Putting all effects of relativistic and free-streaming neutrinos on the temperature and polarization power spectra together, the Planck satellite has resulted in a 6\%~constraint on their energy density of $\Neff = 2.92^{+0.18}_{-0.19}$~\cite{Planck:2018vyg}. Future high-resolution maps of the~CMB could realistically achieve up to a 1\%~constraint of $\sigma(\Neff)= 0.03$ in the coming decade~\cite{SimonsObservatory:2018koc, Abazajian:2019eic, NASAPICO:2019thw}, with additional improvements possible with more futuristic CMB~experiments~(see e.g.~\cite{Baumann:2015rya, Sehgal:2019ewc}). When considering~$\yp$ independent of~BBN or separate bounds on free-streaming and non-free-streaming radiation, these free-streaming density constraints are less stringent~\cite{Planck:2018vyg,Baumann:2015rya, Blinov:2020hmc,Brust:2017nmv} due to the discussed degeneracies. For future experiments like CMB-S4, the constraints on~$\Neff$ are anticipated to relax by a factor of approximately three for free-streaming, and two for non-free-streaming radiation~(see e.g.~\cite{Baumann:2015rya,Brust:2017nmv,CMB-S4:2016ple,Baumann:2017gkg}). At the same time, it will be possible to constrain the helium fraction with a similar sensitivity as the direct measurements of light element abundances, along with much stronger bounds on the non-free-streaming radiation density.

\subsection{Late times and the large-scale structure} 
\label{sec:LSS}

\begin{figure}[ht!]
    \centering
    \includegraphics[width=1\textwidth]{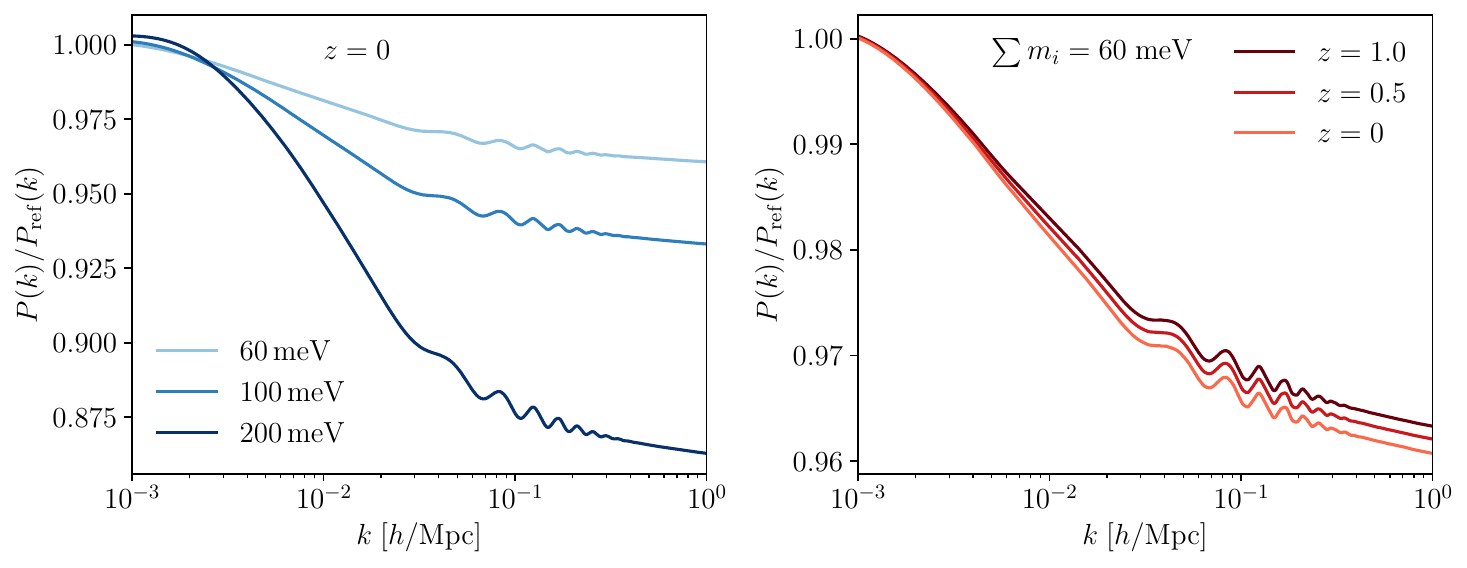}
    \caption{Left: Suppression of the matter power spectrum at $z=0$ for varying neutrino mass relative to a reference cosmology with massless neutrinos. Large-scale structure experiments have little sensitivity at scales $k < 10^{-2} h/\mathrm{Mpc}$ and can mostly resolve the suppressed part of the spectrum.
    Right: Evolution of the suppression of the matter power spectrum for fixed neutrino mass $\sum m_i = 60 \, \mathrm{meV}$ with redshift. The evolution is mild and the suppression slightly increases with time as neutrinos dampen the growth of structures. All comparisons are made keeping physical densities $\Omega_i h^2$ and $h$ fixed.
    }
    \label{fig:Pkmnu}
\end{figure}

For the mass range of ordinary neutrinos consistent with CMB measurements $\sum m_\nu \lesssim 0.2 \: \mathrm{eV}$, neutrinos become non-relativistic around $z_\mathrm{nr} \sim 100$ as the average momentum $\langle p_\nu \rangle \sim 3 T_\nu$ drops below the individual mass $m_\nu$. Therefore neutrinos are almost fully relativistic until the CMB is emitted, but are non-relativistic while the large-scale structure of the Universe is formed. However, as we will see their large thermal velocities still lead to a characteristic phenomenology. On subhorizon scales, the growth of matter perturbations $\delta_m$ is well-described by the Newtonian growth equation (overdots refer to derivatives with respect to conformal time)

\begin{equation}
\label{eq:growth}
\ddot \delta_m + 2 H \dot \delta_m - 4 \pi G \bar \rho \delta_m = 0 \,,
\end{equation}
where we assumed that the pressure vanishes. Neutrinos affect the growth of perturbations in two distinct ways:
\begin{itemize}
\item Through the change in the background expansion discussed above, which enters the drag term. This is a global effect on growth in neutrino cosmologies.
\item Neutrinos have considerable thermal velocities, and move over cosmological distances during the age of the Universe. Concerning the growth of structure, the effect of the large thermal velocities is usually expressed in terms of the neutrino free-streaming scale~\cite{Lesgourgues:2006nd,2013neco.book.....L}
\begin{equation}
    k_{\rm fs} = 0.04 \, h \, {\rm Mpc}^{-1} \times \frac{1}{1+z} \, \left(\frac{\sum m_\nu}{58 \, {\rm meV}}\right) \, . 
    \label{eq:kfs_nu}
\end{equation}
which is comparable to the size of the horizon at the redshift $z_\mathrm{nr}$ when neutrinos become non-relativistic. The free-streaming scale behaves similar to a Jeans length. All perturbations below $k_\mathrm{FS}$ are dampened, since for these wavenumbers neutrino perturbations are erased and gravitational potentials given by the Poisson term in Eq.~\ref{eq:growth} are only sourced by the combined perturbations in the cold dark matter and baryon components, $\delta_c + \delta_b$.
\end{itemize}
The scale-dependent growth for modes $k \ll k_\mathrm{FS}$ and $k \gg k_\mathrm{FS}$ is characteristic for neutrinos and cannot be easily mimicked by other effects~\cite{2013neco.book.....L}. Unfortunately, for a sum of the standard neutrino masses close to the minimal value $\sum m_\nu \approx 0.06~\mathrm{eV}$, the free-streaming scale is too large to be observed by near-future large-scale structure experiments such as Euclid and the Vera Rubin Observatory. However, these experiments will still be sensitive to neutrino masses due to their imprint on scales smaller than the free-streaming scale. The total suppression of power on scales affected by free streaming today is approximately given by~\cite{Green:2021xzn}
\begin{equation}
    P^\nu(k \gg k_\mathrm{FS}, z) \approx \left(1 - 2 f_\nu - \frac{6}{5} f_\nu \log \frac{1+z}{1+z_\mathrm{nr}} \right) P(k \gg k_\mathrm{FS}, z) \,,
\end{equation}
with the fractional neutrino contribution $f_\nu = \Omega_\nu / \Omega_m$, and $P^\nu(k,z)$ and $P(k,z)$ denote the power spectrum in a cosmology with either massive or massless neutrinos, respectively. The suppression has a very mild redshift evolution, which is unlikely to be relevant for most large-scale structure probes. It saturates at z=0 at~\cite{Hu:1997mj}
\begin{equation}\label{eq:suppression}
    P^\nu(k \gg k_\mathrm{FS}, z=0) \approx \left(1 - 8 f_\nu \right) P(k \gg k_\mathrm{FS}, z=0)
\end{equation}
for neutrinos becoming non-relativistic at $z_\mathrm{nr} \approx 100$, as is the case for the three ordinary neutrino species. In Fig.~\ref{fig:Pkmnu} we show
the effect of massive neutrinos on the matter power spectrum. Note that in the above discussion we consider linear evolution to hold up to very small scales. Of course, this is not true in reality. The growth of structures eventually enters the non-linear regime (with small scales being affected earlier than large scales) making it necessary to employ beyond-perturbation-theory approaches. We comment later on this point in the text. For the moment, we just notice that the suppression factor in~\eqref{eq:suppression} is enhanced from 8 to 10 when accounting for non-linear evolution of small scales.

\begin{figure}[b!]
    \centering
    \includegraphics[width=5in]{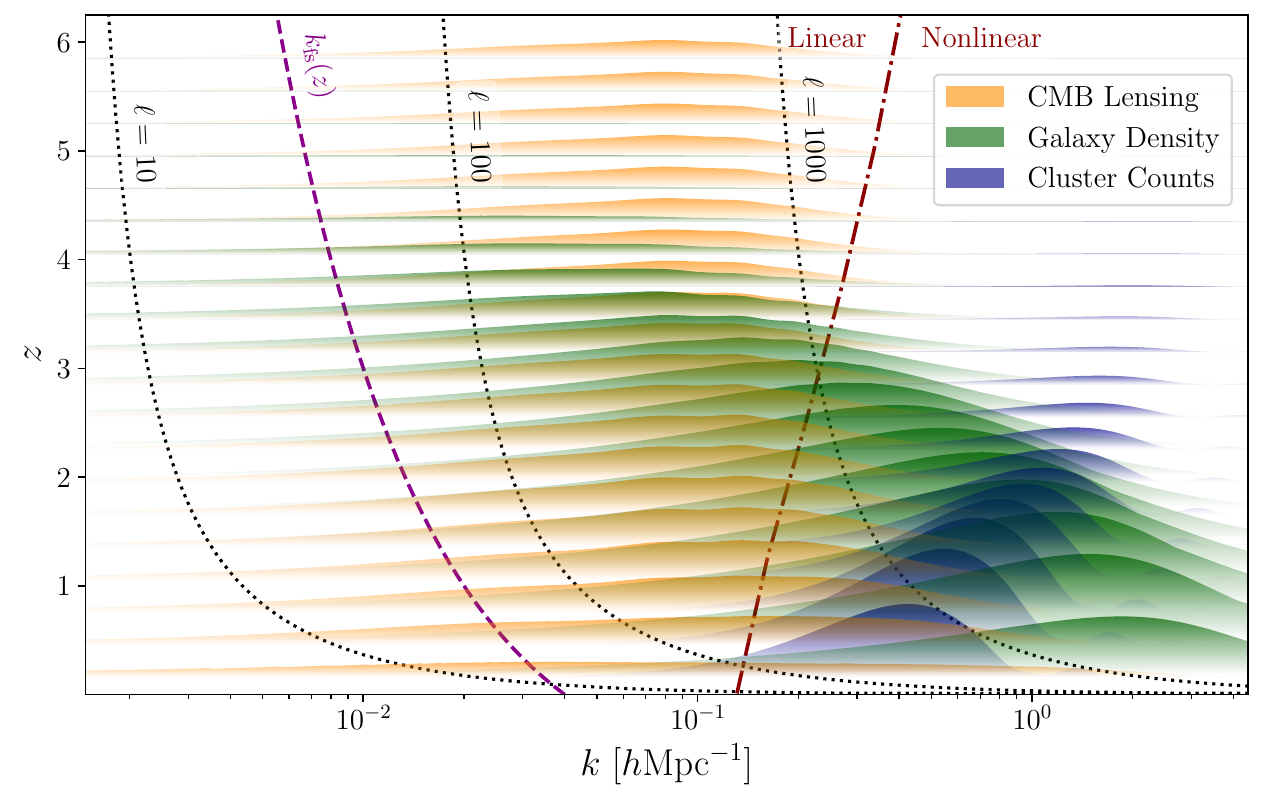}
    \caption{Contributions of the matter power spectrum $P(k,z)$ to various large scale structure observables.  The contributions are weighted by signal-to-noise ratio anticipated for each observable: the CMB lensing power spectrum using the lensing reconstruction expected from CMB-S4, the angular power of galaxy density using observations from the Vera Rubin Observatory gold sample, and number counts of clusters with mass greater than $10^{14}~h^{-1}M_\odot$.  The CMB lensing weighting is multiplied by an additional factor of 3 relative to the others in order to make the CMB lensing contributions more visible despite the very broad lensing redshift kernel.  The values of wavenumber $k$ and redshift $z$ that contribute to a given angular scale $\ell$ in the Limber approximation are shown by the black dotted lines.  The purple dashed line shows the free-streaming scale $k_\mathrm{fs}(z)$ from Equation~(\ref{eq:kfs_nu}) for standard neutrinos with $\sum m_\nu = 58~\mathrm{meV}$; massive neutrinos suppress the amplitude of $P(k,z)$ to the right of that line.  Nonlinear corrections to the matter power spectrum are expected to be non-negligible to the right of the red dash-dot line. Figure reproduced from~\cite{Green:2021xzn}. }
    \label{fig:Pk_stagger}
\end{figure}

Various observables of large scale structure are sensitive to the matter power spectrum on different length scales and at different times, thereby providing multiple methods to search for the suppression of power caused by massive neutrinos.
Sensitivity to the matter power spectrum $P(k,z)$ as a function of wavenumber $k$ and redshift $z$ is shown in Figure~\ref{fig:Pk_stagger} for the CMB lensing power spectrum, the angular power spectra of galaxy density, and number counts of galaxy clusters $N_i$.  
The contributions of $P(k,z)$ to each observable are shown weighted by the signal-to-noise ratio with which those observables will be measured in upcoming surveys.
The weightings are calculated based on forecasts for CMB-S4 lensing reconstruction~\cite{CMB-S4:2016ple,Abazajian:2019eic}, Rubin Observatory galaxy density probes~\cite{LSSTScience:2009jmu}, and counts of clusters with mass greater than $10^{14}~h^{-1}M_\odot$ corresponding roughly to the detection threshold from the thermal Sunyaev-Zeldovich (SZ) effect as observed by CMB-S4.
It can be seen that clusters and low-redshift galaxies are primarily sensitive to the non-linear regime of the matter power spectrum, while CMB lensing and the galaxy density at high redshift are sensitive to the linear matter power spectrum.

\subsubsection{Galaxy clustering}
\label{subsec:gal_clustering}

Galaxies are a biased tracer of the underlying dark matter field. On large scales, the bias allows us to relate the observed power spectrum of galaxies to the power spectrum of matter fluctuations via
\begin{equation}
    P_g(k, z) = b^2(k, z) P_m(k, z) \, ,
\end{equation}
with the bias $b(k, z)$, noting that when including massive neutrinos we should only consider the baryon and dark matter power spectrum $P_{cb}$ instead of $P_m$~\cite{Castorina:2013wga,Villaescusa-Navarro:2013pva,LoVerde:2014pxa, Raccanelli:2017kht,Vagnozzi:2018pwo}. Even in the absence of neutrinos, scale and redshift dependence of $b$ is the major challenge for galaxy clustering surveys. Perturbative treatment of biasing leads to an expansion in terms of local operators formed out of the density and tidal field up to a given order in perturbation theory~\cite{Desjacques:2016bnm}, which gives rise to a number of physically motivated parameters that can be marginalised over when fitting for the shape of the galaxy power spectrum~\citep[e.g.][]{Ivanov:2019hqk}. Note, however, that since the high momenta of neutrinos permit them to travel over cosmological distances, the bias expansion will depend on the history of the matter and neutrino density fields at cosmological distances as well. This fact causes the bias parameters to acquire a scale-dependent feature at scales near and beyond the neutrino free-streaming scale~\cite{LoVerde:2014pxa, Chiang:2017vuk, Chiang:2018laa}. This feature is both a signal and, if not properly accounted for, a systematic to future measurements of neutrino mass from galaxy clustering~\cite{LoVerde:2016ahu, Munoz:2018ajr, Xu:2020fyg}.

Current generation galaxy surveys such as BOSS and DES covered the redshift range up to $z\sim1$, while currently running (DESI) and upcoming surveys (Euclid, Rubin Observatory, Roman Telescope) will map redshift regions as far as $z\sim3$. With increased survey volumes, galaxy surveys are pushing the signal to noise~\cite{PhysRevLett.79.3806} on all scales, including well beyond $k > 0.1~h/\mathrm{Mpc}$. To take full advantage of the information content brought by these surveys (see Fig.~\ref{fig:Pk_stagger}), it thus becomes more and more important to accurately model non-linear scales and baryonic physics. Many approaches exist, making use of perturbative theoretical models~\cite{Chudaykin:2020aoj,DAmico:2019fhj,Cabass:2020jqo,Donath:2020abv,Cabass:2022avo,Ivanov:2022mrd}, simulations~\cite{Bayer:2020tko}, simulation emulator approaches~\cite{Kwan:2013jva,Euclid:2018mlb,Euclid:2020rfv,Angulo:2020vky,Arico:2020lhq}, or hybrid methods based on the halo model with simulation input~\cite{Mead:2020vgs, Philcox:2020rpe}. Note, however, in all cases it is crucial to account for uncertainties in the theoretical modeling in order to avoid biases in the parameter estimation~\cite{Baldauf:2016sjb,Sprenger:2018tdb, Chudaykin:2019ock,Euclid:2020tff,Knabenhans:2021huw}.

\begin{figure}
	\centering
	\includegraphics{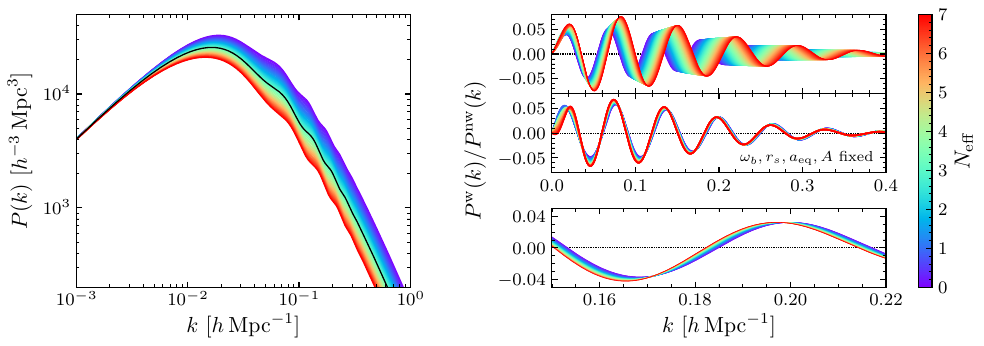}
	\caption{Variation of the matter power spectrum~$P(k)$~(\textit{left}) and the BAO~spectrum~$P^\mathrm{w}(k)/P^\mathrm{nw}(k)$~(\textit{right}) as a function of~$\Neff$~(adapted from~\cite{Wallisch:2018rzj}). The BAO~spectrum is the ratio of the oscillatory part~$P^\mathrm{w}(k)$ of the matter power spectrum and its smooth broadband part~$P^\mathrm{nw}(k) = P(k) - P^\mathrm{w}(k)$. To visualize the shift in acoustic peaks, the physical baryon density,~$\omega_b$, the physical sound horizon at the drag epoch,~$r_s$, the scale factor at matter-radiation equality, $a_\mathrm{eq}$, and the BAO~amplitude~$A$ at the fourth peak are held fixed in the second BAO~panel. This panel and the bottom zoom-in show the remaining phase shift induced by free-streaming relativistic species. We refer to~\cite{Wallisch:2018rzj} for additional details.}
	\label{fig:matterSpectrum_neffVariation}
\end{figure}

Besides the smooth~(broadband) component of the matter power spectrum, significant cosmological information is contained in the oscillatory spectrum of baryon acoustic oscillations~(BAO). The former mainly depends on the background evolution and the latter captures the cosmic sound waves that we also observe in the CMB~anisotropies. In the BAO~spectrum, a change in the radiation density leads to shifts in the frequency, amplitude and phase of the BAO~spectrum. The BAO~frequency corresponds in Fourier space to the BAO~scale, which is the size of the sound horizon at the drag epoch, and, therefore, depends on the background expansion history. This is the quantity that most BAO~analyses extract (see, however, ~\cite{Anselmi:2018vjz,Anselmi:2022exn} for alternative methods) and use to constrain cosmology. As discussed in Sec.~\ref{sec:recombination}, the amplitude and phase shifts originate from the evolution of the neutrino perturbations in the early universe~(see the right panel of Fig.~\ref{fig:matterSpectrum_neffVariation}). While the amplitude is affected by gravitational nonlinearities, the phase shift due to the supersonic propagation of free-streaming species should be robust to these late-time complications~\cite{Baumann:2017lmt,Green:2020fjb}. This allows us to extract a non-zero phase shift from the distribution of galaxies observed by the Baryon Oscillation Spectroscopic Survey~(BOSS)~\cite{Baumann:2019keh,Baumann:2017gkg}, with ongoing and future galaxy surveys\footnote{For an indicative comparison of the redshift coverage (hence available volume) and performance on BAO measurements of current and upcoming surveys, see Fig.~2.9 of~\cite{DESI:2016fyo}.} significantly improving on this first measurement~\cite{Baumann:2017gkg}. At the same time, it provides a way to constrain the free-streaming nature of neutrinos independent of, yet complementary to, the~CMB.

The broadband shape of the matter power spectrum responds to a larger radiation density with a shift in the location of the turn-over towards larger scales and a suppression of power on small scales~(see the left panel of Fig.~\ref{fig:matterSpectrum_neffVariation}). Both effects are due to matter-radiation equality occurring at a later time. In contrast to the BAO~spectrum, the broadband shape therefore cannot distinguish between free-streaming and non-free-streaming radiation. Although these effects are clearly visible in the linear matter power spectrum, they are limited by uncertainties related to gravitational nonlinearities and biasing. This is why a combination of planned spectroscopic large-scale structure (LSS) surveys and Planck could improve the CMB-only constraints on the radiation density by a factor of 2-3, while the potential improvements when combining with a CMB~experiment achieving $\sigma(\Neff) \approx 0.03$, such as CMB-S4~\cite{Abazajian:2019eic,CMB-S4:2022ght}, will be much more modest~\cite{Baumann:2017gkg, Brinckmann:2018owf}. Having said that, very large-volume and high-resolution LSS~surveys, such as the proposed experiments MegaMapper~\cite{Schlegel:2019eqc} and PUMA~\cite{PUMA:2019jwd}, can reach a comparable sensitivity to the CMB if nonlinear effects can be controlled~\cite{Baumann:2017gkg, Ansari:2018ury, Sailer:2021yzm, MoradinezhadDizgah:2021upg}.

In addition, the power spectrum is expected to provide excellent neutrino mass constraints for a survey such as Euclid~\cite{Sprenger:2018tdb,Chudaykin:2019ock} or Rubin Observatory, especially when combined with future CMB surveys such as CMB-S4 or LiteBIRD~\cite{Brinckmann:2018owf}, due to a favorable breaking of degeneracies by complementary surveys~\cite{Archidiacono:2016lnv}. However,the evolved large-scale structure is highly non-Gaussian and contains plenty of information beyond the power spectrum. The imprint of neutrinos on higher order correlations of the galaxy distribution is a promising avenue, and probes such as the bispectrum~\cite{Hahn:2020lou} or the density PDF~\cite{Uhlemann:2019gni} have been shown to be highly sensitive to the total neutrino mass.

\subsubsection{Gravitational lensing}

Weak gravitational lensing allows the direct measurement of the total matter fluctuations. The shapes of galaxies are coherently deformed by the effect of foreground matter, leading to a measurable correlation of galaxy ellipticities. The effect can be expressed in terms of the integrated density along the line of sight, the lensing convergence,
\begin{equation}
    \kappa(\hat{\mathbf{n}}) = \int \mathrm d z \, W^\kappa(z) \delta_m(\chi(z) \hat{\mathbf{n}}, z) \, ,
\end{equation}
with the comoving distance $\chi$ and an appropriate weighting function $W^\kappa(z)$ which depends on survey specifications~\cite{Bartelmann:1999yn, Kilbinger:2014cea}.
Together with the observed galaxy field $\delta_g$, the addition of gravitational lensing allows to construct three distinct two-point functions: cosmic shear $\langle \kappa \kappa \rangle$, the lensing effect of foreground galaxies on background shapes (galaxy-galaxy lensing) $\langle \delta_g \kappa \rangle$ and galaxy-galaxy clustering $\langle \delta_g \delta _g \rangle$ also discussed in Sec.~\ref{subsec:gal_clustering}. 
The combination of the three angular power spectra is known as 3x2 analysis and set tight constraints on cosmological parameters~\cite{Heymans:2020gsg, DES:2021wwk}.

\subsubsection{Galaxy clusters}

Clusters of galaxies form from the highest peaks in the density field and are the largest gravitationally bound objects in the Universe, with masses reaching up to $10^{15} M_\odot$.
The abundance of galaxy clusters is a sensitive probe of the amplitude of matter fluctuations. The total number of clusters detected by a survey in bins of redshift $\Delta z$ can be written as
\begin{equation}
\label{eq:cluster_abundance}
    N(\Delta z, \Delta X) = \int_{\Delta z} \mathrm d z \frac{\mathrm d V}{\mathrm d z} \int_0^\infty \mathrm d M \frac{\mathrm d n}{\mathrm d M} (M, z) \int_{\Delta X} \mathrm d X p(X | M)
\end{equation}
where $X$ is any direct observable used to identify the cluster, such as galaxy richness, SZ amplitude or X-ray emission. The sensitivity to the power spectrum comes through the halo mass function $\mathrm d n / \mathrm d M$, which expresses the density of dark matter halos hosting a galaxy cluster as a function of the total halo mass. Most models for the halo mass function are based on N-body simulations~\cite{Tinker:2008ff, Bocquet:2020tes}. In the high-mass tail, the mass function depends almost exponentially on the variance of the density field $\sigma_8^2$. Since the size of galaxy clusters is below the free-streaming scale for neutrinos, only baryons and dark matter take part in the formation which leads to a stronger suppression of the overall cluster abundance than would be expected from the matter power spectrum alone~\cite{Costanzi:2013bha,Bolliet:2019zuz}.

The main challenge in cluster cosmology is to obtain an accurate mapping between the observable $X$ and the underlying halo mass $M$, characterized by the conditional probability distribution $p(X|M)$ in Eq.~\ref{eq:cluster_abundance}.
Large surveys covering a considerable fraction of the sky also allow to detect the clustering of clusters. Similar to galaxies, they are biased tracers of the underlying matter power spectrum
\begin{equation}
    P_{cc}( M, k, z) \sim b^2_{c}(M, k, z) P_m(k, z) \,
\end{equation}
where the effective cluster bias parameters are linked to derivatives of the mass function~\cite{Sheth:1999mn, Sartoris:2010cr, Desjacques:2016bnm} or have to be determined from N-body simulations.
Utilizing the additional information from the cluster two-point function can greatly help to improve the constraining power of galaxy clusters~\cite{2016MNRAS.459.1764S, Marulli:2020uyy}.

\subsubsection[Late-time \texorpdfstring{$H_0$}{H0} measurements]{Late-time $\mathbf{H}_\mathbf{0}$ measurements}

Since the the neutrino mass parameter $\sum m_\nu$ and the current expansion rate $H_0$ are approximately degenerate for CMB observations, additional measurements of the late-time expansion rate help to shrink the error bars for the total neutrino masses.
The tightest currently available constraints on the total neutrino mass combine CMB measurements with observations of BAOs in the galaxy distribution~\cite{eBOSS:2020yzd}, which mainly help to constrain $H_0$. Direct measurements of $H_0$ with Type Ia supernovae can play a similar role, but are currently in tension~\citep[see e.g.][for a discussion]{Knox:2019rjx} with CMB observations depending on the adopted calibration of the supernova luminosities~\cite{Freedman:2019jwv, Riess:2021jrx}.

Other late-time measurements of the expansion rate exist, e.g. from strong lensing~\cite{Birrer:2018vtm}, from the dispersion of fast radio bursts~\cite{Hagstotz:2021jzu} or from interferometric observations of gravitational wave events~\cite{LIGOScientific:2019zcs}, but are not yet accurate enough to provide an auxiliary $H_0$ prior better than methods based on BAOs.

\subsection{Cosmological constraints on neutrino properties}\label{subsec:cosmo_bounds}
The Planck satellite has been the leading CMB experiment of the past decade. Data collected by Planck in 2009-2014 provided unprecedented measurements of the CMB fluctuations in temperature (cosmic-variance-limited down to ~7\,arcminutes/angular multipole of $\sim1600$) and in polarization over a wide range of angular scales~\cite{Planck:2019nip}. The observations of CMB temperature anisotropies from the Planck satellite, without the addition of any external data, constrain $\Sigma m_\nu$ already at the 0.6\,eV level~\cite{Planck:2018vyg}, which is basically the same as the expected sensitivity of the currently running $\beta$-decay experiment KATRIN~\cite{KATRIN:2022ayy}. Combination of temperature, polarization and CMB lensing data from Planck yields $\Sigma m_\nu<0.24 \,\eV$~\cite{Planck:2018vyg}, at the same level or better than the ones from $0\nu 2\beta$ searches. Planck data are consistent with the standard value of $\Neff=3.044$ and exclude at 95\% CL the presence of light thermal relics decoupling after the epoch of the QCD phase transition ($\sim 200\,\MeV$). BSM neutrino properties have been tested against Planck data, finding no significant deviations from the standard model within current uncertainties.

From the ground, the Atacama Cosmology Telescope (ACT)~\cite{ACT:2020gnv} and the South Pole Telescope (SPT)~\cite{SPT-3G:2021wgf} collaborations have been complementing the reach of satellite missions with high-sensitivity, high-resolution measurements. These latest advances in CMB observations signal that we are getting closer to the stage at which CMB polarization data will become more powerful than temperature measurements in constraining cosmological parameters, including neutrino parameters. The upcoming Simons Observatory~\cite{SimonsObservatory:2018koc} (to start collecting data in 2023) and the next-generation CMB-S4 and LiteBIRD experiments~\cite{Abazajian:2019eic,CMB-S4:2016ple,CMB-S4:2022ght,LiteBIRD:2022cnt} (late 2020's) will revolutionize the field. Moreover, for the first time, current and upcoming LSS surveys (Euclid\hskip1pt\footnote{\url{https://www.euclid-ec.org}}, Rubin\hskip1pt\footnote{\url{https://www.lsst.org}}, DESI\hskip1pt\footnote{\url{https://www.desi.lbl.gov}}, Roman\hskip1pt\footnote{\url{https://roman.gsfc.nasa.gov}}, SPHEREx\hskip1pt\footnote{\url{https://spherex.caltech.edu/index.html}}) have reached or will feature a competitive level of sensitivity with CMB experiments. With more refined polarization measurements, the combination of complementary satellite and ground-based observations, as well as the combination and cross-correlation of CMB and LSS surveys, appear as the most promising path towards unveiling the most mysterious corners of cosmological and particle physics models.

In what follows, we will review the status of current constraints on neutrino properties, as well as provide prospects for the future.

\subsubsection[Cosmological constraints on the sum of neutrino masses \texorpdfstring{$\Sigma m_\nu$}{Sigma mnu}]{Cosmological constraints on the sum of neutrino masses $\mathbf{\Sigma m}_{\boldsymbol\nu}$}  \label{sec:mnu_from_cosmo}

Currently, cosmological neutrino mass constraints can be divided into constraints from CMB data alone, those obtained with CMB observations in combination with other data, and those inferred from low-redshift data alone. The \textit{Planck} Collaboration reports an upper limit of $\sum m_\nu<0.12$ eV (95\% CL) for the combination of CMB and BAO data~\cite{Planck:2018vyg}. Similarly, the ACT collaboration reports an upper limit of $\sum m_\nu<0.27$ eV (95\% CL) for the combination of ACT CMB, \textit{Planck} lensing potential measurements and BAO data~\cite{ACT:2020gnv}. The tightest bound to date on the sum of the neutrino masses is $\sum m_\nu< 0.09$~eV (95\% CL), computed by means of CMB, BAO, SNIa and growth rate measurements of large scale structure~\cite{DiValentino:2021hoh}.

An important role may be played by the phenomenological lensing parameter $A_\mathrm{lens}$, that rescales the amplitude of the lensing-induced smoothing of the acoustic peaks in the primary CMB anisotropies~\cite{Calabrese:2008rt}. Whereas the primary \textit{Planck} dataset preferred $A_\mathrm{lens}\neq 1$ at $2.8\sigma$~\cite{Planck:2018vyg}, ACT found no preference for $A_\mathrm{lens}\neq1$~\cite{ACT:2020gnv}. It has been shown that allowing  $A_\mathrm{lens}$ to vary as a free parameter in the analysis significantly weakens cosmological neutrino mass limits~\cite{PhysRevD.97.123534,RoyChoudhury:2019hls,Sgier:2021bzf,Esteban:2022rjk}. 

It is also possible to constrain the total neutrino mass with cosmological data without including CMB data. For example, several analyses based on the effective field theory of large scale structure reported neutrino mass limits between $\sum m_\nu<0.6$ (95\% CL) and $\sum m_\nu<1.2$ eV (95\% CL)~\cite{Ivanov:2019pdj,Colas:2019ret,Ivanov:2021zmi} from BOSS DR12 and eBOSS data, respectively, combined with a BBN prior for the baryon density. These results are an important cross-check for CMB-independent constraints.

The cosmological neutrino mass bound can be modified for some extensions of the $\Lambda$ Cold Dark Matter ($\Lambda$CDM) model. Some possibilities include a time-varying dark energy~\cite{Vagnozzi:2018jhn}, an effective number of neutrinos different from the canonical one, additional hot dark matter candidates, such as axions~\cite{Giusarma:2014zza,DiValentino:2015wba,Giare:2020vzo}, a  curvature component~\cite{Lorenz:2017fgo,RoyChoudhury:2019hls} or interacting dark sectors~\cite{Yang:2020uga,Mosbech:2020ahp,Stadler:2018dsa}. 

In order to allow for a potential absolute neutrino mass detection at the Karlsruhe Tritium Neutrino (KATRIN) Experiment, the cosmological neutrino bound would need to be relaxed above $\sum m_\nu=600 $ meV, as KATRIN has a projected lower sensitivity of 200 meV (90\% CL) for the electron neutrino mass~\cite{KATRIN:2022ayy}. 
There are only a few cosmological models that can achieve such a high neutrino mass bound, and an absolute neutrino mass detection at KATRIN would therefore have major implications for cosmology. For example, neutrino decays~\cite{Chen:2022idm,Barenboim:2020vrr,Abellan:2021rfq,Escudero:2020ped}, time-varying neutrino masses~\cite{Lorenz:2018fzb,Lorenz:2021alz}, non-standard neutrino distributions~\cite{Oldengott:2019lke,Alvey:2021sji,Farzan:2015pca,Escudero:2022gez}, or long range neutrino interactions~\cite{Esteban:2021ozz} have been discussed in this context. 
Upcoming large scale structure observations are expected to put tight limits on some of the extended models mentioned here.

There are also some analyses in the literature that show a preference for a non-zero value of the neutrino masses~\cite{Sgier:2021bzf,Giusarma:2014zza,Wyman:2013lza,Emami:2017wqa,DiValentino:2021rjj}. A recent combined analysis on the map-level with \textit{Planck} 2018 CMB, KiDS-1000 weak lensing and BOSS DR12 galaxy clustering data finds $\sum m_\nu=0.51^{+0.21}_{-0.24}$ eV at $2.3\sigma$~\cite{Sgier:2021bzf}. In this context, including cross-correlations between CMB (lensing) and weak lensing and galaxy clustering power spectra is a promising avenue to achieve a cosmological neutrino mass detection in the near future~\cite{Lesgourgues:2007ix,Chen:2021vba,Giusarma:2018jei,Tanseri:2022zfe}.

We conclude this section with prospects from upcoming and next-generation surveys. Simons Observatory (data taking to start in 2023) will be able to measure the sum of neutrino masses at the $1\sigma$ sensitivity level or more (depending on the true value of $\Sigma m_\nu$) with three different combinations of probes (i.e., CMB lensing reconstruction, thermal SZ power spectrum, and SZ cluster count, combined with either BAO data or weak lensing measurements)~\cite{SimonsObservatory:2018koc}, therefore providing a robust handle of this important parameter. These figures will be improved once cosmic-variance-limited measurements of CMB large-scale polarization will be available with e.g., the next CMB space mission LiteBIRD (expected launch in 2029)~\cite{LiteBIRD:2022cnt}. In any case, different combinations of next-generation surveys, both CMB-oriented and LSS-oriented, will push the sensitivity of cosmological probes down to $\sim4-5\sigma$ even in the case of $\Sigma m_\nu=0.06\,\mathrm{eV}$ (minimal mass expected in normal ordering)~\cite{Abazajian:2019eic,Brinckmann:2018owf}. 

\subsubsection[Cosmological constraints on the effective number of relativistic degrees of freedom \texorpdfstring{$N_\mathrm{eff}$}{Neff}]{Cosmological constraints on the effective number of relativistic degrees of freedom~$N_\mathit{eff}$}

Estimates of $\Neff$ can be inferred both from CMB and BBN observations. As discussed in Sec.~\ref{sec:recombination}, the most genuine effect of an increased $\Neff$ on the CMB temperature and polarization power spectra is a reduction of power at high multipoles, which correspond to small angular scales on the sky, see~\cite{Hou:2011ec,Lesgourgues:2006nd,Lattanzi:2017ubx}. At present, the most robust bound on $\Neff$ from CMB observations comes from Planck legacy data. Within the framework of $\Lambda$CDM, using the full legacy data including polarization and lensing, the Planck collaboration reports~\cite{Planck:2018vyg}:
\begin{align}
    \Neff = 2.89 \pm 0.19 \quad [\text{68\% CL -- Planck}]\,.
\end{align}
In addition, constraints on $\Neff$ can be improved by adding BAO data. This yields~\cite{Planck:2018vyg}: 
\begin{align}
    \Neff = 2.99 \pm 0.17 \quad [\text{68\% CL -- Planck+BAO}]\,.
\end{align}
Importantly, recent, independent and competitive measurements of $\Neff$ are available from ACT~\cite{ACT:2020gnv} and SPT~\cite{SPT-3G:2022hvq} CMB observations:
\begin{align}
    \Neff &= 2.42 \pm 0.41 \quad [\text{68\% CL -- ACT-DR4}]\,,\\
    \Neff &= 3.55 \pm 0.58 \quad [\text{68\% CL -- SPT-3G\,2018}]\,,
\end{align}
which are consistent with the theoretical prediction at the $1-2\sigma$ level (for a discussion on the low value of $N_\mathrm{eff}$ preferred by ACT, see Ref.~\cite{ACT:2020gnv}). Further improvements are expected from the ongoing analysis of additional data collected by both experiments over the past years and not included in the above analyses.

Finally, the effect of $\Neff$ on the CMB spectra is partially degenerate with that of the Hubble constant, $H_0$. At present, there is a large $4\sigma\!-\!6\sigma$ discrepancy between direct measurements of $H_0$ and the value of $H_0$ predicted within the framework of $\Lambda$CDM~\cite{Riess:2021jrx,Freedman:2021ahq,DiValentino:2021izs}. Although enhancing $\Neff$ cannot explain the Hubble tension~\cite{Vagnozzi:2019ezj,Schoneberg:2021qvd}, it is nevertheless interesting to consider the values of $\Neff$ that one would infer using information from local 
measurements of $H_0$ too. From this exercise (using $H_0 =  73.48\pm 1.66\,{\rm km/s/Mpc}$ as a prior), the Planck collaboration reports~\cite{Planck:2018vyg}:
\begin{align}
    \Neff = 3.27 \pm 0.15 \quad [\text{68\% CL -- Planck+BAO+}H_0\text{ prior}]\,.
\end{align}

From all these measurements we can draw a very important conclusion: current CMB observations are broadly compatible with the Standard Model prediction of  $\Neff$. This is a success of the standard models of both particle physics and of cosmology, and as we discuss below represents a stringent test on many of their extensions.

Taking the very precise baryon energy density from Planck CMB observations, standard BBN predicts the helium abundance to be~\cite{Pitrou:2018cgg}: $\yp \simeq 0.2471(2)$.
Determining the primordial helium abundance with high precision is not easy and its extraction requires modeling various physical quantities such as the electron density and temperature of the regions where $\yp$ is inferred, see e.g.~\cite{Izotov:2003xn,ParticleDataGroup:2022pth}. At present, there are a handful of determinations of $\yp$ with $\sim 1\%$ precision~\cite{ParticleDataGroup:2022pth} and which are in agreement between them.
In this context, the PDG review recommends using $\yp = 0.245 \pm 0.003$, which we can clearly see is in good agreement with the Standard Model prediction for $\yp$.
From the observational perspective, the primordial deuterium abundance is measured with high precision and with systematic uncertainties below current statistical ones. The currently recommended value by the PDG on this quantity is ${\rm D/H} = 10^{-5}\times(2.547 \pm 0.025)$~\cite{ParticleDataGroup:2022pth}. 

By taking into account all relevant nuclear reaction rates governing BBN, the primordial light element abundances can be predicted and several groups present global analyses reporting constraints on $\Neff$~\cite{Fields:2019pfx,Pisanti:2020efz,Pitrou:2018cgg}. By considering the measured values of $\yp$ from~\cite{Aver:2015iza} and ${\rm D/H}$ from~\cite{Cooke:2017cwo}, the constraints on $\Neff$ that can be derived read as follows~\cite{Pisanti:2020efz}:
\begin{align}
    \Neff &= 3.00 \pm 0.22 \quad [\text{68\% CL -- }{\rm D/H}+\omega_b^{\rm CMB}]\,, \label{eq:Neff_DH_omegab} \\
    \Neff &= 2.90 \pm 0.28 \quad [\text{68\% CL -- }\yp+{\rm D/H}]\,.\label{eq:Neff_BBN}
\end{align}
where in the first equation the value of $\omega_b = 0.02224 \pm 0.00022$ as reconstructed by Planck CMB observations is used as an input. Note that the constraint in Eq.~\eqref{eq:Neff_BBN} is governed solely by the $\yp$ abundance while the information from ${\rm D/H}$ is used to constrain $\omega_b$. In addition, combining the direct measurements of $\yp$, ${\rm D/H}$ and CMB observations of $\omega_b$, $\Neff$, and $\yp$, Ref.~\cite{Yeh:2020mgl} finds a combined inference of:
\begin{align}\label{eq:Neff_BBN_CMB}
    \Neff &= 2.91 \pm 0.15 \quad [\text{68\% CL -- }\text{BBN+CMB}]\,.
\end{align}
From these numbers we can clearly see that current BBN determinations of $\Neff$ are compatible with the Standard Model prediction of $N_\mathrm{eff}^\mathrm{SM} = 3.044$. Importantly, from Eq.~\eqref{eq:Neff_DH_omegab} we can appreciate that ${\rm D/H}$ measurements supplemented with CMB determinations of $\omega_b$ can yield competitive $\Neff$ constraints compared to those that can be obtained from the primordial helium abundance.

The agreement between CMB and BBN measurements of $\Neff$ and its Standard Model prediction represents a very powerful constrain of an array of extensions of the Standard Model of particle physics, see~\cite{Allahverdi:2020bys} for a recent comprehensive review. For example, current $\Neff$ measurements preclude the existence of massless particles that were once in thermal contact with the SM plasma at temperatures $T \lesssim 100\,{\rm MeV}$, see e.g.~\cite{Brust:2013ova}. These types of particles are predicted in an array of extensions of the Standard Model that address open problems in fundamental physics, see e.g.~\cite{Berezhiani:2000gw,Cicoli:2012aq,Weinberg:2013kea,Arkani-Hamed:2016rle} for some examples. In addition, $\Neff$ measurements can be used to constrain a myriad of other, not necessarily massless, BSM states. These include, but are not limited to, dark matter particles~\cite{Boehm:2013jpa,Sabti:2019mhn}, new force carriers~\cite{Escudero:2019gzq,Escudero:2019gfk}, axions~\cite{Cadamuro:2010cz}, and eV-scale sterile neutrinos~\cite{Gariazzo:2019gyi,Hagstotz:2020ukm}. These cosmological constraints are highly complementary to those that can be derived from laboratory experiments~\cite{Essig:2013lka}. 

Given that $\Neff$ measurements constrain important aspects of particle physics models, it is relevant to consider how they can be improved in the near future. Regarding CMB observations, the Simons Observatory~\cite{SimonsObservatory:2018koc} is expected to deliver $\Neff$ measurements with $\sigma (\Neff) \simeq 0.07$ precision in $\sim 5$ years. This will represent a precision a factor of $\sim 2.5$ better than current Planck constraints. In order to go beyond this precision, two distinct types of ultrasensitive CMB experiments are on the table: ground based experiments, such as CMB-S4~\cite{CMB-S4:2016ple,Abazajian:2019eic,CMB-S4:2022ght}, or satellite missions, such as PICO~\cite{NASAPICO:2019thw} or CORE~\cite{CORE:2016npo}. In either case, these types of experiments could reach sensitivities at the level of $\sigma(\Neff) \simeq 0.03$.

On the BBN front, it would be desirable to promote further studies of the extraction of the primordial helium abundance as it is the most sensitive element to $\Neff$ and current measurements are dominated by systematic effects. In addition, provided that the baryon energy density is taken from CMB observations, deuterium measurements are already providing relevant constraints on $\Neff$ (see Eq.~\eqref{eq:Neff_DH_omegab}). This is important because ${\rm D/H}$ measurements are not dominated by systematic effects. However, the current bottleneck is on the theory uncertainty in the predictions of ${\rm D/H}$ which is as of today dominated by the experimental knowledge of the $d+d\to n +{}^{3}\text{He}$ and $d+d\to p +{}^{3}\text{H}$ nuclear reaction rates at the energies of interest for BBN~\cite{Pisanti:2020efz,Pitrou:2020etk,Yeh:2020mgl}. Clearly, it is would be desirable to develop experiments that could measure these reactions better~\cite{Mossa:2020gjc,Pitrou:2021vqr}. The pay off will be large, as demonstrated by the recent measurements from the LUNA collaboration~\cite{Mossa:2020gjc}. This would readily yield improved $\Neff$ constraints.

Finally, it appears imperative to improve CMB and BBN measurements on $\Neff$ simultaneously. This is desirable for two main reasons: 1) if a measurement of $\Neff$ that is discrepant with $N_\mathrm{eff}^\mathrm{SM}$ is reported from any of these probes, one would ideally like to have another complementary and equally sensitive probe to test its consistency. This is important and timely, for example, in the context of the Hubble tension as discussed above. 2) CMB and BBN measurements of $\Neff$ provide measures of the expansion rate of the Universe at different epochs. Therefore, it is important to measure $\Neff$ at both epochs with the highest precision possible in order to extract the most information about the early Universe. This is also relevant because many scenarios beyond the Standard Model predict contributions to $N_\mathrm{eff}^\mathrm{CMB}$ but not to $N_\mathrm{eff}^\mathrm{BBN}$, see e.g.~\cite{Chacko:2003dt}.

\subsubsection{Direct detection of the cosmic neutrino background}
\label{sec:CNB_detection}
BBN and CMB observations give us indirect evidence that the Universe should be filled with a Cosmic Neutrino Background (CNB). Its direct detection, however, remains elusive because of the very low energy of such neutrinos, $T_\nu^0 \simeq 1.95\,{\rm K}$. Nevertheless, in recent years the community has taken seriously the possibility to actually detect the CNB. In particular, the PTOLEMY collaboration~\cite{PTOLEMY:2018jst,PTOLEMY:2019hkd} has considered doing so via neutrino capture on beta decaying nuclei, in particular on tritium. Such measurement faces several experimental and physical challenges that may be overcome in the future, see~\cite{Cheipesh:2021fmg,Nussinov:2021zrj,Mikulenko:2021ydo,Brdar:2022wuv} for very recent studies on these issues. Clearly, directly detecting the CNB would be extremely rewarding and could potentially hold surprises, as it would correspond to a laboratory measurement of a cosmological background which need not be the one expected in the Standard Model, see e.g.~\cite{Alvey:2021xmq}. We also mention that recent works have pointed out how a direct measurement of the CNB could help distinguish between the Dirac and Majorana neutrino nature~\cite{Hernandez-Molinero:2022zoo,Hernandez-Molinero:2023jes}. It appears clear that detecting the CNB is a task that merits to be pursued. It also represents a problem where collaboration between cosmologists, particle physicists, nuclear physicists, and material scientists is needed. In parallel, new bold ideas to detect the CNB would be most welcome, see~\cite{Weinberg:1962zza,Stodolsky:1974aq,Weiler:1982qy} for some old suggestions and~\cite{Ringwald:2004te,Yoshimura:2014hfa,Domcke:2017aqj,Akhmedov:2019oxm,Bauer:2021uyj} for some more recent ones. Finally, we note that searches for the CNB have been performed in neutrino mass experiments~\cite{KATRIN:2022kkv,Lobashev:1999dv,Robertson:1991vn}, but they are sensitive only to very exotic scenarios where the neutrino number density on Earth is $\sim 10^{10}$ times larger than the value expected within the standard cosmological model, $n_\nu^{\rm SM} \simeq 112\,{\rm cm}^{-3}$.

\subsubsection{Constraints on nonstandard neutrino cosmologies}
\label{sec:bsm_cosmo_constraints}

In this section, we collect a summary of cosmological constraints on a variety of beyond-standard-picture neutrino models and properties.  In what follows, we will briefly summarize the cosmological bounds for: sterile neutrino properties (mass and mixing angles); neutrino non-standard interactions; neutrino chemical potential; neutrino lifetimes; neutrino magnetic moments; and low-reheating temperature scenarios. For the sake of clarity, we will discuss separately bounds for different ranges of sterile masses, from the lightest to the most massive.  By reading this section, it will be clear how a synergic approach that combines information from multiple observational sources is often key to constrain the parameter space spanned by these BSM models.

\paragraph{Sterile neutrinos in the eV-mass range}
Sterile neutrinos with masses of $\mathcal{O}({\rm eV})$ have been proposed as an explanation for anomalous results observed in short-baseline and reactor neutrino experiments~\cite{LSND:2001aii,MiniBooNE:2018esg, Abdurashitov:2005tb, Abdurashitov:2005tb, Mention:2011rk, Giunti:2011gz,Kopp:2013vaa,Dentler:2018sju, Gariazzo:2017fdh,Diaz:2019fwt,Boser:2019rta}. 
For the mass and mixing parameter preferred by laboratory anomalies, eV sterile neutrinos would be copiously produced in the early universe via oscillations and contribute to $N_\mathrm{eff}$ an additional degree of freedom~\cite{Mirizzi:2013kva}. However one fully thermalized  sterile neutrino is strongly disfavored by BBN computations and observations~\cite{Fields:2019pfx,Cooke:2017cwo,Hagstotz:2020ukm,Grohs:2019cae,Hsyu:2020uqb}. From a purely phenomenological point of view, the inclusion of an additional light sterile neutrino family in the early Universe would impact both estimates of $N_\mathrm{eff}$ (via the contribution to the energy density at early times) and of the total matter density (via the contribution at late times, after transitioning to non-relativistic regime). The combination of the full suite of Planck data and BAO measurements is compatible with the presence of an additional light sterile neutrino provided that $N_\mathrm{eff}<3.34$ (95\% C.L.) and that the effective mass of the sterile\footnote{The effective mass is a phenomenological parameter that quantifies the contribution of the sterile state to the matter density. Similarly to the case of active neutrinos, it is defined in terms of the non-relativistic neutrino energy density, normalized to the case of instantaneous neutrino decoupling, $m_\mathrm{eff}=\Omega_\nu h^2 (94.1)\,\mathrm{eV}$.} is $m_\mathrm{eff}<0.23\,\mathrm{eV}$ (95\% C.L.)~\cite{Planck:2018vyg}. These limits can be converted into constraints on physical properties of the sterile neutrinos (physical mass and mixing parameters) -- thus allowing comparison with terrestrial constraints from e.g., flavor oscillation experiments -- once a production mechanism for populating the sterile state is specified. Recently, a consistent framework to set limits on light sterile-three active neutrino couplings in the early Universe has been developed~\cite{Gariazzo:2019gyi}. With the full suite of Planck data, it was found that the 3+1 active-sterile mixing matrix elements $|U_{a4}|^2$, with (a=e,$\mu$,$\tau$) must be smaller than $10^{-3}$~\cite{Hagstotz:2020ukm}, confirming severe tension between cosmology and a subset of anomalous short-baseline neutrino oscillation results. 

Further investigations are needed, as the tension between the the eV-scale anomalies and cosmological bounds could imply yet other manifestations and hints of new physics. Different mechanisms have been proposed  to  suppress the sterile abundance  and consequently their  thermalization. A class of solutions involve the existence of non-standard interactions in the neutrino sector ($\nu$NSI). Constraints on $\nu$NSI are presented in the dedicated paragraph below. Another proposed mechanism involves a neutrino-antineutrino asymmetry L$_\nu$ in the evolution equation for the active-sterile system, via a  in-medium  suppression of the mixing angle~\cite{Chu:2006ua,Abazajian:2004aj}.  In recent studies, it was found that a value of  L$_\nu \sim 10^{-2}$ can block  the active-sterile  flavor  conversions, keeping the sterile contribution to $\Neff$ more in agreement with the value preferred by BBN. However, the inclusion of the neutrino asymmetry shifts the active-sterile oscillation at lower temperatures. For values of L$_{\nu}\sim 10^{-2}$, the conversions start about the time of the active neutrino decoupling~\cite{Mirizzi:2012we}. This causes a depletion of active neutrinos since the active species oscillated in the sterile one are not repopulated by collisions anymore. Distortions in the electron (anti-)neutrino spectra then emerge, impacting the production of the BBN yields. As a result, the tension between BBN predictions and the eV interpretation of the oscillations anomalies remains~\cite{Saviano:2013ktj}. More so, cosmology prior to BBN is unknown and in motivated theories could be significantly distinct than the standard cosmology typically assumed. As demonstrated in Ref.~\cite{Gelmini:2019clw,Gelmini:2019esj,Gelmini:2019wfp,Gelmini:2020duq,Chichiri:2021wvw}, the cosmological sterile neutrino bounds are not robust to these uncertainties and laboratory sterile neutrino searches may constitute a sensitive probe of the pre-BBN epoch. 
 
\begin{figure}[t!]
\centering
\includegraphics[width=5in]{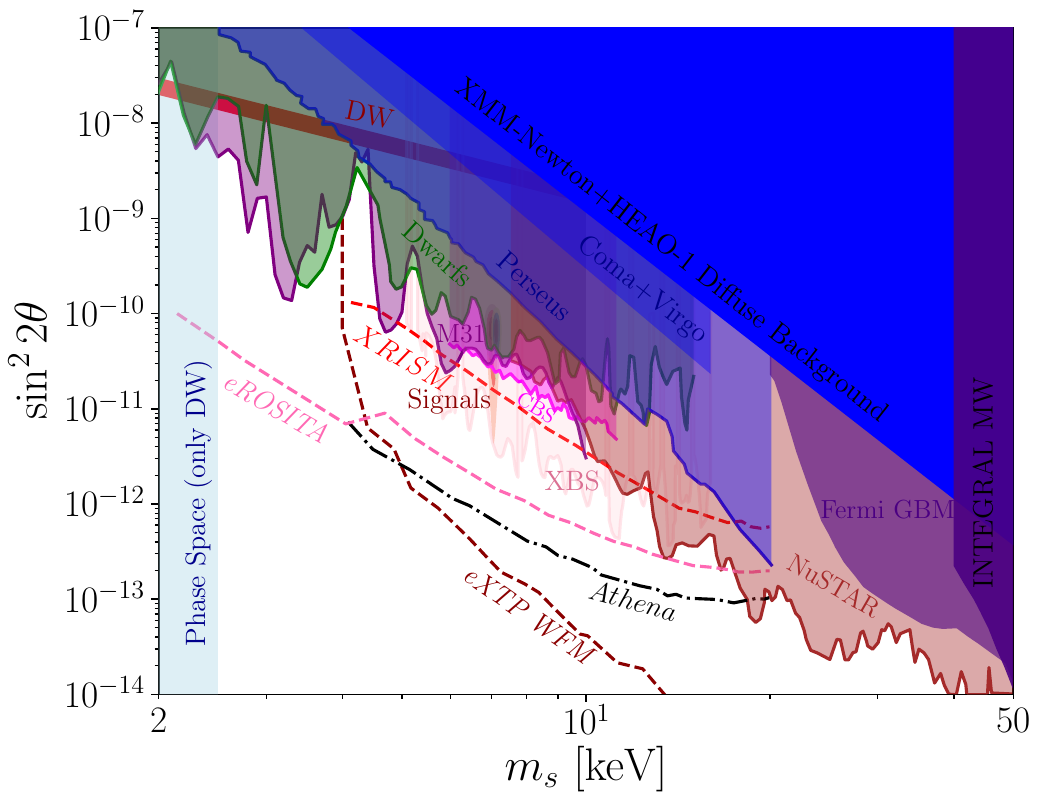}
\caption{Shown is the full parameter space for sterile neutrino dark matter,
  in the case of it comprising all of the dark matter, independent of its model of production. Among the most
  stringent constraints at low energies and masses are constraints
  from X-ray observations M31 Horiuchi et al.~\cite{Horiuchi:2013noa},
   stacked dwarfs~\cite{Malyshev:2014xqa}, 99\% upper limit from 51 Ms of blank sky data from \textit{Chandra}~\cite{Sicilian:2020glg}, and the claimed  95\% limit from blank sky \textit{XMM-Newton} data~\cite{Foster:2021ngm}. Also shown are
  constraints from the diffuse X-ray background
 ~\cite{Boyarsky:2005us}, and individual clusters ``Coma+Virgo''
 ~\cite{Boyarsky:2006zi}. At higher masses and energies, we show the
  limits from NuSTAR~\cite{Roach:2019ctw}, Fermi GBM~\cite{Ng:2015gfa} and INTEGRAL
 ~\cite{Boyarsky:2007ge}. The signals near 3.55 keV from M31 and
  stacked clusters are also shown
 ~\cite{Bulbul:2014sua,Boyarsky:2014jta}. Also shown is the forecast
  sensitivity of the planned {\it XRISM} and {\it Athena X-ray Telescope}
 ~\cite{Ando:2021fhj}, and the potential optimistic-case reach of the WFM instrument aboard the {\it eXTP X-ray Telescope}~\cite{Zhong:2020wre,Malyshev:2020hcc}. The vertical constraint at low masses comes from observations of small-scale structures and
  specifically applies, as shown, to the Dodelson-Widrow model being all of the
  dark matter, labeled ``DW''. Together with the model-independent X-ray bounds, this shows how Dodelson-Widrow sterile neutrinos are now excluded as all of the
  dark matter. The Dodelson-Widrow model could however still produce sterile
  neutrinos as a fraction of the dark matter. Other production mechanism can be similarly constrained by observations of small-scale structures, see e.g.~\cite{Zelko:2022tgf,Schneider:2016uqi,Cherry:2017dwu,DES:2020fxi} (not shown in the figure).}
  \label{fig:sterileDM}
\end{figure}

\paragraph{Sterile neutrinos in the keV-mass range}
$\mathcal{O}({\rm keV})$ sterile neutrinos became of interest as dark matter candidates~\cite{Dodelson:1993je,Shi:1998km} and also in relation to a possible mechanism for generating pulsar kicks~\cite{Kusenko:1997sp,Fuller:2003gy}. 
Constraints on sterile neutrinos strongly benefit from a multi-probe approach; see e.g.~\cite{Boveia:2022syt,Adhikari:2016bei} and refs. therein. Potential signals of keV sterile neutrino dark matter arise from their radiative decay~\cite{Abazajian:2001vt} as observable by current and future X-ray space telescopes. Complementary, albeit model-dependent, bounds can be obtained from probes of cosmological structure formation~\cite{Abazajian:2005gj,Abazajian:2005xn}. Indeed, as warm dark matter candidates, keV sterile neutrinos would suppress structure formation on scales that are accessible via observations of Lyman-alpha forests, strong lensing and dwarf/satellite galaxies; see e.g.~\cite{Zelko:2022tgf,Schneider:2016uqi,Cherry:2017dwu,DES:2020fxi}.
The model dependency of these bounds stems from the fact that their interpretation in terms of the sterile neutrino parameters relies on the knowledge of the distribution function, and thus ultimately on the mechanism producing sterile neutrinos in the early Universe. A model-independent constraint $m_s > 0.13\,\mathrm{eV}$ (95\% CL) can however be obtained from observations of cosmic structures through phase-space arguments~\cite{Alvey:2020xsk}, in the spirit of the classic Tremaine and Gunn argument.\footnote{Due to the exclusion principle, the number of fermions in a given phase-space volume is limited. Therefore, light neutrinos cannot account for the total virial mass of cosmic structures.} Figure~\ref{fig:sterileDM} shows the parameter space of mass and mixing angle relevant for sterile neutrino dark matter where it comprises all of the dark matter~\cite{Abazajian:2017tcc,Abazajian:2021zui}. The prediction of the Dodelson-Widrow (DW)~\cite{Dodelson:1993je} model, in which sterile neutrino dark matter is produced through non-resonant oscillations from the thermal bath of active neutrinos, is shown. The DW mechanism is now exclude by a combination of X-ray observations and probes of structure formation. Other models --- either based on production through oscillations, or through other mechanisms---exist in the remainder of the parameter space.

\begin{figure}[t]
\centering
\includegraphics[scale=0.33]{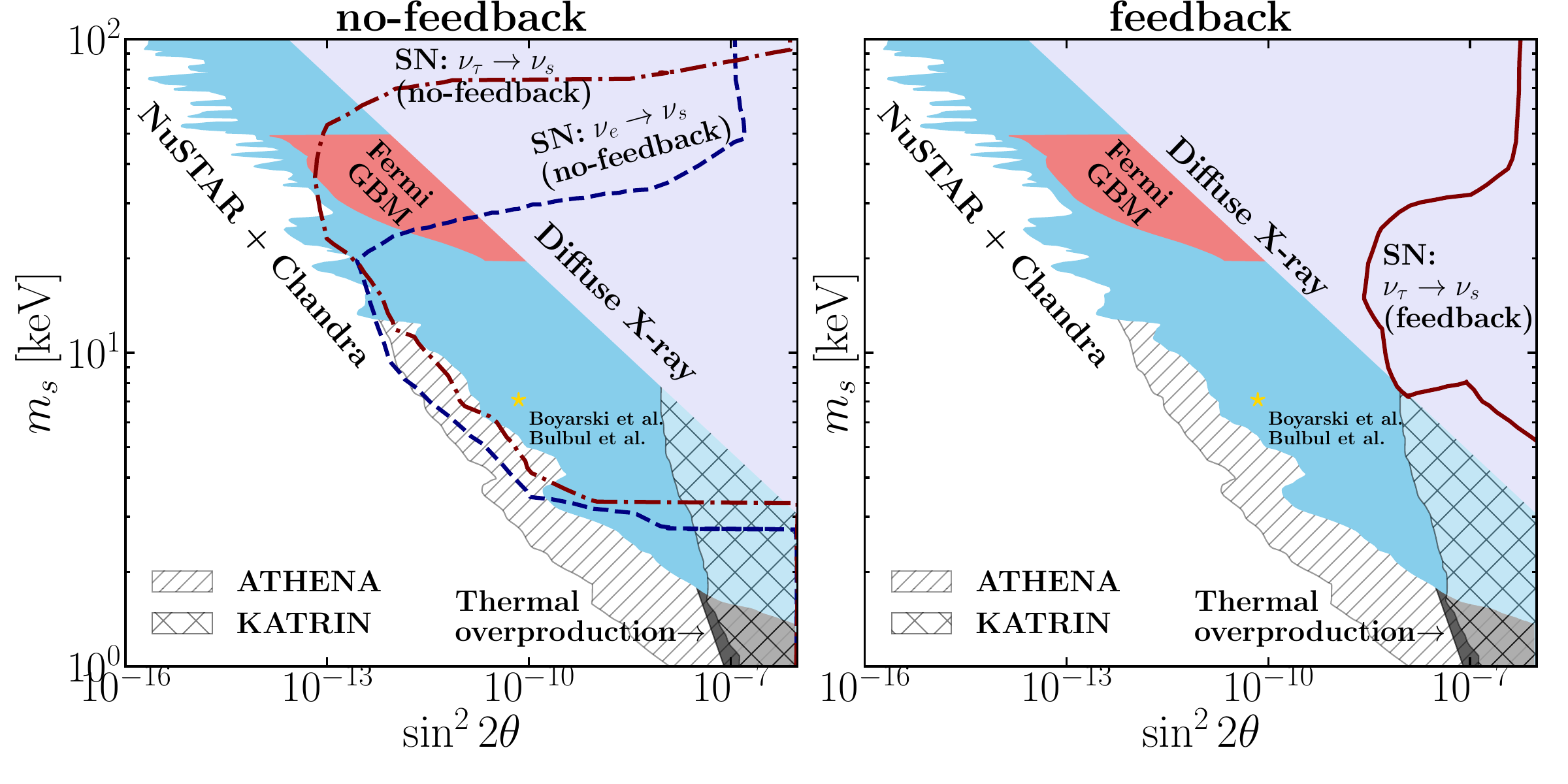}
\caption{Existing and prospective bounds on the sterile neutrino dark matter. The existing limits come from the constraints set by the observations of the M31 galaxy in X-rays~\cite{Ng:2019gch,Horiuchi:2013noa} (NuSTAR$+$Chandra - blue region), the constraints from the observations of the diffuse X-ray background~\cite{Boyarsky:2005us,Abazajian:2006jc} (solid light purple region), the Fermi Gamma-Ray Burst Monitor all-sky spectral analysis~\cite{Ng:2015gfa} (solid red region), the overproduction of the sterile neutrinos according to the Dodelson-Widrow mechanism~\cite{Dodelson:1993je,Abazajian:2017tcc} (solid grey region); The prospective sensitivities of ATHENA~\cite{Neronov:2015kca} and KATRIN/TRISTAN~\cite{Mertens:2018vuu} experiments are depicted as the hatched regions. The SN exclusion limits derived in Refs.~\cite{Suliga:2019bsq,Suliga:2020vpz} without dynamical feedback are plotted as a blue dashed (red dash-dotted) line for the $\nu_e-\nu_s$ ($\nu_\tau - \nu_s$) mixing on the left panel and on the right panel -- the results obtained in case of the calculations with the feedback. The ($\sin^2 2\theta, m_s$)  parameter space turns out to be unconstrained for the $\nu_e-\nu_s$ mixing case (and nearly unconstrained for the $\nu_\tau - \nu_s$ mixing) once the dynamical feedback originating from the production of sterile neutrinos is included in the simulations.}
\label{fig:Fig-intro}
\end{figure}

Constraints can also be placed on the keV-mass scale from supernovae dynamics, thanks to: 1) the feasibility of depleting the energy supply stored in active neutrino flavors via the sterile-active neutrino conversions that could thwart an explosion~\cite{Shi:1993ee,Nunokawa:1997ct,Abazajian:2001nj,Hidaka:2006sg,Hidaka:2007se,Raffelt:2011nc,Arguelles:2016uwb,Warren:2014qza,Warren:2016slz,Syvolap:2019dat,Suliga:2019bsq,Suliga:2020vpz}; 2) modification of the active neutrinos, electrons, and nucleons chemical potentials~\cite{Hidaka:2006sg,Hidaka:2007se,Raffelt:2011nc,Syvolap:2019dat,Suliga:2019bsq,Suliga:2020vpz}. Figure~\ref{fig:Fig-intro} illustrates the prospective bounds on the keV sterile neutrinos from core-collapse supernovae from a multi-zone simulations~\cite{Suliga:2019bsq,Suliga:2020vpz} together with several other limits~\cite{Horiuchi:2013noa,Boyarsky:2005us,Ng:2015gfa,Ng:2019gch,Abazajian:2006jc} and future sensitivities~\cite{Neronov:2015kca} from astrophysical observations, as well as, the region potentially excluded by the future experiment KATRIN/TRISTAN~\cite{Mertens:2018vuu}. The left panel shows the limits on the sterile neutrino dark matter from the SN1987a cooling argument ($\nu_\tau - \nu_s$ mixing limits are shown with a dash-dotted red line and $\nu_e - \nu_s$ with a dashed blue line) obtained without incorporating of the feedback coming from the sterile-active neutrino conversions, in the core of a collapsing star. The right panel shows the limits simulated including the feedback effects. 
The inclusion of the feedback effects challenges some of the previous sterile neutrino bounds and leaves the sterile neutrino mass-mixing angle parameter space relevant for dark matter searches unconstrained.

\paragraph{Sterile neutrinos in the MeV-mass range and above}
Sterile neutrinos with larger masses emerge naturally in theories beyond the Standard Model, like low-scale seesaw models in the Neutrino Minimal Standard Model in connection with the origin of the neutrino mass, and with the baryon asymmetry in the early Universe~\cite{Asaka:2005an,Asaka:2005pn}. Depending on their mixing with the active species, the parameter space of sterile neutrino in MeV mass range is  strongly constrained by collider and beam-dump experiments~\cite{Alekhin:2015byh},  searches of decays of D mesons and $\tau$ leptons~\cite{Chun:2019nwi,Orloff:2002de}. Additional constraints can also come from astrophysical environments such as core-collapse supernova~\cite{Dolgov:2000jw,Fuller:2008erj,Mastrototaro:2019vug} and  from cosmological observations and in particular from BBN data. Indeed, sterile neutrinos produced in the early universe via  mixing with  active neutrinos and in presence of collisions, can decay into lighter species injected into the primordial plasma with consequent effects on  both   $\Neff$ and the abundance of the primordial yield~\cite{Dienes:2018yoq,Poulin:2016anj,Salvati:2016jng,Mastrototaro:2021wzl,Fuller:2011qy,Gelmini:2019deq,Gelmini:2020ekg}. Early decays of sterile neutrinos with $\sim\MeV$ mass have been also studied in connection with tensions in cosmological data. As an example~\cite{Salvati:2016jng,Poulin:2015woa}, a sterile neutrino with mass $M_S = 4.35\pm 0.13$ MeV (at 95\% c.l.) and a decay time $\tau_S = 1.8\pm2.5 \cdot 10^5$ s (at 95\% c.l.) is allowed by a combination of cosmological, astrophysical and laboratory data and is able to account for the so-called Lithium problem (the disagreement between predicted and measured abundance of primordial $^7 \mathrm{Li}$). The required abundance of sterile neutrinos in this scenario is however at odd with the standard thermal history of the Universe, requiring e.g., a low reheating temperature scenario. To make another example, additional radiation produced just before BBN by decay of sterile neutrinos with mass $m_s ~ O(10)$ MeV would increase the value of $\Neff$ and thus assist in alleviating the Hubble parameter tension~\cite{Gelmini:2019deq}. Such sterile neutrinos, if primarily coupled to $\nu_\mu$ and/or $\nu_\tau$, would be within reach of the Super-Kamiokande, NA62 and DUNE experiments.

\begin{figure}[t]
\centering
\includegraphics[scale=0.7]{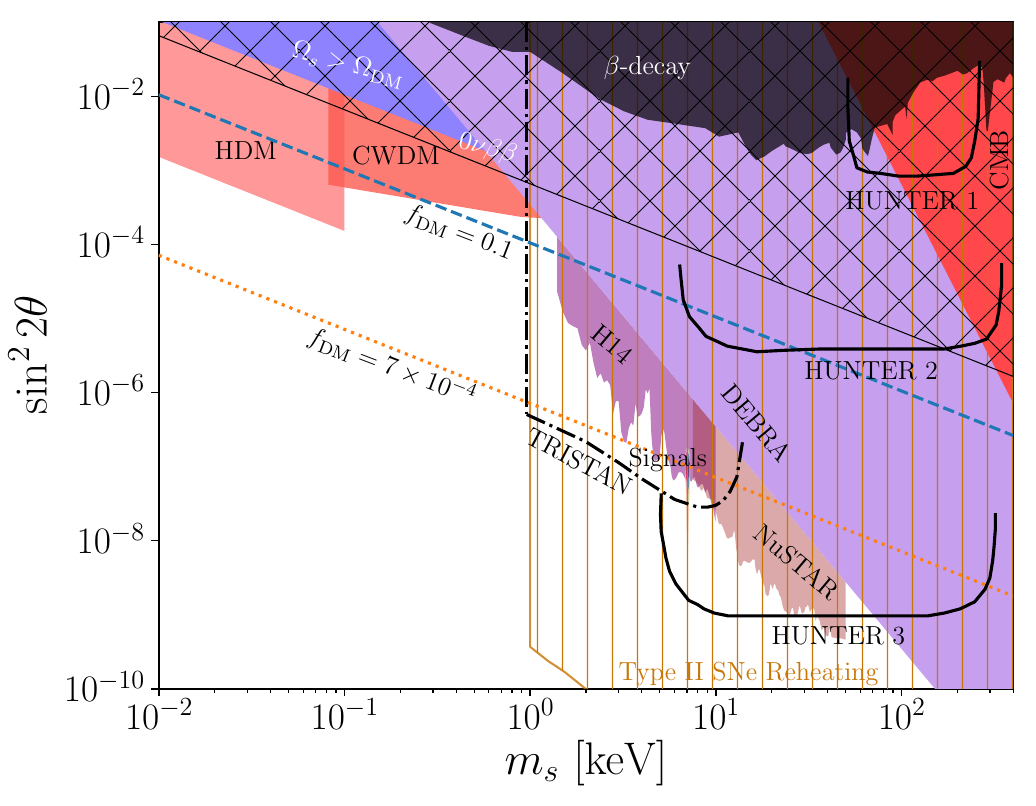}
\caption{The parameter space of interest for low-reheating temperature Universes and keV-scale neutrinos is shown ($T\approx 5\rm~MeV$) . This is an update to the limits from~\cite{Abazajian:2017tcc}, with the X-ray limits reflecting production at the given mass and mixing angle parameter space, with projected limits from the HUNTER experiment's proposed phases~\cite{Martoff:2021vxp}, and from KATRIN/TRISTAN~\cite{Mertens:2018vuu}. The diagonally hatched region refers to the double beta decay limit mentioned in the text. The red region in upper right is from CMB distortion limits. The
purple region is from diffuse x-ray background observations.}
\label{fig:hunter_lowR}
\end{figure}

The absolute reheating scale of the Universe could be as low as $\sim$5~MeV~\cite{deSalas:2015glj}. In such a scenario, the keV-scale sterile neutrino can have a much larger mixing angle and still be consistent with X-ray and structure formation bounds, while being detectable in the laboratory with HUNTER or KATRIN/TRISTAN experiments. See Fig.~\ref{fig:hunter_lowR}. In that figure, the constraints from all of the underlying processes are cosmological model-dependent, depending on
the unknown temperature of post-inflation reheating, the magnitude of any primordial lepton number
asymmetry, the assumed content of the sterile neutrino sector, and other factors. See for example the
five models considered in Figs. 1 and 2 of Ref.~\cite{Gelmini:2020duq}. The reach of HUNTER Phase 3 is free from cosmological constraints in several early Universe models, and even the reach of Phase 2 and possibly Phase 1 could be
free of constraints (e.g., in cosmologies with large lepton asymmetry, low reheating temperature and/or neutrino non-standard interactions) considering all the relevant uncertainties~\cite{Gelmini:2019clw,Gelmini:2019esj,Gelmini:2019wfp,Chichiri:2021wvw}.

Heavy neutral leptons with masses greater than $\mathcal{O}({\rm MeV})$ are a common ingredient in many neutrino mass models. Their admixture with the three light active neutrinos would alter the flavor structure and strength of neutral-current and charged-current interactions with matter. In turn, this would modify the picture of neutrino decoupling~\cite{Gariazzo:2022evs}.

\paragraph{Cosmological constraints on ${\boldsymbol\nu}$NSI}
Cosmological observables are not only sensitive to the background evolution of neutrinos but also to their perturbations, which are affected by neutrino free-streaming, as explained in Sections~\ref{sec:recombination} and~\ref{sec:LSS}. At present, measurements are broadly in agreement with the SM value of $\Neff = 3.044$ and the three SM neutrino species are constrained to be non-interacting. 
Nonetheless, a multitude of studies of non-standard neutrino interactions have been performed in the context of cosmological data.
These studies have addressed, for example, models with neutrinos interacting with a majoron~\cite{Escudero:2019gvw,EscuderoAbenza:2020egd,Escudero:2021rfi}, neutrino self-interactions~\cite{Oldengott:2017fhy,Kreisch:2019yzn,RoyChoudhury:2020dmd,Brinckmann:2020bcn,Taule:2022jrz}, and self-interacting dark radiation~\cite{Baumann:2015rya,Brust:2017nmv,Blinov:2020hmc,Ghosh:2021axu,Brinckmann:2022ajr}. 
A major conceptual difference between these works are the assumptions about the temperature scaling of the interaction, the number of neutrinos undergoing interactions and the presence or absence of additional radiation. In the following, we frame the discussion of these results around the mass of the particle mediating these interactions.\footnote{While we focus on interactions among neutrinos, we note that additional free-streaming and non-free-streaming radiation that is relevant for cosmological observables can easily arise independently of neutrinos (see e.g.~\cite{Buen-Abad:2015ova, Chacko:2015noa, Choi:2018gho} for illustrative examples).} We refer the reader to the dedicated Snowmass white paper on non-standard neutrino interactions~\cite{Berryman:2022} for further discussions. 

The scattering reactions mediated by a particle with mass larger than about \SI{1}{keV} can be described by a new Fermi interaction with an effective Fermi constant $G = g_\phi^2 / m_\phi^2$ for the purposes of CMB~physics, as described in Sec.~\ref{sec:recombination}. The constant~$G$ can be vastly larger than the electroweak Fermi constant~$G_F$ for small~$m_\phi$. This kind of interaction was first analyzed in~\cite{Cyr-Racine:2013jua}, with several follow-up studies using more sensitive CMB~data and additional cosmological datasets~\cite{Lancaster:2017ksf,Oldengott:2017fhy,Kreisch:2019yzn,RoyChoudhury:2020dmd,Brinckmann:2020bcn,Archidiacono:2013dua,Mazumdar:2020ibx,Das:2020xke}. The latest results from~\cite{RoyChoudhury:2020dmd,Das:2020xke} constrain $G < 10^{-3.4}\,\si{MeV^{-2}}$ when all three neutrinos interact. This corresponds to neutrino decoupling at $z > \mathrm{few} \times 10^5$. Interestingly, it was already found in~\cite{Cyr-Racine:2013jua} that the Planck CMB~temperature data supports a ``strongly-interacting''~(SI) neutrino mode, a region of parameter space with $G \approx \SI{e-2}{MeV^{-2}}$ (corresponding to neutrino decoupling at $z\sim 10^4$) and significant departures in the other cosmological parameters from the $\Lambda$CDM best-fit values. The SI~mode is preferred compared to the non-interacting regime if allowing for additional free-streaming radiation and only considering the Planck temperature power spectrum (in conjunction with a prior on the Hubble constant from recent local distance ladder measurements, which is significantly discrepant with CMB~$\Lambda$CDM~measurements, but without polarization data)~\cite{Kreisch:2019yzn} and would provide an interesting avenue for resolving the Hubble tension. It is to be noted that cosmologies with SI neutrinos significantly deviate from the preferred fit in $\Lambda$CDM. This signals that the effects of SI neutrinos on the CMB temperature spectrum can be compensated for by appropriately adjusting other parameters with similar impacts (e.g., the scalar spectral index $n_s$, the number of relativistic species $N_\mathrm{eff}$, the baryon density $\omega_b h^2$). CMB temperature alone is not able to efficiently break the degeneracy chain between parameters that arise in the presence of the SI~mode. Analyses that include Planck polarization data however find no such preference~\cite{RoyChoudhury:2020dmd, Brinckmann:2020bcn,Das:2020xke}. A recent analysis of ACT-DR4 data~\cite{Kreisch:2022zxp} finds an analogous preference for a SI~mode driven by intermediate-scale E-mode data. As commented in the text, this can be explained with an upward fluctuation of residuals in that region of angular scales. The final ACT-DR6 data release will be able to either confirm or rule out the significance of the SI~mode preference. In any case, it is worth stressing that minimal models that implement such strong neutrino self-interactions are strongly constrained by various laboratory probes and~BBN~\cite{Blinov:2019gcj}. 

So far, we have assumed that all three neutrino species are subject to the novel self-interaction. It is however plausible that certain flavors or mass eigenstates have significantly stronger or weaker scattering rates. This possibility was recently considered in~\cite{Brinckmann:2020bcn,Das:2020xke}. Since fewer neutrinos do not free-stream, the departure from the standard cosmology is less drastic, leading to weaker constraints on the interaction strength and the decoupling redshift.

The temperature dependence of the neutrino interaction rate determines which multipoles in the CMB spectra are impacted by the novel dynamics. In the heavy-mediator scenario considered above, all distance scales smaller than the horizon size at neutrino decoupling (equivalently all multipoles larger than some~$\ell_\nu$) are affected by the non-free-streaming of neutrinos, while larger scales evolve as in the standard model. The situation is qualitatively different if the mediator mass is light. If the scattering is larger than the Hubble rate at one time, it will remain so until the neutrino temperature becomes comparable to the mediator mass. If this occurs well after recombination, then the neutrinos are always fluid-like for the purposes of the~CMB and~BAO. This limiting scenario was considered in~\cite{Baumann:2015rya, Brust:2017nmv, Blinov:2020hmc, Brinckmann:2020bcn}, where free-streaming and non-free-streaming radiation scenarios were distinguished at high significance~(cf.~\cite{Bell:2005dr, Friedland:2007vv} for earlier works), with the latest Planck and BAO~scale data implying that at least 80\% of the neutrino energy density must be free-streaming/non-interacting~\cite{Blinov:2020hmc}.\footnote{In the past, there have been attempts to parameterize and constrain free-streaming radiation in terms of a viscosity parameter~(see e.g.~\cite{Hu:1998kj, Archidiacono:2013lva, Gerbino:2013ova, Audren:2014lsa, Planck:2015fie}), but the fiducial choice is not equivalent to free-streaming radiation and significantly differs from $\Lambda$CDM~\cite{Oldengott:2017fhy,Sellentin:2014gaa,Oldengott:2014qra}.}

A more complicated situation arises if the mediator is light, but the coupling is so small that $\Gamma < H$ at early times. In this regime, the neutrinos can recouple with the rest of the SM~bath as $\Gamma/H$ grows over time, eventually becoming cosmologically relevant. This model was explored in~\cite{Basboll:2008fx, Forastieri:2015paa,Forastieri:2019cuf}. Unlike the ``heavy'' mediator case, the posterior is unimodal with no hints of a strongly-interacting mode. The reason is that larger scales are affected by the neutrino non-free-streaming after recoupling, whereas they are $\Lambda$CDM-like in the heavy-mediator scenario. The current constraint on the dimensionless coupling in the scattering rate $\Gamma\propto g^4_\phi T$ is $g_\phi \lesssim 2\times 10^{-7}$~\cite{Forastieri:2019cuf}. This corresponds to neutrinos ceasing to free-stream only after matter-radiation equality, when they are already a subdominant component of the universe.

A final note on the ability of cosmological models with $\nu$NSI to alleviate the Hubble tension. In this context, models with an eV-scale majoron interacting with neutrinos right before recombination~\cite{Escudero:2019gvw,EscuderoAbenza:2020egd,Escudero:2021rfi} appear to be a good fit to the CMB observations while being able to substantially relax the Hubble tension~\cite{Schoneberg:2021qvd}. Along similar lines, strongly interacting dark radiation models also seem to provide a good fit to CMB observations while are somewhat less successful in ameliorating the Hubble tension~\cite{Schoneberg:2021qvd}. On the other hand, it has been shown that self-interacting neutrinos cannot ameliorate the Hubble tension~\cite{Schoneberg:2021qvd,RoyChoudhury:2020dmd,Brinckmann:2020bcn}. Importantly, in either of these scenarios, one expects substantial differences in the CMB power spectrum as compared with $\Lambda$CDM, particularly at small angular scales, or alternatively high $\ell$. This clearly highlights the relevance of upcoming ultrasensitive CMB experiments in testing these models, which could well be related to the origin of the small neutrino masses and baryogenesis~\cite{Escudero:2021rfi}.

A model based on new ``secret'' interactions among sterile neutrinos only has been proposed ~\cite{Chu:2015ipa,Hannestad:2013ana} in order to
reconcile cosmological observations with the sterile neutrino interpretation of the anomalies observed in oscillation experiments. This is achieved through the suppression of sterile abundance due to a matter term. Different choices for the mediator of the new interaction can be probed with cosmological data.
In particular,  BBN information strongly constrain the mass $M_X$ of a vector mediator. Indeed, in analogy with the case of neutrino asymmetry discussed before,  while the active–sterile neutrino mixing in the early universe is suppressed down to lower temperatures lowering the sterile contribution to to $\Neff$, the momentum spectra of active neutrinos will be  distorted due to delayed oscillations affecting the production of deuterium~\cite{Saviano:2014esa}. 
The new interaction would also change the free-streaming properties of sterile neutrinos, leaving distinct imprints in cosmological observables, as detailed in the previous sections. In fact, further considering additional information from  CMB and LSS observations, it is found that this model is strongly disfavored with respect to $\Lambda$CDM~\cite{Mirizzi:2014ama,Forastieri:2017oma,Chu:2018gxk}. 

Non-standard interactions between neutrinos and charged leptons are also a general prediction in extensions of the Standard Model accounting for neutrino masses. In particular, NSI with electrons would significantly alter the decoupling of neutrinos from the cosmic plasma, either advancing or delaying it and thus modifying~$\Neff$~\cite{Mangano:2006ar,deSalas:2016ztq,deSalas:2021aeh}.

\paragraph{Cosmological constraints on neutrino chemical potential}
While the baryon and charged lepton asymmetries are tightly constrained by observations of light element abundances and the charge-neutrality of the universe to be of $\mathcal{O}(\num{e-9})$~\cite{Kolb:1990vq}, the cosmic neutrino-antineutrino asymmetry is much less well known.\footnote{The chemical potential of a species is non-vanishing if there is a difference between the number density of particles and antiparticles.} This means that the neutrino chemical potential (parameterized by the dimensionless quantity $\xi = \mu/T_\nu$) can be large. A large chemical potential modifies the neutrino phase-space distribution and this has potentially observable effects on~BBN and the~CMB. The effects of a potential neutrino degeneracy on~BBN were already considered by Wagoner, Fowler and Hoyle~\cite{Wagoner:1966pv}, with many updated analyses since then~(see e.g.~\cite{Iocco:2008va,Pitrou:2018cgg,Freese:1982ci, Kang:1991xa, Esposito:2000hh, Esposito:2000hi, Barger:2003rt, Cuoco:2003cu, Kneller:2004jz, Simha:2008mt, Mangano:2011ip, Grohs:2016cuu}). At the beginning of~BBN, a neutrino chemical potential shifts the equilibrium ratio of protons and neutrons, and changes the neutrino energy density at a given temperature which modifies the expansion rate. The former effect is linear in the chemical potential, while the latter is quadratic which implies that the modification to~$n/p$ dominates for small values of~$\xi$. This ratio is key in determining the primordial ${^4}$He~abundance,~$\yp$, whose measurement should therefore provide a strong constraint on the neutrino asymmetry. The observed light element abundances constrain $|\xi| < 0.032$~($2\sigma$)~\cite{Pitrou:2018cgg}. The~CMB is also sensitive the neutrino chemical potential~\cite{Lesgourgues:1999wu,Lattanzi:2005qq,Castorina:2012md}. A recent analysis yields the constraint $|\xi| < 0.11$~(95\%\,c.l.)~\cite{Oldengott:2017tzj}. These results from BBN~and CMB~analyses correspond to bounds on the neutrino asymmetry $L_\nu = (n_\nu - n_{\bar{\nu}})/n_\gamma$ of $|L_\nu| < 0.24$ ($2\sigma$) and $|L_\nu|< 0.085$~(95\%\,c.l.), respectively. Finally, very recently, the EMPRESS collaboration has analyzed a new set of extremely metal poor systems and reported a value of the primordial helium abundance that is smaller than the one predicted in the Standard Model by $3\sigma$~\cite{Matsumoto:2022tlr}. Although this measurement needs to be confirmed by independent groups it could be interpreted as signal of a large lepton asymmetry in the electron neutrino flavor, $L_\nu \sim 0.01$,  see~\cite{Burns:2022hkq,Escudero:2022okz}. Improved determinations of the helium abundance, such as those projected from CMB-S4~\cite{Fields:2019pfx,Abazajian:2019eic} and direct measurements from metal-poor galaxies~\cite{Aver:2020fon} will shed light on this issue and further constrain the lepton asymmetry.

\paragraph{Cosmological constraints on neutrino lifetime}
Neutrinos are stable particles in the standard models of cosmology and particle physics. We in particular assume that neutrinos have lifetimes larger than the age of the universe when cosmologically measuring their masses~(see Sec.~\ref{sec:mnu_from_cosmo}). However, decaying neutrinos are actually a characteristic feature of many BSM~models that describe the origin of neutrino masses. This includes the minimal SM~extension in which the non-renormalizable Weinberg operator generates the masses~\cite{Petcov:1976ff, Goldman:1977jx, Marciano:1977wx, Lee:1977tib, Pal:1981rm}. While minimal scenarios typically exhibit neutrino lifetimes that are much longer than the age of the universe, this is not necessarily the case in more general models where neutrino masses are related to the spontaneous breaking of global symmetries~\cite{Gelmini:1980re, Chikashige:1980ui, Schechter:1981cv,Georgi:1981pg, Valle:1983ua, Gelmini:1983ea} (see also~\cite{Escudero:2020ped,Dvali:2016uhn, Funcke:2019grs}). This means that neutrinos could potentially decay or annihilate into lighter states on a shorter timescale than the age of the universe~\cite{Farzan:2015pca,Beacom:2004yd, Serpico:2007pt, Serpico:2008zza}, which would evade the standard cosmological neutrino mass limits.

The astrophysical, cosmological and terrestrial limits on the neutrino lifetime depend on the final states, but are generally dominated by cosmic measurements. If the decay products contain photons, the leading bounds come from limits on CMB~spectral distortions and are in excess of the age of the universe~\cite{Aalberts:2018obr}. On the other hand, decays to invisible final states are less constrained. While limits have been placed using astrophysical and terrestrial data from Supernova~1987A~\cite{Frieman:1987as}, solar neutrinos~\cite{Joshipura:2002fb, Beacom:2002cb, Bandyopadhyay:2002qg, Berryman:2014qha}, astrophysical neutrinos measured at IceCube~\cite{Baerwald:2012kc, Pagliaroli:2015rca, Bustamante:2016ciw, Denton:2018aml, Abdullahi:2020rge, Bustamante:2020niz}, atmospheric neutrinos and long-baseline experiments~\cite{GonzalezGarcia:2008ru, Gomes:2014yua, Choubey:2018cfz, Aharmim:2018fme}, these bounds are much weaker than cosmological bounds. When neutrinos decay into dark radiation while they are relativistic, the decay and inverse decay processes prevent neutrinos from free-streaming which results in a limit on the lifetime of $\tau_\nu \geq \SI{4e6}{s}\,(m_\nu/ \SI{0.05}{eV})^5$ from CMB~anisotropies~\cite{Chen:2022idm,Barenboim:2020vrr,Escudero:2019gfk,Archidiacono:2013dua,Basboll:2008fx,Hannestad:2005ex}. In the case of non-relativistic decay, the current neutrino mass limits from CMB~and LSS~observations are relaxed to $\sum m_\nu < \SI{0.42}{eV}$~(95\%\,c.l.)~\cite{Abellan:2021rfq,Lorenz:2021alz,Chacko:2019nej,Chacko:2020hmh}. In all cases, future cosmological measurements of the CMB~spectrum, the CMB~anisotropies and the large-scale structure of the universe are forecasted to result in improvements by orders of magnitude over the current bounds on the neutrino lifetime.

\paragraph{Cosmological constraints on neutrino magnetic moments}
In the Standard Model of particle physics, neutrinos are electrically neutral and do not possess electric or magnetic dipole moments. However, in many extensions of the theory, neutrinos acquire non-zero (and even sizeable) electromagnetic properties via quantum loop effects. In particular, the interactions resulting from non-vanishing dipole moments flip chirality and have the potential to significantly alter the cosmological history of the universe. For instance, in the presence of non-zero magnetic moments, neutrino interactions with charged leptons would be enhanced, delaying neutrino decoupling, leading to a larger \Neff~and altering the abundance of primordial elements~\cite{Morgan:1981psa,Morgan:1981zy,Fukugita:1987uy,Vassh:2015yza}. For Dirac neutrinos, magnetic moment-mediated interactions such as electron-positron annihilation and elastic scattering on electrons would result in a population of (sterile) right-handed neutrinos~\cite{Dolgov:2002wy,Brdar:2020quo,Elmfors:1997tt}. In the early universe, it is also relevant to consider the spin-flip process undergone by left-handed neutrinos, which results from non-vanishing magnetic moments and external electromagnetic fields~\cite{Carenza:2022ngg,Li:2022dkc}. Finally, non-vanishing magnetic moments would lead to neutrino radiative decays too (see previous discussion on neutrino lifetime).

\paragraph{Cosmological constraints on low-reheating temperature scenarios}
The constraints discussed so far assume that the Universe underwent a standard thermal history; in particular, that
it was radiation-dominated since well before the time of neutrino decoupling. This implies, among other things, that neutrinos had enough time 
to come to equilibrium with the electromagnetic plasma. 

This situation can be modified in low-reheating scenarios, in which the latest 
reheating\footnote{In this context, the term denotes generically the thermalization of the decay product of a massive particle, and is not necessarily related to the reheating process at the end of inflation.} episode in the history of the Universe results in a reheating temperature $T_\mathrm{RH}$, i.e. the temperature at the beginning of the radiation-dominated era, as low as a few MeV~\cite{deSalas:2015glj,Kawasaki:1999na,Kawasaki:2000en,Giudice:2000ex,Giudice:2000dp,Hannestad:2004px,Ichikawa:2005vw}.
For values of $T_\mathrm{RH}$ close to the neutrino decoupling temperature, the thermalization of the neutrino background would be incomplete, and neutrino spectra would not present an equilibrium form at the same temperature as photons.
In particular, the energy density of neutrinos would be smaller with respect to the standard scenario, resulting in $\Neff < 3.044$. This change in the effective number of relativistic species, as well as, more in general, distortions in the neutrino spectra caused by the incomplete thermalization, can be used to constrain $T_\mathrm{RH}$ from cosmological observations. Measurements of the abundances of light elements limit the reheating temperature to be $> 4.1\,\mathrm{MeV}$, while CMB anisotropies provide the slightly tighter bound $T_\mathrm{RH}> 4.7\,\mathrm{MeV}$~\cite{deSalas:2015glj}.

Another interesting feature of low-reheating scenarios is that the constraints on neutrino masses can in principle be relaxed. The number
density $n_\nu$ of neutrinos is reduced if thermalization is incomplete, and since cosmological observations do in fact constrain
$\Omega_\nu h^2$, i.e. the product $n_\nu m_\nu$, this allows for larger values of the neutrino mass. Ref.~\cite{deSalas:2015glj} has however shown that for Planck 2015 data the effect, while present, is marginal for the values of $T_\mathrm{RH}$ allowed by observations.

%% file: lab_probes.tex
\section{Laboratory probes of neutrino properties}\label{sec:lab_probes}

Laboratory experiments allow for a more direct and controlled probe of neutrino properties.  In this sense, they are complementary to cosmological observations.

Flavor oscillation experiments have provided the first evidence that neutrinos have a mass.
To date, the best information that we have on the neutrino mixing matrix, as well as the squared neutrino mass differences, comes from oscillation experiments.
These properties can hardly, if not at all, be measured through cosmological observations.
Despite providing information about the mass splittings of neutrinos, oscillation experiments paradoxically do not provide any information on the absolute scale of neutrino masses other than imposing a lower limit.

This absolute mass scale can be probed, for example, by measuring the endpoint in the energy distribution of electrons emitted in the $\beta$-decay of tritium.
Albeit less sensitive than other probes, such as cosmology or searches for neutrinoless double $\beta$-decay, such a ``direct measurement'' has the attractive feature of being model independent, relying simply on energy conservation.
Searches for the lepton-number violating $0\nu2\beta$ decay, on the other hand, provide more stringent limits on the neutrino mass scale, at the price of model dependency, as will be detailed more in the following.
An observation of $0\nu2\beta$ decay would also indicate that neutrinos are Majorana particles.
Note that even if $\beta$-decay experiments, $0\nu2\beta$ searches, and cosmological observations are all able to probe the absolute mass scale, nevertheless they do so by measuring distinct combinations of the mass eigenvalues and of the elements of the mixing matrix.
This adds another fundamental layer of complementarity to these classes of experiments.

More exotic possibilities can also be explored in the laboratory.
For example, signatures of nonstandard neutrino interactions can possibly contribute to flavor oscillations, to the amplitude for $0\nu2\beta$ decay, or to coherent neutrino scattering.
Similarly the existence of a sterile neutrino might affect the pattern of neutrino oscillations (and indeed, some anomalies observed in oscillation experiments have been interpreted as a hint in this direction), the kinematics of $\beta$-decay, or the $0\nu2\beta$ signal.
These searches for nonstandard interactions and sterile neutrinos are complementary to cosmological probes, as they explore different ranges in terms of mass, energy, and couplings.

In the following sections we will review in more detail these laboratory probes of neutrino properties.

\subsection{Neutrino flavor oscillations} 
\label{sec:oscpreliminary}

Neutrino flavor oscillations have been robustly
established by the data from solar, atmospheric, reactor and long baseline neutrino
experiments, unambiguously proving that neutrinos
are massive particles and thus \textit{providing the first laboratory departure from  the SM of particle physics}.
For this reason, the Royal Swedish Academy of Sciences awarded the 2015 Nobel Prize in Physics to
Takaaki Kajita and
Arthur B.\ McDonald
\textit{for the discovery of neutrino oscillations, which shows that neutrinos have mass.
[...] New discoveries about the deepest neutrino secrets are expected to change our current understanding of the history, structure and future fate of the Universe},
see
Refs.~\cite{Super-Kamiokande:1998kpq,SNO:2002tuh,SNO:2001kpb,KamLAND:2002uet,DayaBay:2012fng,T2K:2011ypd}.

The most economical way to accommodate these observations is via the three-neutrino oscillation framework. Neutrino oscillations are described by seven parameters: two mass squared differences\footnote{$\Delta m_{ij}^{2} \equiv m_i^2 -m_j^2$} ($\Delta m_{21}^2$ and $\Delta m_{31}^2$), three Euler angles ($\theta_{12}$, $ \theta_{23}$ and $\theta_{13}$), one Dirac CP phase ($\delta_\mathrm{CP}$), and the mass ordering. The neutrino flavor transition probabilities have an oscillatory behavior: the oscillation length is $L\sim 4\pi E/\Delta m^2$ and the amplitude is proportional to the elements of the three-neutrino PMNS mixing matrix in Eq. (\ref{eq:PMNS}). Notice that the neutrino oscillation physics is only sensitive to the squared mass differences $\Delta m^2_{ij}$ and not to the total neutrino mass scale, as is the case for cosmological or direct kinematic searches.

Depending on the neutrino energy ($E$) and the distance between the source and the detector ($L$), oscillation experiments focus on one or several possible neutrino sources,  such that $E/L\sim \Delta m^2$, see Tab.~\ref{tab:exps1}. Terrestrial (accelerator and reactor) neutrino oscillation experiments are usually further classified as short-baseline experiments (SBL) and long-baseline experiments (LBL). SBL experiments are characterized by detection distances of $L\sim$ one hundred meters to several kilometers, while LBL experiments instead make use of distances $L\sim$ several hundred or thousand of kilometers.

In many extensions to the SM, the lepton sector may violate CP.
If SM leptons do not obey CP symmetry, then processes involving neutrinos and antineutrinos will have measurable differences in the laboratory. If only three neutrinos exist with distinct masses and unitary mixing, then all CP violation in the lepton sector is described by three complex phases, two of which are unphysical (i.e., they can be re-absorbed with field redefinitions) if neutrinos are Dirac fermions.\footnote{
In the case of Majorana neutrinos, all three phases are instead physical. However, two of the CP-violating phases can only be measured in processes where the neutrino mass is relevant and where lepton number is not preserved, for instance if neutrinoless double-beta decay is observed. Thus, there is still only one phase that is relevant for oscillation phenomenology.} In the PMNS parameterization of the neutrino mixing matrix, the third phase is indeed the Dirac $\delta_\mathrm{CP}$ discussed above.

\begin{table}[h!]\centering
\renewcommand{\arraystretch}{1.4}
\begin{tabular}{c|c|c}
\hline
Experiment & $L$ (km) & $E$ (MeV) \\
\hline \hline
solar & $10^7$ & $1$ \\
atmospheric & $10-10^4$& $10^2-10^5$\\
reactor & $10^{-1}-10$ & $1$ \\
LBL& $10^2-10^3$& $10^4$\\
\hline
\end{tabular}
\caption{Order of magnitude of the neutrino energy, $E$, and baseline of the experiment, $L$, for different neutrino sources and/or experiments.}
\label{tab:exps1}
\end{table}
The standard way to connect the solar, atmospheric, reactor and accelerator data with some of the oscillation parameters listed above is to identify the two mass splittings and the two mixing angles which drive the solar and atmospheric transitions with ($\Delta m_{21}^2$, $\theta_{12}$) and ($|\Delta m_{31}^2|$, $\theta_{23}$), respectively.
Table~\ref{tab:exps} (extracted from Ref.~\cite{deSalas:2020pgw}) show where our current experimental knowledge of the oscillation parameters is coming from, namely, from solar (Sol), atmospheric (Atm), short-baseline reactor (Reac), long-baseline accelerator experiments (LBL) as well as the  long-baseline reactor experiment KamLAND~\cite{KamLAND:2008dgz}.
\begin{table}[t!]\centering
\renewcommand{\arraystretch}{1.4}
\begin{tabular}{c|c|c}
\hline
Parameter & Main contribution & Other contributions \\
\hline \hline
$\theta_{12}$ & Sol & KamLAND \\
$\theta_{13}$ & Reac & Atm+LBL and Sol+KamLAND \\
$\theta_{23}$ & Atm+LBL & - \\
$\delta_\mathrm{CP}$ & LBL & Atm  \\
$\Delta m^2_{21}$ & KamLAND & Sol \\
$|\Delta m^2_{31}|$ & LBL+Atm+Reac & - \\
MO &  LBL+Reac and Atm & - \\ \hline
\end{tabular}
\caption{Summary of the set of experiments contributing to the determination of each of the oscillation parameters in the three neutrino picture.}
\label{tab:exps}
\end{table}
As a result of matter effects in the Sun, we know that $\Delta  m_{21}^2>0$.\footnote{Note that the observation of matter effects in the Sun constrains the product $\Delta m_{21}^2\cos 2\theta_{12}$ to be positive. Therefore, depending on the convention chosen to describe solar neutrino oscillations, matter effects either fix the sign of the solar mass splitting $\Delta m_{21}^2$ or the octant of the solar angle $\theta_{12}$, with $\Delta m_{21}^2$ positive by definition.} In practice, the atmospheric mass splitting $\Delta m_{31}^2$ is only  measured  via neutrino oscillations in vacuum. These oscillations are only sensitive to the absolute value of the atmospheric mass gap. Consequently, the sign of $\Delta m_{31}^2$ is still unknown and the two possibilities have been dubbed as  \textit{normal ordering} (NO, $\Delta m_{31}^2>0$) and \textit{inverted ordering} (IO, $\Delta m_{31}^2<0$). Current oscillation data can be remarkably well described with a solar mass splitting of $\Delta m_{21}^2\sim 7.5 \cdot 10^{-5}$~eV$^2$ and an atmospheric mass splitting of $|\Delta m_{31}^2|\sim 2.5 \cdot 10^{-3}$eV$^2$~\cite{deSalas:2020pgw,Esteban:2020cvm,Capozzi:2021fjo}, see Sec.~\ref{sec:global} for details. 

Since neutrino oscillation data measure the two above  distinct mass gaps, we know that there must be, at least, two massive neutrinos in nature: these two neutrinos should have a mass above $\sqrt{\Delta m_{21}^2}>0.008$~eV. In addition, one of these two neutrinos should have a mass above $\sqrt{|\Delta m_{31}^2|}>0.05$~eV. Consequently, neutrino oscillation measurements impose a bound on the sum of the neutrino masses, which reads as
\begin{eqnarray}
\sum m_\nu^{\rm{NO}}
&=&
m_1
+ \sqrt{m_1^2 + \Delta m_{21}^2}
+ \sqrt{m_1^2 + \Delta m_{31}^2 }\gtrsim 0.06\ \textrm{eV} \, ,\\ \nonumber
\sum m_\nu^{\rm{IO}}
&=&
m_3
+ \sqrt{m_3^2+ |\Delta m_{31}^2| }
+ \sqrt{m_3^2+ |\Delta m_{31}^2| + \Delta m_{21}^2} \gtrsim 0.10\ \textrm{eV} ~.
\end{eqnarray}
Despite the high precision of current measurements of the neutrino oscillation parameters, there are still a number of crucial unknowns in the leptonic mixing sector. Specific remaining unknowns are the value of the leptonic CP violating phase ($\delta_\mathrm{CP}$), the ordering of the neutrino mass spectrum (MO), and the octant of the atmospheric mixing angle. For a discussion of these and other aspects of the current status of our knowledge, see Section \ref{sec:global}.

\subsection{Absolute scale of neutrino masses}
As we have seen in Sec.~\ref{sec:oscpreliminary}, neutrino oscillations are not sensitive to the absolute mass scale. Other laboratory probes can be employed to get a handle of this fundamental property of neutrinos, as we shall see in the following.

\subsubsection{Kinematic measurements}
Kinematic measurements of weak decays involving a neutrino or anti-neutrino provide the only model-independent information on the absolute neutrino mass scale~\cite{Formaggio:2021nfz}.
These measurements earn the term ``direct measurements'' as they only rely on energy and momentum conservation to derive their neutrino mass constraint.
The effect of neutrino mass is most clearly exhibited at the endpoint of the electron energy spectrum of the decay, as demonstrated in the early 1930's by the work of Perrin~\cite{Perrin1933} and Fermi~\cite{Fermi1934}.

The relevant observable is the electron-weighted neutrino mass:
\begin{equation}\label{eq:mbeta}
    m_{\beta} = \sqrt{\sum_{i=1}^3 \left| U_{ei} \right|^2 m_i^2} \quad ,
\end{equation}
where $U_{ei}$ are elements of the $3\times 3$ unitary neutrino PMNS mixing matrix and $m_i$ are the masses of the individual neutrino mass eigenstates.
This parameter, $m_{\beta}$, is also commonly referred to as the effective electron neutrino mass or just the electron neutrino mass, the latter of which is a misnomer as the flavor or interaction eigenstates lack well-defined mass.

The incoherent sum over the mass states naturally yields a lower bound for the allowed values of $m_{\beta}$ based on the oscillation measurements of $\Delta m_{ij}^2$.
This bound is dependent on the mass ordering, but sets the ultimate target for direct searches of $m_\beta \geq 9$\,meV/$c^2$ ($m_\beta \geq 40$\,meV/$c^2$) for the normal (inverted) mass ordering~\cite{ParticleDataGroup:2022pth}.
Due to CPT symmetry, no distinction is made between the neutrino and anti-neutrino masses, as these are assumed to be equivalent,\footnote{For a CPT-violating analysis of neutrino oscillation data see Refs.~\cite{Barenboim:2017ewj,Tortola:2020ncu}} although this can be experimentally verified by measuring different appropriate decays.

Direct neutrino mass experiments measure the decay electron spectrum in the region of the endpoint.
Here the beta spectrum can be approximated:
\begin{equation}
    \frac{dN}{d\epsilon} \propto \epsilon \sqrt{\epsilon^2-m^2_\beta} \quad ,
\end{equation}
where $\epsilon$ is defined as the energy away from the endpoint ($E_0 - E$).
Two experimental signatures become evident: a shift in the endpoint energy and a distortion in the spectrum shape.

In the experimental limit of exceptional energy resolution, the spectrum will exhibit distinct contributions from each neutrino mass state individually.
These ``kinks'' in the spectrum can be utilized to constrain or validate the mass ordering simultaneously with measuring the mass scale~\cite{AshtariEsfahani:2020bfp}.
Additional physics searches for sterile neutrinos is also possible by searching for kinks at relevant energies in the spectrum, either in the vicinity of~\cite{KATRIN:2020dpx} or more distant from~\cite{Mertens:2014nha} the endpoint.

\subsubsection{Neutrinoless double-beta decay searches \label{sec:0nu2b}}

Neutrinoless double beta decay ($0\nu2\beta$) provides a model-dependent constraint on the neutrino mass scale, but with the powerful interpretation of the fundamental nature of the neutrino~\cite{Dolinski:2019nrj,DellOro:2016tmg}.
Neutrinos are unique among the fundamental fermions in their ability to be Majorana fermions, due to their lack of electric charge~\cite{Schechter:1981bd}.
Demonstration of the Majorana nature of neutrinos~\cite{Majorana:1937vz} would be a profound discovery, both shedding light on the disparate mass generation mechanism of neutrinos and introducing lepton number violation ($\Delta L = 2$).

The experimental signature of $0\nu2\beta$ is a mono-energetic peak at $Q_{\beta\beta}$, the total decay energy of a nucleus that can undergo double-beta decay.
This arises from the kinematics of the decay and is in stark contrast to the Standard Model allowed process, $2\nu2\beta$ (double beta decay accompanied by the emission of two anti-neutrinos)~\cite{Goeppert-Mayer:1935uil}, wherein the neutrinos carry away energy giving rise to a broad continuum spectrum akin to the standard single $\beta$-decay.
An experimental measurement of $0\nu2\beta$ will measure the rate of decays in the $Q_{\beta\beta}$ peak, or place a limit thereon.

Interpreting this rate into a constraint on the neutrino mass scale introduces theoretical model dependence.
Any observation of $0\nu2\beta$ does require new physics, and generically the decay rate is related to the mass scale of that physics.
The standard paradigm involves light Majorana neutrino exchange, giving rise to the relation:
\begin{equation}
    T_{1/2}^{-1} = G^{0\nu} \left| M_{0\nu} \right|^2 \left( \frac{m_{\beta\beta}}{m_e} \right)^2 \quad ,
\end{equation}
where $T_{1/2}$ is the measured decay half-life, $G^{0\nu}$ is the phase space factor, $M_{0\nu}$ is the nuclear matrix element (NME), and $m_{\beta\beta}$ is the effective Majorana mass of interest.
The Majorana mass is the coherent sum over the neutrino mass eigenstates:
\begin{equation}\label{eq:mbetabeta}
    m_{\beta\beta}=\biggr|\sum_{i=1}^3 U^2_{ei} m_i\biggr| \quad ,
\end{equation}
where the PMNS matrix must now be extended to add two additional Majorana CP-violating phase terms.
Additional theoretical uncertainties are also introduced in this interpretation due to discrepancies in calculations of the nuclear matrix element ($M_{0\nu}$)~\cite{Engel:2016xgb} and treatment of $g_A$ quenching~\cite{Dolinski:2019nrj,DellOro:2016tmg,Agostini:2022zub}.\footnote{The ``quenching'' is a renormalization in the weak axial-vector coupling $g_A$ from its free value of 1.27. The introduction of this ``effective'' $g_A$ is needed to account for discrepancies between calculation and measurement, particularly of $\beta$ and $2\nu2\beta$ decays~\cite{Suhonen:2017krv}.} The $M_{0\nu}$ variation leads to at least a factor of two to four range in $m_{\beta\beta}$ based on a $T_{1/2}$ measurement.  The $g_A$ impact is typically excluded from quoted $m_{\beta\beta}$ values, with the free value assumed.

Again the oscillation measurements of $\Delta m_{ij}^2$ impose bounds on the allowable range of values.  In the inverted ordering scenario, $m_{\beta\beta} \geq 18$\,meV, although the unknown Majorana phases broaden the allowed $m_{\beta\beta}$ values for a given set of neutrino masses.
Due to the coherent nature of the summation, $m_{\beta\beta}$ may become arbitrarily small in a small mass range allowed in the normal mass ordering for appropriate fine-tuning of the Majorana phases~\cite{ParticleDataGroup:2022pth}.

\subsection{Beyond the three-neutrino paradigm }
Going beyond the standard three-massive-neutrinos paradigm, laboratory-based experiments are sensitive to a number of SM extensions, including whether neutrinos are subject to additional interactions with SM matter beyond the weak interactions and whether additional light, neutrino-like particles exist in nature. In the following, we summarize the current understanding of these questions and how upcoming experiments plan on improving on this understanding and -- if fortunate -- making a new discovery.

\subsubsection{Non-standard neutrino interactions}
Many theories of physics beyond the SM introduce new force mediators, including those that couple neutrinos to other SM fermions. If they exist, they give rise to new neutrino interactions other than the weak interactions, often characterized as Non-standard Neutrino Interactions (NSI). Neutrino oscillation experiments have exquisite sensitivity to NSI due to the many possibilities for neutrino interactions with matter along the path of their propagation.
Other laboratory experiments, specifically those studying coherent elastic neutrino-nucleus scattering (CEvNS), are also sensitive to neutral-current NSI. We refer the reader to the dedicated Snowmass white paper on non-standard neutrino interactions~\cite{Berryman:2022} for further discussions.

\subsubsection{Sterile Neutrinos}
As we have already seen in Sec.~\ref{sec:intro_cosmo}, the possibility that additional neutral fermions exist beyond the three neutrinos of the SM has been studied for decades now, with a variety of motivations. While cosmology is sensitive to such additional species through their effects on the background and perturbation evolution, terrestrial experiments have the possibility of discovering such species through their coherent interactions with active neutrinos and production in rare decay processes. The discovery of such a new particle would drastically modify our understanding of the universe and require re-evaluation of our knowledge of early-universe physics.

If the new sterile states are light enough so that they can be produced in the neutrino source, i.e. if kinematically accessible, they can directly participate in flavor oscillations. The simplest and most popular scenario considers one additional sterile neutrino and is referred to as the 3+1 picture. In such case, they can manifest in the form of additional oscillation frequencies determined by the mass-squared differences between the active and sterile states and with amplitudes determined by the mixing. Additionally, depending on their mass, sterile states can also manifest as additional kinks in the beta-decay spectrum probed by kinematic measurements~\cite{Mertens:2018vuu,KATRIN:2020dpx,Dekens:2021qch,KATRIN:2022ith}. If neutrinos are Majorana particles, sterile states can also contribute to the definition of the allowed parameter space for $0\nu2\beta$ searches~\cite{Dekens:2021qch,Agostini:2020cpz,Bolton:2019pcu}. 

Finally, if heavy sterile neutrinos exist, they can resonantly contribute to the decay of $\tau$ leptons and heavy mesons~\cite{Goudzovski:2022vbt,Helo:2011yg}. Additional laboratory signatures of heavy sterile neutrinos can be identified in hadron collisions~\cite{DeVries:2020jbs} and high-energy electron-muon scattering. We refer to Ref.~\cite{Shrock:1980vy} for earlier studies.

\subsection{Summary of experimental results}

In this section we summarize the current status of laboratory measurements of neutrino properties. We will mainly focus on the parameters of the three-neutrino framework, namely: mass squared differences, elements of the PMNS mixing matrix and absolute mass scale. We will also briefly comment on measurements probing physics beyond the three-neutrino framework, e.g., the existence of sterile neutrino states.
Before reporting constraints from specific experiments, let us briefly recall what information each class of experiments provides.

In Sec~\ref{sec:global}, we summarize the determination of neutrino masses and mixing from the global analysis of solar, reactor, atmospheric, and accelerator neutrino experiments performed in the context of three-neutrino framework. Neutrino oscillation experiments are only sensitive to a subset of the parameters involved in the phenomenon. Global fits to oscillation data exploit the complementarity between the different experiments in order to break degeneracies and hence provide more precise oscillation parameter determinations~\cite{deSalas:2020pgw, Esteban:2020cvm, Capozzi:2021fjo}. Such analyses can be extended with the addition of information from other laboratory probes, mainly beta decay and neutrinoless double beta decay experiments\cite{deSalas:2020pgw,Capozzi:2021fjo}. 

The three neutrino picture provides a description of flavor oscillations which is consistent with data. As a result, the mass splitting, mixing angles and the Dirac CP-phase have been determined with increasing precision in the last years. Nonetheless, as it is discussed in Section \ref{sec:global}, there are still some open questions which are being targeted by current and next-generation experiments. 

Kinematic measurements, such as those based on observations of single $\beta$ decay discussed in Sec.~\ref{sec:expbeta}, will soon provide an extremely robust constraint on the absolute scale of neutrino masses. If neutrinos are assumed to be Majorana particles, the non-observation of $0\nu2\beta$ can be used to  provide complementary information on the mass scale, see Sec.~\ref{sec:expbetabeta}. Constraints from kinematic measurements are model-independent, but less tight than those
from $0\nu2\beta$ (and cosmology). Conversely, inferences on neutrino masses from $0\nu2\beta$ searches have to rely on theoretical assumptions, like the already-mentioned Majorana nature of neutrinos, or the fact that there is a direct relationship between the mass mechanism and the $0\nu2\beta$ decay rate.

Oscillation results and direct neutrino mass probes, such as $\beta$ decay, neutrino-less double $\beta$ decay and cosmological observations, can all be used to obtain information on the mass ordering. In the case of flavor oscillations, the ordering affects the oscillation pattern, as detailed in Sec~\ref{sec:global} below.
In the case of probes of the absolute mass scale, the sensitivity comes from the fact that part of the relevant parameter space is only available in the case of NO. 

Beyond the three-neutrino paradigm, there are anomalous results that could be interpreted as a consequence of the existence of eV-scale sterile neutrinos, with potentially significant consequences in particle physics and cosmology.
These are originating from measurement of neutrinos from accelerators, nuclear reactors, and from radioactive sources.
In analogy with the three-flavor neutrino results, there  are sterile neutrino global fits
designed to incorporate relevant experimental evidence. These are discussed in further detail in Sec.~\ref{sec:beyond3}. In the same section, we also address current constraints on nonstandard neutrino interactions.

\subsubsection{Neutrino masses and mixing from oscillation experiments}
\label{sec:global}

Global fits to experimental data~\cite{deSalas:2020pgw,Esteban:2020cvm,Capozzi:2021fjo,Gonzalez-Garcia:2021dve} combine solar, atmospheric, short-baseline reactor, long-baseline accelerator experiments as well as the long-baseline reactor experiment KamLAND, see Table \ref{tab:exps}. As previously detailed (see Sec.~\ref{sec:oscpreliminary}), flavor oscillations are described, within the three-neutrino picture, in terms of six parameters: two mass-squared differences, $\Delta m^2_{21}$ and $\Delta m^2_{31}$, three mixing angles, $\theta_{12}$, $\theta_{13}$ and $\theta_{23}$, and a phase accounting for possible CP violation in the neutrino sector, $\delta_\mathrm{CP}$. The sign of the mass splitting $\Delta m^2_{31}$ determines the mass ordering (MO), which can be normal (NO) or inverted (IO), for $\Delta m^2_{31} > 0$ and $\Delta m^2_{31} <0 $, respectively. 
Among these oscillation parameters, four of them are currently well measured. First of all, the solar parameters $\theta_{12}$ and $\Delta m^2_{21}$, are well determined through the combination of the data from KamLAND~\cite{KamLAND:2008dgz} and solar neutrino experiments~\cite{Cleveland:1998nv, Kaether:2010ag, SAGE:2009eeu, Bellini:2011rx, Borexino:2013zhu, Super-Kamiokande:2005wtt, Super-Kamiokande:2008ecj, Super-Kamiokande:2010tar,Nakano:PhD, SNO:2011hxd}. KamLAND is sensitive to $\sin^2 2\theta_{12}$ and thus, it is nearly insensitive to the octant of the solar mixing angle~$\theta_{12}$.\footnote{Actually, the small matter effects in KamLAND vaguely lift the degeneracy, but the  $\theta_{12}$ octant sensitivity from KamLAND-only analyses is very mild.} Nonetheless, solar neutrino experiments are sensitive to $\sin^2\theta_{12}$ through the adiabatic flavor conversion that takes place in the Sun, allowing to exclude the solution in the second octant. Regarding the mass splitting $\Delta m^2_{21}$, its determination is dominated by KamLAND. The preferred value of $\Delta m^2_{21}$ from solar neutrino experiments used to be in slight tension with the one from KamLAND. However, with the latest solar neutrino results from Super-Kamiokande, the agreement has improved~\cite{yasuhiro_nakajima_2020_4134680}. The next-generation medium baseline experiment JUNO will lead to a more accurate determination of both parameters~\cite{JUNO:2021vlw}.

Two other parameters, $\theta_{13}$ and $\Delta m^2_{31}$, have already been measured with great precision in oscillation experiments. Our current knowledge of the value of the reactor mixing angle $\theta_{13}$ is dominated by short-baseline reactor data~\cite{RENO:2018dro, jonghee_yoo_2020_4123573, DayaBay:2018yms}. Actually, the contribution from KamLAND, long-baseline, atmospheric and solar experiments is unimportant. Concerning the absolute value of the atmospheric mass-splitting $\Delta m^2_{31}$, its measurement comes from the combination of reactor and long-baseline experiments~\cite{alex_himmel_2020_3959581, patrick_dunne_2020_3959558, MINOS:2014rjg, K2K:2006yov}, although atmospheric experiments~\cite{Super-Kamiokande:2017yvm, IceCube:2017lak, IceCube:2019dqi} also contribute significantly.

Currently, there are still three open questions in the three neutrino picture. The first one is the octant of the atmospheric mixing angle $\theta_{23}$. Its measurement results mainly from the study of muon (anti)neutrino disappearance in atmospheric and long-baseline experiments. Since this channel depends on $\sin^2 2 \theta_{23}$, it is not sensitive to whether $\sin^2 \theta_{23} < 0.5$ or $\sin^2\theta_{23} > 0.5$, i.e. to the octant of the mixing angle. Actually, the degeneracy is not exact due to matter effects, which can manifest as a preference for one octant over the other in atmospheric experiments and, to a smaller extent, in long-baseline experiments too. Both kinds of experiments are also sensitive to electron (anti)neutrino appearance, which is directly sensitive to $\sin^2 \theta_{23}$. From these two facts, global fits to oscillation data used to show a preference for the second octant ($\sin^2\theta_{23} > 0.5$). Nevertheless, recent Super-Kamiokande atmospheric results seem to have shifted the preference to the first octant~\cite{newSkatm}. 

The second unknown parameter is the CP phase $\delta_\mathrm{CP}$, which induces opposite shifts in the electron neutrino and antineutrino appearance probabilities, $\nu_\mu \rightarrow \nu_e$ and $\bar{\nu}_\mu \rightarrow \bar{\nu}_e$.\footnote{Due to neutrino-electron and antineutrino-electron interactions along the path of propagation, these two oscillation probabilities can be different even if $\delta= 0,\, \pi$ and CP is preserved in the lepton sector.} Consequently, it is accessible in long-baseline and atmospheric neutrino experiments. Currently, the best measurements of $\delta_\mathrm{CP}$ come from long-baseline (hundreds of kilometers) neutrino oscillation experiments with ${\sim}$ few GeV muon-neutrino beams.\footnote{Measurements of atmospheric neutrinos with Super-Kamiokande have demonstrated mild sensitivity to $\delta_\mathrm{CP}$ to date~\cite{Super-Kamiokande:2019gzr}.}
In 2019, the Tokai to Kamioka (T2K) experiment announced a measurement of CP violation at nearly $3\sigma$ confidence~\cite{T2K:2019bcf}. However, with more recent data included, this confidence has decreased to $2\sigma$ confidence~\cite{T2K:2021naz}. Concurrently, the NuMI Off-Axis $\nu_e$ Appearance (NOvA) experiment has demonstrated capability of measuring long-baseline $\nu_\mu \to \nu_e$ oscillation probabilities. NOvA results to date have no strong preference for any specific value of $\delta_\mathrm{CP}$~\cite{NOvA:2021nfi}. Combined analyses of these long-baseline experiments (and those including other oscillation results) find no strong preference for either CP conservation or violation~\cite{deSalas:2020pgw,Esteban:2020cvm,Kelly:2020fkv} and an accurate determination of $\delta_\mathrm{CP}$ is not possible yet. Moreover, whereas for inverted ordering both experiments point to similar values close to $\delta_\mathrm{CP}\sim 1.5\pi$, their preferred values of $\delta_\mathrm{CP}$ for normal ordering show an important disagreement. 

This fact is related to the last unknown of the picture: the mass ordering, i.e. the sign of $\Delta m^2_{31}$.  Atmospheric neutrino experiments are sensitive to the mass ordering through matter effects\footnote{The transition probability for propagation in matter has a resonance when the matter potential is equal to a given combination of the ($\Delta m^2_{31}$) mass splitting and mixing angle (MSW effect). For the resonance to appear in the (anti)neutrino transition probability, $\Delta m^2_{31}$ must be positive (negative). Therefore, the observation of a resonance in the (anti)neutrino transition probability allows to unveil the mass ordering.} and both Super-Kamiokande and DeepCore prefer normal ordering. Concerning long-baseline experiments,  T2K and NOvA are modestly sensitive to the ordering, also due to matter effects, and the individual analyses show a small preference for normal ordering too. Nonetheless, due to the above mentioned disagreement in the determination of $\delta_\mathrm{CP}$, which happens only for normal ordering, the overall preference for normal ordering over inverted ordering is penalised, though still present. 

\begin{figure}[t!]
\includegraphics[width = 0.98\textwidth]{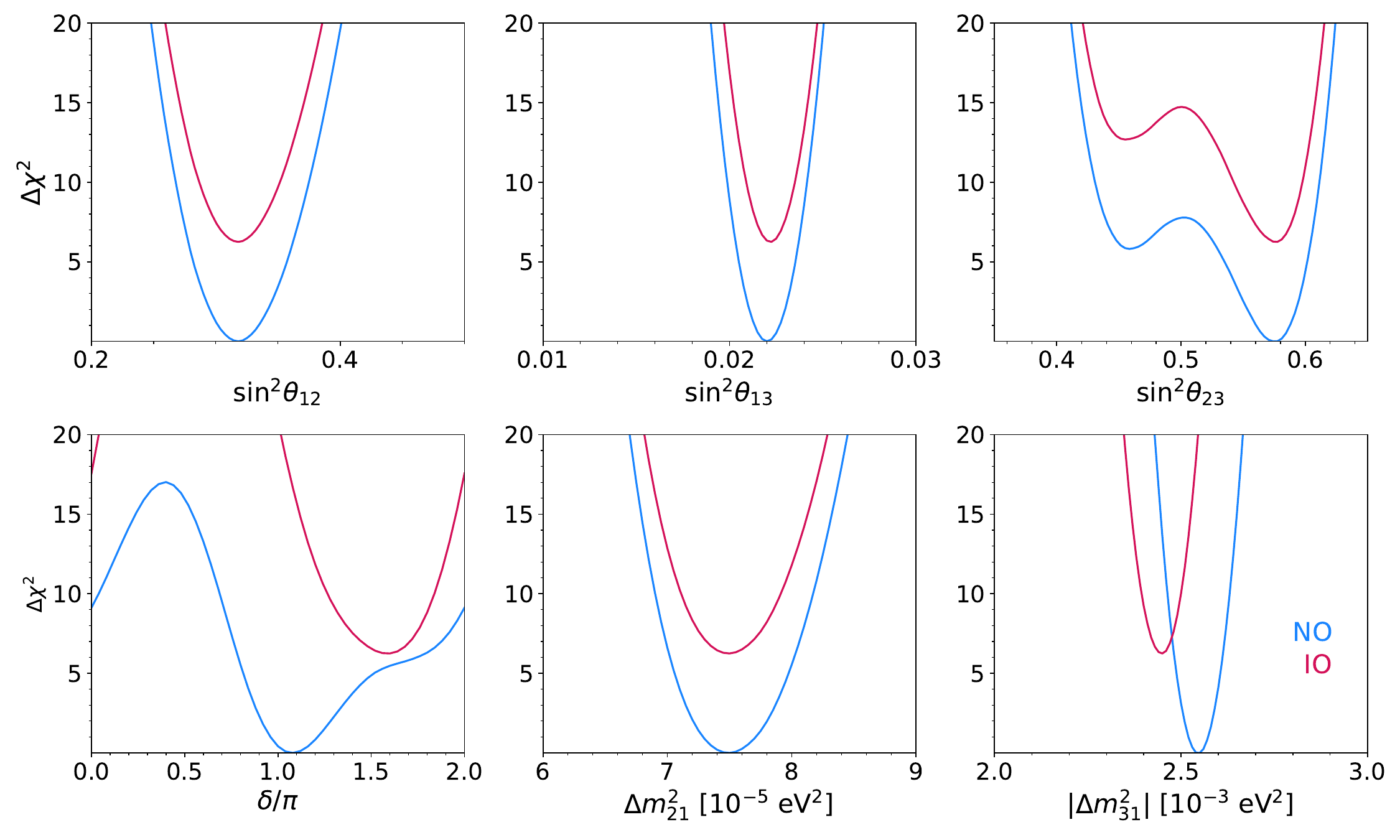}
\caption{$\Delta \chi ^2$ - profiles summarizing the status of the determination of the oscillation parameters. Blue and pink lines correspond to normal and inverted mass ordering, respectively. Adapted from~\cite{deSalas:2020pgw}.}
\label{fig:chip2020}
\end{figure}

In any case, determinations of the octant of $\theta_{23}$, $\delta_\mathrm{CP}$ and the mass ordering remain inconclusive to date and they will be targeted by next-generation experiments and forthcoming efforts to perform global fits to neutrino data. As for the current status~\cite{deSalas:2020pgw}, the best fit values for the oscillation parameters, together with the confidence intervals, are summarized in Table \ref{tab:sum-2020}. For completeness, the $\Delta \chi^2$ profiles for each of the parameters are shown in Figure \ref{fig:chip2020} for normal and inverted ordering, normalised to the overall minimum of the fit. 

 Future challenges to address the neutrino oscillation unknowns are briefly described in what follows. Large underground and underwater neutrino observatories, such as Hyper-Kamiokande, DUNE and the future extensions of KM3Net and IceCube are currently under construction or design. The approved Deep Underground Neutrino Experiment (DUNE)~\cite{DUNE:2020ypp,DUNE:2020lwj} in the USA will use accelerator neutrinos to study neutrino oscillations with unprecedented accuracy. Similarly, the Hyper-Kamiokande~\cite{Hyper-Kamiokande:2018ofw} detector in Japan will also be exposed to accelerator neutrinos.

In the longer term, DUNE will provide far better mass ordering determination, thanks to its 1300~km long baseline between neutrino production and detection locations. After 2 (10) years of operation, DUNE alone will determine the neutrino mass ordering with a significance of at least 5 (10) standard deviations. Likewise, the determination of the CP phase will be a specific goal for the next-generation experiments DUNE~\cite{DUNE:2020ypp,Kelly:2019itm}, Hyper-Kamiokande and IsoDAR~\cite{Bungau:2012ys}. If operated as planned, $\delta_\mathrm{CP}$ will be measured at high confidence in the next decade or so.

\begin{table}[t!]\centering
\renewcommand{\arraystretch}{1.4}
\begin{tabular}{c|ccc}
\hline
parameter & best fit $\pm$ $1\sigma$ & \hphantom{x} 2$\sigma$ range \hphantom{x} & \hphantom{x} 3$\sigma$ range \hphantom{x}
\\ \hline \hline
$\Delta m^2_{21} [10^{-5}$eV$^2$]  &  $7.50^{+0.22}_{-0.20}$  &  7.12--7.93  &  6.94--8.14 \\[3mm]
$|\Delta m^2_{31}| [10^{-3}$eV$^2$] (NO)  &  $2.55^{+0.02}_{-0.03}$  &  2.49--2.60  &  2.47--2.63 \\
$|\Delta m^2_{31}| [10^{-3}$eV$^2$] (IO)  &  $2.45^{+0.02}_{-0.03}$  &  2.39--2.50  &  2.37--2.53   \\[3mm]
$\sin^2\theta_{12} / 10^{-1}$         &  $3.18\pm0.16$  &  2.86--3.52  &  2.71--3.69 \\[3mm]
$\sin^2\theta_{23} / 10^{-1}$       (NO)  &  $5.74\pm0.14$  &  5.41--5.99  &  4.34--6.10 \\
$\sin^2\theta_{23} / 10^{-1}$       (IO)  &  $5.78^{+0.10}_{-0.17}$  &  5.41--5.98  &  4.33--6.08  \\[3mm]
$\sin^2\theta_{13} / 10^{-2}$       (NO)  &  $2.200^{+0.069}_{-0.062}$  &  2.069--2.337  &  2.000--2.405   d \\
$\sin^2\theta_{13} / 10^{-2}$       (IO)  &  $2.225^{+0.064}_{-0.070}$  &  2.086--2.356  &  2.018--2.424   \\[3mm]
$\delta/\pi$                        (NO)  &  $1.08^{+0.13}_{-0.12}$  &  0.84--1.42  &  0.71--1.99 \\
$\delta/\pi$                        (IO)  &  $1.58^{+0.15}_{-0.16}$  &  1.26--1.85  &  1.11--1.96  \\

\hline
\end{tabular}
\caption{
Neutrino oscillation parameters summary determined from the frequentist global analysis in~\cite{deSalas:2020pgw} (the Bayesian analysis also discussed in the same Ref. gives very similar constraints).
The intervals quoted for inverted ordering refer to the local minimum for this neutrino mass ordering. See also~\cite{Capozzi:2021fjo,Gonzalez-Garcia:2021dve,Bergstrom:2015rba} for similar analyses in both Bayesian and frequentist frameworks.}
\label{tab:sum-2020}
\end{table}

\subsubsection[\texorpdfstring{$m_\beta$}{mbeta} in tritium end-point experiments]{$\mathbf{m}_{\boldsymbol\beta}$ in tritium end-point experiments\label{sec:expbeta}}
The experimental challenge of an endpoint measurement is to acquire sufficient statistics in the endpoint region of the spectrum.
Due to the spectral shape, the neutrino mass sensitivity scales approximately as the fourth root of the number of events.
The energy resolution must be excellent to not smear out the spectral distortion, and backgrounds must be low so as not to bias the neutrino mass extraction~\cite{Formaggio:2021nfz}.

Tritium has been the workhorse isotope for direct neutrino mass measurement for over seven decades~\cite{Formaggio:2021nfz}.
Tritium has an 18.6\,keV endpoint energy and a 12.32\,yr half-life.
The decay of tritium is super-allowed,\footnote{A super-allowed decay (also known as Fermi decay) is a transition with the two emitted particles having anti-parallel spin ($S=0$), implying conservation of the total angular momentum ($\Delta I = 0$).} resulting in a spectral shape that is exactly calculable with no theoretical uncertainty.
Even for an excellent candidate isotope like tritium, the branching ratio for decays into the last eV of the spectrum is only $2 \times 10^{-13}$.

The current state-of-the art in the field is the KATRIN experiment~\cite{KATRIN:2005fny, KATRIN:2022ayy}.
KATRIN employs a magnetic adiabatic collimation with electrostatic (MAC-E) filter spectrometer with an intense windowless gaseous molecular tritium source.
KATRIN released their first results from tritium commissioning data in 2019~\cite{KATRIN:2019yun} and have since achieved the first sub-eV neutrino mass limit, $m_\beta < 0.8$\,eV/$c^2$~\cite{KATRIN:2021uub}.
With the five (calendar) year run fully underway, KATRIN will continue to drive down towards its ultimate design sensitivity of 0.2\,eV/$c^2$.

Project 8 is a next-generation experimental concept targeting sensitivity down to 40\,meV/$c^2$~\cite{Project8:2017nal}, covering the inverted mass ordering allowed region.
The collaboration is developing the frequency-based Cyclotron Radiation Emission Spectroscopy (CRES) technique~\cite{Monreal:2009za}, which relies on measuring the $\sim 1$\,fW of radiated cyclotron power (at 18\,keV and 1\,T field strength) to extract the decay electron energy with high precision.
This will be combined with an atomic source to circumvent the systematic associated with molecular final state uncertainty, which presently limit at the $m_\beta \sim 0.1$\,eV/$c^2$ level.
Project 8 has demonstrated CRES both on the mono-energetic calibration electrons from $^{83m}$Kr~\cite{Project8:2014ivu} and at low-statistics on the last few keV of the tritium spectrum~\cite{Project8:2022hun,Project8:2023jkj}.
The coming years are dedicated to R\&D demonstrating the key technologies for scaling the technique and establishing the atomic production, cooling, and trapping in preparation for a conceptual design of the ultimate experiment~\cite{Project8:2022wqh}.

An alternate isotope under investigation is $^{163}$Ho, which decays via electron capture with a Q-value of 2833(34)\,eV~\cite{ECHo:2015qgh}.
A neutrino is produced in the final state, the mass of which modifies the available phase space akin to beta decay, and thus a distortion is observed near the endpoint.
Microcalorimeters are the chosen technology to study this decay process, which collect all the energy avoiding final state effects, but at the cost of some pileup background.
The ECHo~\cite{Gastaldo:2017edk} and HoLMES~\cite{HOLMES:2019ykt} collaborations are pursuing multiplexed arrays to validate this concept towards a next-generation experiment.

\subsubsection[\texorpdfstring{$m_{\beta\beta}$}{mBetaBeta} in \texorpdfstring{$0\nu2\beta$}{0Nu2Beta} experiments]{$m_{\beta\beta}$ in $0\nu2\beta$ experiments\label{sec:expbetabeta}}

The experimental challenge of neutrinoless double beta decay searches is to measure the mono-energetic peak at $Q_{\beta\beta}$ to high significance.
With every successive generation of experiments, increased exposure and decreased backgrounds are required to optimize the sensitivity.
In the extreme low-background limit, sensitivity scales linearly with exposure, whereas at even modest background levels sensitivity only scales with the square root of exposure~\cite{Agostini:2017jim}.

The first experimental constraint is that the isotope of interest must be compatible with the detector technology employed.
The $2\nu2\beta$ process has been measured directly in only nine isotopes with half-lives of $\sim 10^{20}$\,yrs~\cite{Barabash:2020nck}; competitive sensitivity to the rarer $0\nu2\beta$ process requires an integrated source as detector configuration.
The experiments discussed here employ between 10\,kg (for smallest current-generation) and 10\,tonnes (for largest next-generation) of the isotope of interest, typically enriched to $\sim 90\%$.
Reduction of backgrounds is also critical, with particular attention paid to cleanliness of materials and employing self-shielding where possible.
Finally the energy resolution plays an important role in background rejection by reducing the region of interest thereby limiting the impact of the surrounding background spectral features.
Additionally the $2\nu2\beta$ may become an irreducible background for technologies with poorer energy resolution.

The best published limits to date exceed a half-life of $10^{26}$\,yrs, with both KamLAND-Zen and GERDA crossing that threshold.
KamLAND-Zen employs a large enriched $^{136}$Xe-loaded liquid scintillator detector to achieve the most stringent limit\footnote{The interval of values in the quoted upper limit reflects the 
uncertainty on the calculation of the nuclear matrix elements, see Sec.~\ref{sec:0nu2b}.} on $m_{\beta\beta}$ of $36 -  156$\,\meV~\cite{KamLAND-Zen:2022tow}.
GERDA employs an array of enriched point contact $^{76}$Ge detectors to achieve the greatest half life sensitivity and limit, $T_{1/2} > 1.8 \times 10^{26}$\,yr~\cite{GERDA:2020xhi}.
Other experiments crossing the $10^{25}$\,yr sensitivity threshold include CUORE operating a bolometer array measuring the decay of $^{130}$Te~\cite{CUORE:2019yfd}, EXO-200 operating a single-phase enriched $^{136}$Xe time projection chamber (TPC)~\cite{EXO:2017poz}, and the \textsc{Majorana Demonstrator} operating an enriched $^{76}$Ge detector array~\cite{Majorana:2022udl}.

Several additional experimental efforts are in the commissioning phase or taking data targeting sensitivity beyond $10^{26}$\,yrs.
KamLAND-Zen 800, an upgrade of the previous detector, has been taking data since early 2019~\cite{KamLAND-Zen:2021aha}.
LEGEND-200, a successor to the GERDA and \textsc{Majorana} $^{76}$Ge programs, is commissioning a new detector array to begin operation in spring 2022~\cite{LEGEND:2021bnm}.
SNO+, a liquid scintillator detector utilizing the former SNO detector will begin loading the $^{130}$Te compound later in 2022~\cite{SNO:2021xpa}.

The next generation experiments target a sensitivity covering the inverted mass ordering allowed range, with half-life sensitivities up to and exceeding $10^{28}$\,yrs~\cite{Agostini:2021kba}.
These ``tonne-scale'' experiments are necessarily built on the success of current generation results.
CUPID exchanges the TeO$_2$ bolometers of CUORE for scintillating Li$_2$MoO$_4$ ($^{100}$Mo-enriched) bolometers with dual signal readout~\cite{CUPID:2019imh}.
LEGEND-1000 takes a modular upgrade path allowing a staged deployment of larger-mass detectors in an enhanced underground liquid argon active veto~\cite{LEGEND:2021bnm}.
nEXO scales to 5\,t of xenon, taking advantage of the self-shielding of the monolithic xenon volume and improved material cleanliness~\cite{nEXO:2021ujk}.

Beyond next generation concepts are being pursued to chart the future sensitivity into the normal mass ordering band.
Drastically scaling the sensitive exposure of the experiments (kton-scale detectors) is critical to achieve order(s)-of-magnitude improvement in the half-life reach~\cite{Avasthi:2021lgy, Theia:2019non}.
Optimizing the discovery potential requires incredible suppression of all backgrounds, with daughter-ion identification a key enabling technology~\cite{Rivilla:2020cvm}.

\subsubsection{Constraints on neutrino properties beyond the standard paradigm} \label{sec:beyond3}

\paragraph{Laboratory constraints on sterile neutrinos}

Several anomalous experimental results in the last decades have motivated the search for sterile neutrinos with masses in the eV range in the 3+1 framework. For instance, the observation of a deficit in the flux of electron neutrinos from radioactive sources of  $^{51}$Cr and $^{37}$Ar at GALLEX and SAGE~\cite{Abdurashitov:2005tb, Kaether:2010ag,SAGE:1998fvr} has been recently confirmed by the BEST experiment~\cite{Barinov:2021asz} and can be interpreted as due to the existence of a sterile neutrino with a mass of the order of $\mathcal{O}$(1-10 eV$^2$). These results are normally referred to as the Gallium anomaly. A mismatch between the observed reactor antineutrino flux and the prediction from several theoretical models has also be regarded as an indication of electron antineutrino disappearance due to oscillations involving eV sterile neutrinos. This is the so-called reactor antineutrino anomaly (RAA)~\cite{Mention:2011rk}. Nonetheless, recent studies have shown that, depending on the theoretical input, there is no such anomaly concerning reactor neutrinos~\cite{Giunti:2021kab}.

Note that the amplitude of the oscillations involved in both the RAA and the Gallium anomaly is set by the parameter $|U_{e4}|^2$ where $U$ is the extended mixing matrix. However, $|U_{e4}|^2$ is already constrained by solar experiments~\cite{Giunti:2021iti}. Aiming to further explore the 3+1 picture, several very-short baseline reactor experiments have searched for electron antineutrino disappearance. These include STEREO~\cite{STEREO:2019ztb}, PROSPECT~\cite{PROSPECT:2020sxr}, NEOS~\cite{NEOS:2016wee, RENO:2020hva}, DANSS~\cite{Danilov:2021oop} and Neutrino-4~\cite{Serebrov:2020kmd}. Over the last years, there have been claims of a preference for the light sterile neutrino hypothesis over the three neutrino picture. However, not only they do not agree on their results but also the preferred regions in parameter space reported were excluded by other experiments. As a consequence, the statistical methods required and the possible overestimation of the significance of the results have been the subject of heated discussion, see~\cite{Giunti:2021iti,PROSPECT:2020raz, Serebrov:2020yvp, Danilov:2020rax}. 

These are not the only anomalies motivating the 3+1 scenario. Two short-baseline reactor experiments, LSND~\cite{LSND:2001aii} and MiniBooNE~\cite{MiniBooNE:2020pnu}, reported an excess of $\nu_e$-like (and $\bar{\nu}_e$-like) events which, if addressed as arising from $\nu_\mu \to \nu_e$ (and $\bar{\nu}_\mu \rightarrow \bar{\nu}_e$) oscillations, would be pointing to the existence of a sterile neutrino with a mass $m_4^2 \sim $ 0.1-1 $ {\rm eV}^2 $ and a large mixing with the active states. In particular, both experiments are sensitive to the product $|U_{\mu 4}|^2|U_{e4}|^2$, which would be required to be considerably large. Nonetheless, the parameter space involved has been scrutinised by experiments looking at muon (anti)neutrino disappearance, which are sensitive to $|U_{\mu 4}|^2$. Long baseline accelerator experiments, for instance MINOS/MINOS+~\cite{MINOS:2016viw, MINOS:2017cae}, have set strong limits on $|U_{\mu 4}|$. Similarly, atmospheric neutrinos studied at IceCube~\cite{IceCube:2017ivd} have also constrained active-sterile mixing. \footnote{Note, however, that in their latest analysis~\cite{IceCube:2020phf}, a very mild preference for non-zero mixing is found.} It is important to remark that the study of muon (anti)neutrino disappearance has also allowed to constrain $|U_{\tau 4}|^2$.

Summing up, the 3+1 scenario is in crisis due to tensions between the existing datasets~\cite{Dentler:2018sju,Gariazzo:2017fdh}. On the one hand, muon (anti)neutrino datasets constrain $|U_{\mu 4}|^2$, and so do electron (anti)neutrino datasets for $|U_{e4}|^2$. It follows from those limits, that the product $|U_{\mu4}|^2|U_{e4}|^2$ is not large enough to explain the electron neutrino appearance anomaly reported by LSND and MiniBooNE. This incompatibility between the three datasets is often referred to as the `appearance - disappearance tension' and strongly undermines the sterile neutrino hypothesis. Nonetheless, important efforts are still being made by the scientific community to shed light in this issue, like the Short Baseline Neutrino Program, which includes MicroBooNE. \footnote{MicroBooNE recently presented their first results which were found to be consistent with the nominal electron neutrino interactions expected~\cite{MicroBooNE:2021rmx}. Its impact on the fits to the 3+1 scenario has been explored in~\cite{Arguelles:2021meu, MiniBooNE:2022emn}.}

We now turn to the alternative case of heavy ``sterile'' neutrinos. This is the preferred scenario in view of the seesaw picture of neutrino mass generation~\cite{Schechter:1980gr}. It is important to note that some of the experimental bounds derived on active-sterile mixing, though generally discussed for light sterile neutrinos with masses of the order of the eV, can also constrain heavier ones. However, the physical picture 
for neutrinos with masses so large that they can not be produced in the source is completely different. In such case, their existence would be inferred through the non-unitarity of the $3\times3$ lepton mixing matrix~\cite{Escrihuela:2015wra}. 
Strong bounds on this scenario have been derived from global analysis of  neutrino oscillation data~\cite{Escrihuela:2016ube,Blennow:2016jkn,Forero:2021azc}. The parameter space of heavy sterile neutrinos has been more or less severely constrained via searches for rare leptons and mesons decay, as well as via hadron collisions. A summary of experimental constraints can be found in, e.g.,~\cite{Helo:2011yg,Cvetic:2019shl,Atre:2009rg,Drewes:2015iva,Kusenko:2004qc}.

Finally, we refer to~\cite{Hagstotz:2020ukm,Mertens:2018vuu,KATRIN:2020dpx,KATRIN:2022ith,Mertens:2015ila} for results from searches of sterile neutrinos in beta-decay and to~\cite{Hagstotz:2020ukm,Bolton:2019pcu,Jha:2021oxl,Huang:2019qvq} for studies of the sensitivity of neutrinoless double-beta decay experiments to sterile neutrinos.

\paragraph{Laboratory constraints on non-standard neutrino interactions}
To date, modulo a couple of possible hints~\cite{Denton:2020uda,Chatterjee:2020kkm}, oscillation data provide no strong preference for additional neutral-current neutrino interactions~\cite{Esteban:2018ppq,IceCube:2022ubv}. Similar to prospects for CP-violating measurements, next-generation oscillation experiments, especially DUNE~\cite{deGouvea:2015ndi,Coloma:2015kiu}, will improve the current picture drastically with excellent discovery potential.

The COHERENT collaboration, with their first demonstrated measurements of CEvNS~\cite{COHERENT:2017ipa,COHERENT:2020iec}, has gone on to place complementary constraints on NSI to those from oscillations~\cite{Esteban:2018ppq,Coloma:2019mbs}. With more measurements of CEvNS (including those on different target nuclei) expected on the horizon, these constraints will continue to improve. We refer the reader to the dedicated Snowmass white paper on non-standard neutrino interactions~\cite{Berryman:2022} for further discussions.

%% file: synergy.tex
\section{Synergy between cosmology and laboratory experiments \label{sec:synergy}}

In the preceding sections, we have reviewed the constraints that cosmological and laboratory probes can
provide, each on its own regard, about neutrino properties. In this section, we aim at giving an insight
on how much can be gained by combining the information coming from different observational/experimental channels.
We will first concentrate on what we dub ``concordance'' scenarios, in which the signals from the different experiments
are in agreement, when interpreted under a minimal set of assumptions (to be better specified in the following).
We will see that in this case one can expect, for example, to have precise measurements of the neutrino masses and/or information
on the neutrino hierarchy. Then we will discuss the, perhaps more interesting, case of signals that cannot be interpreted
consistently\hskip1pt\footnote{It should be understood that, in the following discussion, we are implicitly assuming that experimental 
systematics are under control and that tensions should be ascribed to a failure in the theoretical modeling of the observed phenomena.}
unless one gives up some of the usual assumptions about, e.g., the standard cosmological model, or the mechanism behind $0\nu2\beta$ decay,
just to make two examples. This would signal the presence of new physics -- or, to put it more precisely, of \textit{unexpected} new physics.

\subsection{General considerations}

\begin{figure}
	\centering
	\includegraphics[width=0.95\textwidth]{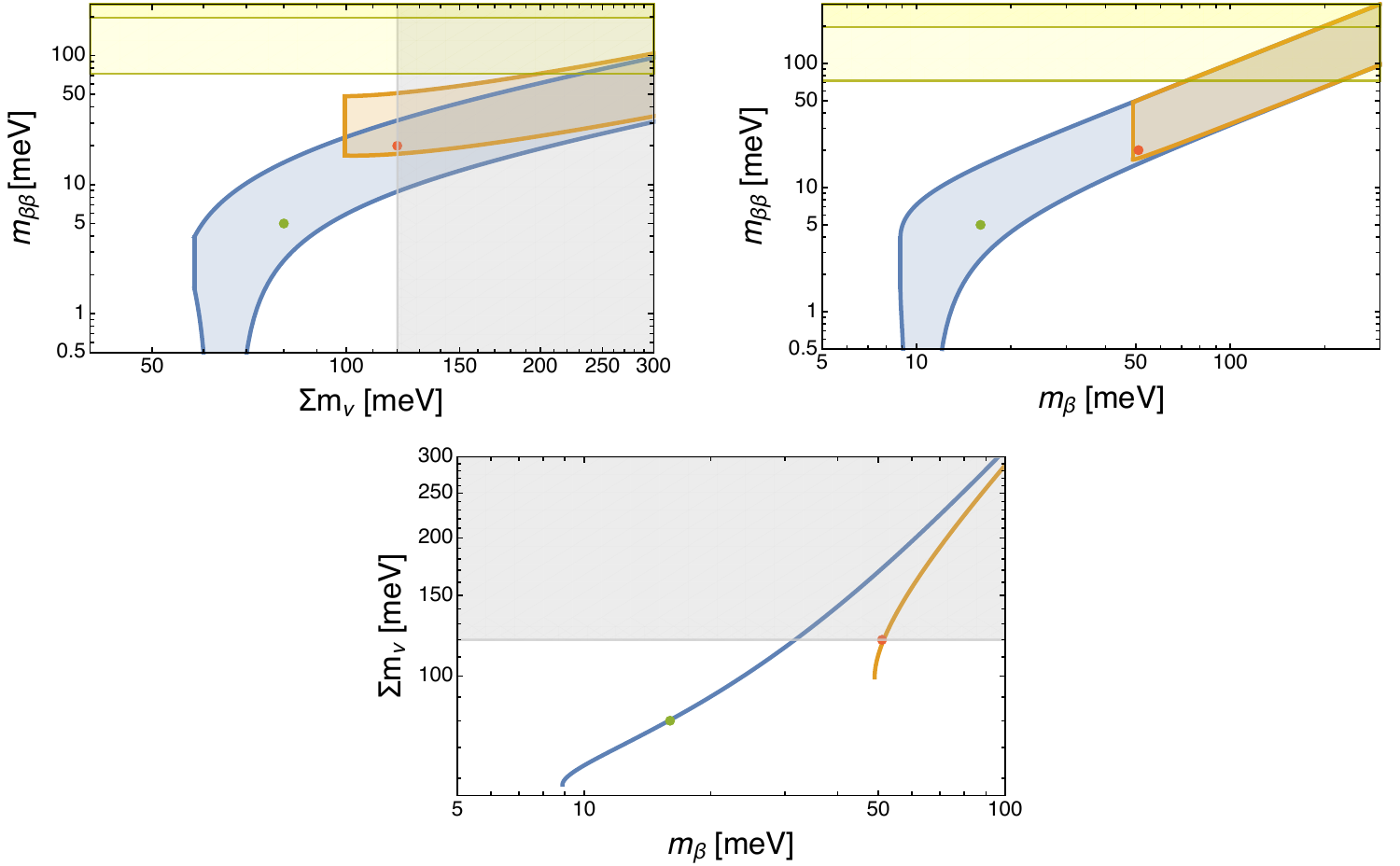}
	\caption{Neutrino mass parameters $\Sigma m_\nu$ (Eq.~\ref{eq:summnu}), $m_{\beta\beta}$ (Eq.~\ref{eq:mbetabeta}) and $m_\beta$ (Eq.~\ref{eq:mbeta}) for NO (blue) and IO (orange). The shaded gray and yellow regions are excluded by cosmological observations (Planck + BAO~\cite{Planck:2018vyg}) and $0\nu2\beta$ searches (KamLAND-Zen~\cite{KamLAND-Zen:2016pfg}), respectively. Note that these constraints are in both cases model-dependent. The red and green dots correspond to the two ``concordance'' scenarios discussed in Sec.~\ref{sec:conc}. The widening of the contours in the panels involving the effective Majorana mass $m_{\beta\beta}$ is due to the uncertainty on the Majorana phases. The oscillation parameters are fixed to their best-fit values from Ref.~\cite{deSalas:2020pgw}.}
	\label{fig:numassparams}
\end{figure}

We start by making some general considerations on the sensitivity of next-generation experiments, and their implication for neutrino properties. For
the following discussion, the reader might want to refer to Fig.~\ref{fig:numassparams}, showing the possible combinations of the neutrino absolute mass parameters
introduced in the previous sections. In more detail, the figure shows the possible values of $\Sigma m_\nu$, $m_\beta$ and $m_{\beta\beta}$ once the oscillation parameters (mass
squared differences and mixing angles) are fixed to their best-fit values. Note that taking into account the uncertainty associated to the oscillation parameters
would slightly widen the curves shown in Fig.~\ref{fig:numassparams}; this is however not relevant given the qualitative nature of the following discussion.
The widening of the curves in the plots involving $m_{\beta\beta}$ is instead due to the uncertainty in the Majorana phases. In the figure
we also show the constraints $\Sigma m_\nu < 0.120\,\meV$ (95\% CL) from Planck + BAO~\cite{Planck:2018vyg} and $m_{\beta\beta} < 61 - 165 \,\meV$ (90\% CL)\footnote{For the sake of coherence, in Fig.~\ref{fig:numassparams} we show the corresponding 95\% CL, calculated assuming a Gaussian distribution.} from KamLAND-Zen, where the interval reflects the 
uncertainty on the calculation of the nuclear matrix elements. Finally, we indicate with dots two models that might lead to different detection scenarios discussed further below.

Next-generation cosmological observations are expected to provide a statistically significant measurement of the absolute scale of neutrino masses, even in the worst case of normal ordering and vanishing mass for the lightest eigenstate (yielding $\Sigma m_\nu \simeq 60\,\meV$). In particular, a combination of satellite observations of the CMB anisotropies at large scales with satellite measurements of galaxy clustering and cosmic shear will allow for a $4\sigma $ detection of $\Sigma m_\nu = 60\,\meV$, i.e., $\sigma(\Sigma m_\nu) = 15\,\meV $ in the framework of the simple one-parameter extension $\Lambda$CDM+$\Sigma m_\nu$ model~\cite{Brinckmann:2018owf}. Considering also CMB observations at small-scales
from the ground, or intensity mapping data from radio telescopes, will further improve the sensitivity. In particular, adding either the small-scale CMB or intensity mapping datasets will bring the sensitivity down to $\sigma(\Sigma m_\nu) = 12\,\meV $, while adding both would yield $\sigma(\Sigma m_\nu) = 8\,\mathrm{meV} $, corresponding to a $7.5\sigma$ detection of $\Sigma m_\nu = 60\,\meV$~\cite{Brinckmann:2018owf}. As mentioned in Sec. \ref{sec:mnu_from_cosmo}, the sensitivity can degrade in extended cosmological models. Brinckmann et al.~\cite{Brinckmann:2018owf} consider some simple extensions of the base $\Lambda$CDM model and find that the worst sensitivity is obtained
for a time-varying dark energy equation of state. Even in this case, however, the combination of datasets described above will allow to reach a detection
at the $>3\sigma$ level. The implication of these numbers is that a non-detection of neutrino masses in next-generation cosmological experiments
will imply, by itself, a failure in our theoretical assumptions. This failure might be related to our modeling of the cosmological dark sector, albeit as we have seen simple modifications to $\Lambda$CDM are not necessarily able to explain the absence of a cosmological signal of neutrino masses; or it could be
more directly related to the neutrino sector, requiring for example the existence of nonstandard interactions.

Another interesting issue to address is whether cosmological observations will be able to pinpoint the mass ordering. The analyses of, e.g.,~\cite{Gerbino:2016ehw,Archidiacono:2020dvx,Mahony:2019fyb,DeSalas:2018rby,Heavens:2018adv,Gariazzo:2018pei,Schwetz:2017fey,Vagnozzi:2017ovm,Gariazzo:2023joe}
shows that the sensitivity of next-generation experiments on the ordering mostly comes from the fact that the region $\Sigma m_\nu <100\,\mathrm{meV}$ is only allowed 
in the case of normal ordering. This creates an asymmetry between the two orderings: the determination of the normal hierarchy would be a byproduct of a measurement of $\Sigma m_\nu <100\,\mathrm{meV}$, with a significance that becomes stronger the closer $\Sigma m_\nu$ is to $60\,\mathrm{meV}$, while if $\Sigma m_\nu >100\,\mathrm{meV}$ cosmology cannot discriminate between normal and inverted ordering. 

In any case, the neutrino mass ordering will be determined by neutrino oscillation experiments without relying on any other experimental facility, thanks to matter effects and/or precise determination of the electron antineutrino survival probability~\cite{DeSalas:2018rby}. As an example, we expect DUNE to provide a $5\sigma$ detection of the neutrino mass ordering with 7 years of data~\cite{DUNE:2020ypp}. 

Next-generation $0\nu2\beta$ searches, exploiting different isotopes, aim at reaching sensitivities to the half-life $T_{1/2}$ in the $10^{27}-10^{28}\,\mathrm{yrs}$ ballpark. An observation of $0\nu2\beta$ decay would imply, through the black-box theorem~\cite{Schechter:1981cv}, that neutrinos are Majorana particles. Inputs from nuclear theory are required to translate $T_{1/2}$, and the corresponding sensitivity, into a Majorana mass $m_{\beta\beta}$, see Sec. \ref{sec:0nu2b}. This conversion, and in particular the computation of reliable nuclear matrix elements, is thus nontrivial and affected by theoretical uncertainties.
In spite of this, an approach combining experiments using different isotopes is expected to cover all the region in parameter space spanned by the inverted ordering ($m_{\beta\beta} > 18 \,\mathrm{meV}$), as well as a significant fraction of the parameter space for normal ordering~\cite{Agostini:2021kba}. Exploring the smallest values of $m_{\beta\beta}$ allowed by normal ordering will however
likely require an improvement in sensitivity only achievable by beyond-next-generation experiments. Concerning theoretical uncertainties, significant advances have been made in the last few years in reducing the nuclear physics uncertainties, and further advances would certainly help in increasing the constraining power of $0\nu2\beta$ searches.

As far as kinematic measurements are concerned, the currently running KATRIN experiment is expected to reach its sensitivity threshold of $m_\beta<0.2\,\mathrm{eV}$ (90\% C.L.) or a $5\sigma$ discovery of $m_\beta=0.35\,\mathrm{eV}$ in the next few years (i.e., after $\sim 5$ years of data). If taken at face value, current cosmological bounds are already a factor of $5-6$ more stringent than the expected KATRIN sensitivity. Therefore, as we shall comment in more details in the next sections, we do not expect an observed signal from kinematic measurements if the the total mass sum is confirmed to be $\leq 0.12\,\mathrm{eV}$. Next-generation experiments, such as Project8, aims to reach a sensitivity of $m_\beta<40\,\meV$ (90\% C.L.), fully covering the range allowed in inverted hierarchy. With this sensitivity, we might expect a detection from kinematic experiments in some of the scenarios described in the following sections.

To conclude, we would like to acknowledge that a more precise assessment of the viability of the scenarios to be discussed in the next sections would obviously require a more detailed statistical analysis, taking into account the sensitivities of next generation cosmological, $0\nu2\beta$ and oscillation experiments.

\subsection{Concordance scenarios \label{sec:conc}} 
Let us start by discussing the possibility that all observations can be interpreted in terms of three families of active neutrinos with only weak interactions.
Whenever needed for the discussion, we will also assume that the $\Lambda$CDM model correctly describe the evolution of the Universe, and that the amplitude for neutrinoless double $\beta$ decay is dominated by the mass mechanism, i.e., by the exchange of massive neutrino states. In this framework, these are some of the possible scenarios:
\begin{enumerate}
    \item Cosmological observations and $0\nu2\beta$ searches allow to measure nonzero $\Sigma m_\nu$ and $m_{\beta\beta}$, respectively.
    The two determinations are consistent once information on the mixing matrix and mass differences from flavor oscillation experiments is taken into account. A model example of
    this scenario is indicated with a red dot in Fig.~\ref{fig:numassparams}.
    The observation of $0\nu2\beta$ decay implies that neutrinos are Majorana particles. If cosmological observations indicate, with enough statistical significance, that $\Sigma m_\nu < 100\,\mathrm{meV}$, hierarchy is normal. If instead $\Sigma m_\nu > 100 \mathrm{meV}$ (a possibility presently challenged by cosmological data, but still not excluded), as in the model shown in Fig.~\ref{fig:numassparams}, cosmology alone will not be able to discriminate the hierarchy, but the measured value of $m_{\beta\beta}$ might give a hint in either direction. In this case, the mass hierarchy has to be determined by oscillation experiments. $\Neff$ is measured to be $3.044$ within uncertainty. 
    
    \item A nonzero value $\Sigma m_\nu \ge 60\,\mathrm{meV}$ is inferred from cosmological observations, but no signal is observed in $0\nu2\beta$ decay experiments. This could happen either because i) neutrinos are Dirac, or ii) neutrinos are Majorana, but the Majorana phases arrange to reduce the $0\nu2\beta$ transition amplitude below the sensitivity of $0\nu2\beta$ experiments. The latter possibility only exists in the case of normal ordering, and for small enough values of the mass of the lightest eigenstate, or, equivalently, of $\Sigma m_\nu$. A model example is indicated with a green dot in Fig.~\ref{fig:numassparams}.
    This can be excluded if oscillation experiments determine that the hierarchy is inverted; it can thus be concluded that neutrinos
    are Dirac particles. If instead hierarchy is found to be normal, the Majorana/Dirac nature is undetermined. A caveat to this statement is that, if $\Sigma m_\nu$ is large enough\footnote{The precise value of $\Sigma m_\nu$ that would make this explanation viable depends on the sensitivity of future experiments and might lie in a region already disfavored, albeit not excluded, by present data.}, it might not be possible to bring the $0\nu2\beta$ signal below the detection threshold; in that case we would again be forced to conclude that neutrinos are Dirac particles. $\Neff$ is measured to be $3.044$ within uncertainty. 
    
\end{enumerate}

In both scenarios, a signal in a $\beta$ decay experiment with a $90\%$ sensitivity of $40\,\meV$ is expected in the case of inverted mass ordering, or in the case of normal ordering if $\Sigma m_\nu \gtrsim 140\,\meV$.

\subsection{Beyond concordance} 
Another possibility is that measurements of different neutrino properties will somehow be in tension, i.e., that the complete set of observations cannot
be consistently interpreted in terms of three families of active neutrinos with weak interactions, of the $\Lambda$CDM model for cosmology and of the mass mechanism for $0\nu2\beta$ decay. Such discrepant measurements might point to nonstandard scenarios in either the particle physics or cosmological sector, or in both.
In the following, we discuss a few interesting examples:

\begin{enumerate}
    \item A signal is observed in $0\nu2\beta$ searches (implying that neutrinos are Majorana particles), but there is no detection of $\Sigma m_\nu \ne 0$ from cosmological observations. As explained above, the non-observation of $\Sigma m_\nu$ from cosmology is a problem \emph{per se}. It is thus likely that the problem lies in the fact that $\Lambda$CDM with massive neutrinos is not the right cosmological model. Assuming that the mass mechanism is behind the $0\nu2\beta$ signal, the measurement of the Majorana mass can be used, together with information from oscillation experiments, to infer an allowed range for $\Sigma m_\nu$. A successful alternative cosmological model (for example one involving modifications to gravity) should be in agreement with this value. Alternative cosmological models involving modifications to the neutrino sector, like e.g., models introducing new interactions that might lead to a ``neutrinoless Universe'', can be further tested in the laboratory. The new interaction might itself contribute to the $0\nu2\beta$ signal, or be probed by coherent neutrinos scattering. Provided that the true value of the mass scale is large enough, a measurement of $m_\beta$ from next-generation $\beta$-decay experiments will strengthen evidence for a failure in the $\Lambda$CDM model.
    
    \item Signals are observed from both cosmology and $0\nu2\beta$, but they are discordant, i.e., they lie outside the contours in the $\Sigma m_\nu$, $m_{\beta\beta}$ plane defined by oscillation experiments. The discordance might origin from either the cosmological model or the assumptions on the mechanism behind $0\nu2\beta$.
    Kinematic measurements might be the key to understanding where the incorrect assumptions are more likely to lie. 
    For example, a light sterile neutrino eigenstate could contribute to the amplitude for $0\nu2\beta$, altering significantly the prediction for the Majorana mass~\cite{Giunti:2015kza}. Such a sterile neutrino might also affect cosmological observables, for example giving a detectable contribution to $\Neff$.  Both oscillation 
    experiments and kinematic measurements could be able to confirm or rule out this hypothesis. If instead a heavy sterile neutrino dominates the $0\nu2\beta$ rate~\cite{Bolton:2019pcu,Lopez-Pavon:2012yda}, this could also explain the observed tension with cosmology. In this case, one does not expect a nonstandard value of $\Neff$, although the new state might still leave an imprint on cosmological observables, nor in flavor oscillation experiments.
    In general, a discordant $0\nu2\beta$ signal might appear in several exotic scenarios, since the exchange of massive Majorana neutrinos is not the only element of physics
    beyond the Standard Model that can induce $0\nu2\beta$ decay. Such exotic contributions might arise, to make just a few examples, in left-right symmetric models, R-parity violating super symmetry or in models with leptoquarks. See e.g. the reviews~\cite{Rodejohann:2011mu, Deppisch:2012nb} for discussion of specific models, including the ones just mentioned, as well as~\cite{Bonnet:2012kh} for a systematic study of the possible diagrams contributing to the $0\nu2\beta$ amplitude. 
    
\end{enumerate}

In general, discordant signals between two or more of the different probes of neutrino masses - cosmology, $0\nu2\beta$ and single $\beta$ decays - would support the hypothesis that BSM mechanisms are at play in $0\nu2\beta$ decay, and/or that the currently accepted cosmological model has to be revised.

%% file: conclusions.tex
\section{Conclusions}\label{sec:concl}

In this article, we have stressed the complementarity of cosmological, astrophysical and laboratory probes of neutrino properties. By reviewing the state of the art of the constraints on neutrino properties from a wide range of experimental searches, we have shown that each search is sensitive to specific imprints of neutrino properties on the relevant observable. As such, a complete picture of the neutrino sector can be obtained by combining information from multiple sources. This will not only allow to overcome limitations and systematic effects of individual searches, but will also increase our confidence on the robustness of the constraints. 

In fact, by looking at future prospects in all the aforementioned areas of research, it is clear that significant -- if not transformational -- improvements are expected in the coming decade from both cosmology and laboratory searches. The constraining power of both cosmological and terrestrial searches will be competitive. This is key to allow for cross-checks of the results. If the different probes described in this article are to provide results in agreement with each other, our confidence on a concordance scenario
in the neutrino sector will be strong. On the other hand, if any of those probes is to provide unexpected results that are in tension with the rest, this will signal a need for a revision of our current understanding of the neutrino sector.

We note that none of the two scenarios described above (i.e., concordance versus disagreement) can be explored with one experimental probe. On the contrary, only the synergic approach can enable such a comprehensive investigation of the neutrino sector.

Therefore, with this article, we advocate for pursuing such a synergy in the next decade. Interactions between different communities (cosmological and terrestrial, theorists and instrumentalists) have to be fostered. Networks of researchers with diverse background must be welcomed in order to favor cross-cutting projects. Critical discussions as well as out-of-the-box interpretations of results are needed as a testbed of putative detection claims and/or controversial results. As concrete examples, we strongly encourage the organization (and participation) of extended workshops on topical teams open to different communities (in the vein of the Aspen workshops or GGI workshops); we advise funding agencies on continuing supporting specific calls devoted to multi-disciplinary approaches to topical teams, like e.g., the COST-EU Actions or the ERC-Synergy grants. On a smaller scales, we strongly support individual agreements between institutions to establish dedicated exchange programs for students and researchers, as well as joint PhD programs to create a culture of collaboration across groups and disciplines.

We hope that this manuscript can serve the purpose of bridging between different communities and catalyze joint efforts towards the understanding of neutrinos.